\documentclass[twocolumn,twocolappendix]{aastex631}
\usepackage{amsmath}
\usepackage{verbatim}
\usepackage{longtable}
\usepackage{xspace}
\newcommand{\LCDM}{$\Lambda$CDM\xspace}
\newcommand{\Planck}{\textit{Planck}\xspace}
\newcommand{\XMM}{\textit{XMM}\xspace}

\begin{document}

\title{Simulation-Based Cosmological Mass Calibration of XXL Galaxy Clusters using HSC Weak Lensing}

\author[0000-0002-6724-833X]{Sut-Ieng Tam}
\affiliation{Institute of Physics, National Yang Ming Chiao Tung University, No. 1001, Daxue Rd. East Dist., Hsinchu City, Taiwan}

\author[0000-0002-7196-4822]{Keiichi Umetsu}
\affiliation{Academia Sinica Institute of Astronomy and Astrophysics (ASIAA),
No. 1, Sec. 4, Roosevelt Road, Taipei 106216, Taiwan}

\author{Adam Amara}
\affiliation{School of Mathematics and Physics, University of Surrey, Guildford, Surrey GU2 7XH, UK}

\author{Dominique Eckert}
\affiliation{Department of Astronomy, University of Geneva, ch. d’Ecogia 16, 1290 Versoix, Switzerland}

\author{Manon Regamey}
\affiliation{Department of Astronomy, University of Geneva, ch. d’Ecogia 16, 1290 Versoix, Switzerland}

\author{Nicolas Cerardi}
\affiliation{
Institute of Physics, Laboratory of Astrophysics, École Polytechnique Fédérale de Lausanne (EPFL), Observatoire de Sauverny, Versoix, 1290, Switzerland}

\author[0000-0002-5819-6566]{I-Non Chiu}
\affiliation{Department of Physics, National Cheng Kung University, No.1,
University Road, Tainan City 70101, Taiwan}

\author[0000-0003-0302-0325]{Mauro Sereno}
\affiliation{INAF - Osservatorio di Astrofisica e Scienza dello Spazio di Bologna, via Piero Gobetti 93/3, I-40129 Bologna, Italy}
\affiliation{INFN, Sezione di Bologna, viale Berti Pichat 6/2, I-40127 Bologna, Italy}

\author{Florian Pacaud}
\affiliation{Argelander-Institut für Astronomie (AIfA), Universität Bonn, Aufdem Hügel 71, 53121 Bonn, Germany}

\author{Sunayana Bhargava}
\affiliation{AIM, CEA, CNRS, Université Paris-Saclay, Université Paris Diderot, Sorbonne Paris Cite, F-91191 Gif-sur-Yvette, France}

\author{Christian Garrel}
\affiliation{Max-Planck-Institut für extraterrestrische Physik (MPE), Giessenbachstrasse 1, D-85748 Garching bei München, Germany}
\author{Fabio Gastaldello}
\affiliation{INAF - IASF Milan, via A. Corti 12, I-20133 Milano, Italy}

\author{Elias Koulouridis}
\affiliation{Institute for Astronomy \& Astrophysics, Space Applications \& Remote Sensing, National Observatory of Athens, GR-15236 Palaia Penteli, Greece}

\author{Ben Maughan}
\affiliation{School of Physics, HH Wills Physics Laboratory, Tyndall Avenue, Bristol, BS8 1TL, UK}

\author{Rogério Monteiro-Oliveira}
\affiliation{Observatório Nacional (ON/MCTI), Rua General José Cristino, 77, Rio de Janeiro 20921-400, Brazil}

\author{Marguerite Pierre}
\affiliation{AIM, CEA, CNRS, Université Paris-Saclay, Université Paris Diderot, Sorbonne Paris Cite, F-91191 Gif-sur-Yvette, France}




\begin{abstract}
We present a cosmological analysis of the X-ray-selected galaxy cluster sample from the XXL survey, employing a simulation-based inference (SBI) framework to jointly constrain cosmological parameters and X-ray scaling relations through forward modeling of cluster counts, X-ray observables, and weak-lensing measurements. Our analysis combines X-ray data from the \XMM-XXL survey with shear measurements from the three-year shape catalog of the Hyper Suprime-Cam Subaru Strategic Program. The analysis focuses on the XXL C1 sample, comprising 171 clusters for abundance modeling, a subset of 86 clusters located within the XXL-N region for lensing-based mass calibration, and 162 clusters with X-ray temperature and luminosity measurements used to constrain scaling relations. Using the density-estimation likelihood-free inference (DELFI) algorithm, we construct a forward model with 12 parameters that incorporates the XXL selection function and cluster population modeling and accounts for key systematic effects including cluster miscentering, photometric redshift bias, and mass-dependent weak-lensing bias. Our SBI analysis yields a constraint on the cosmological parameter $S_8 \equiv \sigma_8\sqrt{\Omega_{m}/0.3} = 0.867 \pm 0.063$, with an additional 3\% systematic uncertainty from neural network stochasticity. The result is consistent with \Planck and recent cluster-based measurements. The inferred temperature--mass relation is consistent with self-similar expectations within uncertainties, whereas the luminosity--temperature relation exhibits a slope steeper than the self-similar prediction. From the resulting posterior distribution of the forward model, we derive lensing-calibrated mass estimates for all individual XXL clusters with measured X-ray temperatures or luminosities. These results provide a self-consistent mass calibration for future multi-probe cosmological analyses of the XXL sample.
\end{abstract}


\section{Introduction} \label{sec:intro}
As the most massive gravitationally bound structures formed through hierarchical growth, galaxy clusters are key observational probes of cosmology.
Their abundance as a function of mass and redshift is highly sensitive to the amplitude and growth of matter density fluctuations, providing strong constraints on cosmological parameters such as the matter density parameter, \( \Omega_{m} \), and the amplitude of linear density fluctuations, \( \sigma_8 \). Consequently, the number counts of galaxy clusters across a wide redshift and mass range have been extensively used to constrain cosmological models \citep[e.g.,][]{2016ApJ...832...95D,2017MNRAS.471.1370S,2018A&A...620A..10P,2019ApJ...878...55B,2021PhRvD.103d3522C,2021PhRvL.126n1301T,2023MNRAS.522.1601C,Sunayama2024,Chiu2024,2024A&A...689A.298G, DES2025, 2025arXiv250714285L}.

One of the significant challenges in the field is the so-called ``\( S_8 \) tension,'' where \( S_8 \equiv \sigma_8 (\Omega_{m}/0.3)^\alpha \), and \( \alpha \sim 0.3\text{--}0.5 \) quantifies the degeneracy between \( \Omega_{m} \) and \( \sigma_8 \). This tension arises from discrepancies between the high-redshift cosmic microwave background (CMB) constraints from the \Planck satellite \citep{Planck2018VI} and low-redshift probes such as cosmic shear and cluster abundances \citep[e.g.][]{ 2014MNRAS.443.1973V,2015MNRAS.449..685H,2016A&A...594A..24P,2018PhRvD..98d3528T,2021A&A...646A.140H,2022PhRvD.105b3514A}.  
This discrepancy may indicate limitations in the standard cosmological model, $\Lambda$CDM, or it may arise from residual systematic uncertainties. For example, recent results from the KiDS-Legacy analysis \citep{Wright2025} suggest that the $S_8$ tension could be driven by residual uncertainties in redshift calibration for the cosmic shear analysis. In the low-redshift measurements, the modeling and calibration of cluster masses can also introduce systematic biases that affect cosmological inference.

Low-redshift cosmological probes, particularly galaxy groups and clusters, pose unique challenges due to the complexity of their observational measurements and the nonlinear nature of structure formation. These systems are further affected by baryonic processes \citep{Pratt2019}---including AGN feedback, radiative cooling, star formation, and merger-driven nonthermal gas motions---that can significantly impact the observables used to infer cluster masses. Consequently, unbiased mass calibration is essential for deriving robust cosmological constraints from cluster samples. In the era of precision cosmology, such calibration must be accompanied by statistically rigorous inference frameworks capable of accounting for observational systematics and calibration uncertainties in the modeling process.

Accurate calibration of cluster mass measurements is a fundamental requirement in cluster cosmology \citep[e.g.][]{Pratt2019,vonderLinden2014, 2015MNRAS.449..685H,2015MNRAS.450.3633S,2022A&A...661A..11C,2024A&A...687A.178G,2025arXiv250309952O}. Although the majority of a cluster's mass resides in dark matter, which cannot be observed directly, total mass estimates can be inferred through observable mass proxies---such as the X-ray gas temperature \citep{2022A&A...661A..11C}, optical richness \citep{2019MNRAS.488.4779C,2021PhRvL.126n1301T}, the Sunyaev--Zel'dovich effect (SZE) signal \citep{2016ApJ...832...95D,2019ApJ...878...55B,2021PhRvD.103d3522C}, and velocity dispersion measurements \citep{2025A&A...693A...2S}---that correlate with the underlying halo mass. 
These correlations, including their intrinsic scatter, depend on the dynamical state of the systems, the degree of hydrostatic or virial equilibrium, and the extent to which light traces mass in the selected sample. As a result, cluster mass estimates based on such proxies are subject to systematic biases arising from a variety of sources, including merger histories, non-gravitational processes (e.g., AGN feedback), and instrumental calibration uncertainties \citep{2014ApJ...794..136D,2014MNRAS.438...49R, 2015A&A...575A..30S,2024A&A...688A.107M}.

To overcome these limitations, cosmological analyses of galaxy cluster samples often rely on external mass calibration via weak gravitational lensing \citep[e.g.,][]{vonderLinden2014,Umetsu2014,2015MNRAS.449..685H,Okabe+Smith2016,2019MNRAS.483.2871D,2020MNRAS.496.4032T,2022A&A...661A..11C,2024PhRvD.110j3006S,2025arXiv250401076C}. Weak lensing provides a direct probe of the projected mass distribution in galaxy clusters, independent of their physical or dynamical state \citep{Bartelmann+Schneider2001,Umetsu2020rev}. By measuring the coherent distortions in the shapes of background galaxies, it enables unbiased---or accurately calibratable---estimates of cluster masses, which are essential for precision cosmology. Weak-lensing estimates of three-dimensional (spherical) cluster masses are subject to several sources of systematic uncertainty, including shear calibration bias, photometric-redshift (photo-$z$) errors, cluster miscentering, and modeling-related mass biases. Properly accounting for these effects is critical for both accurate mass modeling and robust cosmological inference.
In cluster cosmology, constraints on cosmological parameters are traditionally derived using Bayesian inference frameworks that rely on explicit evaluations of the likelihood function. However, the nonlinear nature of cluster observables, combined with the complexity of analysis pipelines linking raw data to summary statistics, often renders the likelihood function analytically intractable or computationally prohibitive to evaluate directly. As a result, practical implementations often adopt simplified likelihood approximations, such as Gaussian or Poisson statistics, to simplify the inference process.

However, this assumption may fail to capture non-Gaussian contributions arising from complex measurement processes and the intrinsically nonlinear statistical fluctuations in cluster properties, potentially leading to biased cosmological constraints. To obtain reliable inferences from cluster data, more flexible statistical frameworks are required—capable of accommodating complex, non-analytic likelihood structures.

Simulation-based inference (SBI), also known as likelihood-free or implicit-likelihood inference, is an emerging statistical methodology that enables parameter estimation without requiring explicit likelihood evaluation. By generating synthetic data through forward simulations, SBI facilitates robust inference even in cases where the likelihood is intractable. In recent years, SBI has gained growing attention and has been increasingly applied across a range of astrophysical problems, including galaxy population modeling \citep{SBI-cosmo}, cosmic shear analysis \citep{SBI-kids}, 21-cm signal modeling \citep{SBI-21cm}, and cluster cosmology analysis \citep[e.g.][]{2025A&A...701A.110C,2025A&A...693A..46K, 2025arXiv250410230Z}.

In  \cite{Tam2022}, we developed a forward-modeling framework integrated with SBI to study cluster cosmology, focusing on the redshift evolution of cluster abundance and weak-lensing mass calibration. 
The flexibility of the forward model demonstrated its potential for robust cosmological inference from cluster counts and weak-lensing measurements. While the methodology in \citet{Tam2022} was validated using synthetic cluster catalogs as a proof-of-concept, this study applies the approach to real observations from the \XMM-XXL survey \citep{XXL}, extending the SBI framework to perform weak-lensing mass calibration using wide-field shear measurements from the Hyper Suprime-Cam Subaru Strategic Program (HSC-SSP; \citealt{2018PASJ...70S...4A, 2018PASJ...70S...8A}). This marks a significant step toward applying SBI techniques to observational datasets, providing a statistically robust, likelihood-free framework that complements and extends traditional methods in cluster cosmology.

The \XMM-XXL survey \citep{XXL} is one of the largest observational programs conducted with the \textit{XMM-Newton} satellite, designed to deliver precise cosmological constraints using a sample of X-ray-selected galaxy groups and clusters extending out to redshift $z\sim 1$. The survey covers approximately 25~deg$^2$ in each of two separate sky fields, referred to as XXL-N and XXL-S. Through an extensive suite of multiwavelength follow-up campaigns, the survey has assembled a catalog of 365 galaxy clusters with a well-defined selection function and spectroscopic confirmation \citep{XXL_XX}, publicly released as part of its second data release (DR2). This large and well-characterized cluster sample offers a valuable dataset for deriving cosmological constraints complementary to those obtained from other cluster-based and large-scale structure surveys.

To enable accurate mass calibration of the XXL survey sample for cosmological applications, we rely on an external survey of the HSC-SSP. Hyper Suprime-Cam, mounted on the 8.2-meter Subaru Telescope, is a wide-field optical imager covering 1.77~deg$^2$ field of view (\citealt{2018PASJ...70S...1M, 2018PASJ...70S...2K, 2018PASJ...70S...3F, 2018PASJ...70...66K}).The HSC-SSP has conducted a deep, high-resolution optical imaging survey optimized for weak lensing studies \citep{2018PASJ...70S..25M, 2018PASJ...70S..28M, 2019PASJ...71...43H, 2020PASJ...72...16H}, covering five broad bands (\textit{grizy}) across three survey layers (Wide, Deep, and Ultradeep). The HSC-\XMM field overlaps with the XXL survey footprint, enabling joint analysis of X-ray and lensing data to facilitate precise mass calibration.

In this study, we extend our simulation-based forward-modeling framework by incorporating the XXL survey selection function and by implementing key observational effects, including cosmic noise from uncorrelated large-scale structure, cluster miscentering, photo-$z$ bias, and weak-lensing mass modeling bias. We apply this enhanced forward simulator within the SBI framework to the joint XXL–HSC survey dataset. This enables us to constrain cosmological parameters and the scaling relations of the XXL clusters, and to derive robust, lensing-calibrated mass estimates for individual systems.

This paper is organized as follows. In Section~\ref{sec:obs}, we describe the XXL cluster catalog and the HSC weak lensing dataset. Section~\ref{sec:obs_measure} outlines the formalism of cluster–galaxy weak lensing and presents the blinded weak-lensing measurements for the XXL clusters. In Section~\ref{sec:modeling}, we detail the construction of our forward modeling framework. Section~\ref{sec:SBI} introduces the SBI methodology and the neural network architecture developed for this analysis. The results of the likelihood-free mass calibration based on synthetic surveys and the joint XXL–HSC observations are presented and discussed in Section~\ref{sec:result}. The advantages, limitations, and possible extensions of the SBI framework are discussed in Section~\ref{sc:Discussion}. Finally, we summarize our conclusions in Section~\ref{sec:conclusions}.

Throughout this paper, we assume a flat \LCDM cosmology with a Hubble constant of $H_0=100~h$~km~s$^{-1}$~Mpc$^{-1}$ and adopt $h=0.7$. The cosmological parameters $\Omega_{m}$ and $\sigma_8$ are treated as free parameters in this study.
We denote the critical density of the universe at a particular redshift $z$ as $\rho_\mathrm{c}(z)=3H^2(z)/(8\pi G)$, where $H(z)$ represents the redshift-dependent Hubble function. Mass enclosed within a sphere of radius $r_\Delta$, where the mean enclosed density equals $\Delta \times \rho_\mathrm{c}(z)$, is denoted as $M_\Delta$.  
Three-dimensional cluster-centric distances are denoted by \( r \), while projected (two-dimensional) cluster-centric distances are represented by \( R \), both expressed in comoving units.
We use ``$\log$" to indicate the base-10 logarithm and "$\ln$" for the natural logarithm. Unless otherwise specified, intrinsic scatters in scaling relations are expressed in natural logarithmic units.

\begin{figure}[htbp]
 \centering
 \includegraphics[width=0.45\textwidth,trim={0cm 0cm 0cm 0cm},clip]{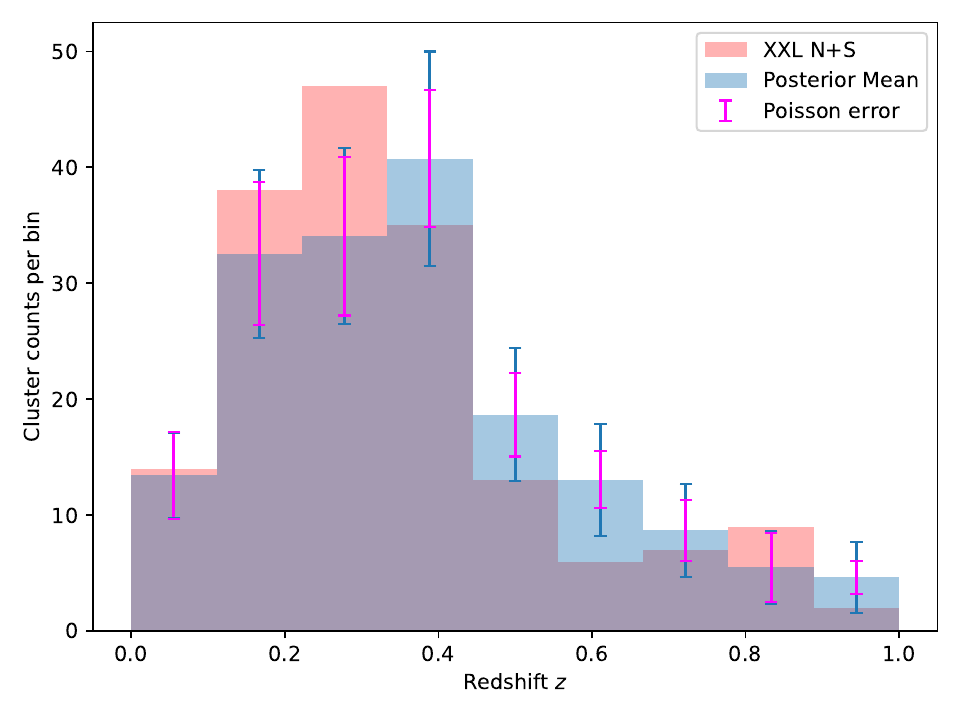}

\caption{Number counts of galaxy clusters as a function of redshift. The red histogram displays the observed XXL C1 clusters from the combined northern and southern regions. The blue histogram represents the mean redshift distribution of synthetic XXL-detected clusters, generated from approximately 5,000 posterior samples inferred via SBI. The blue error bars indicate the confidence intervals marginalized over all parameter uncertainties, while the magenta error bars represent the expected Poisson noise of the observed data. Note that the posterior uncertainties (blue) generally exceed the Poisson noise (magenta), reflecting the additional variance propagated from scaling relation and cosmological uncertainties.}
 \label{fig:Num_vs_z}
\end{figure}

\begin{figure}[htbp]
 \centering
 \includegraphics[width=0.55\textwidth,trim={0cm 0cm 0cm 1cm},clip]{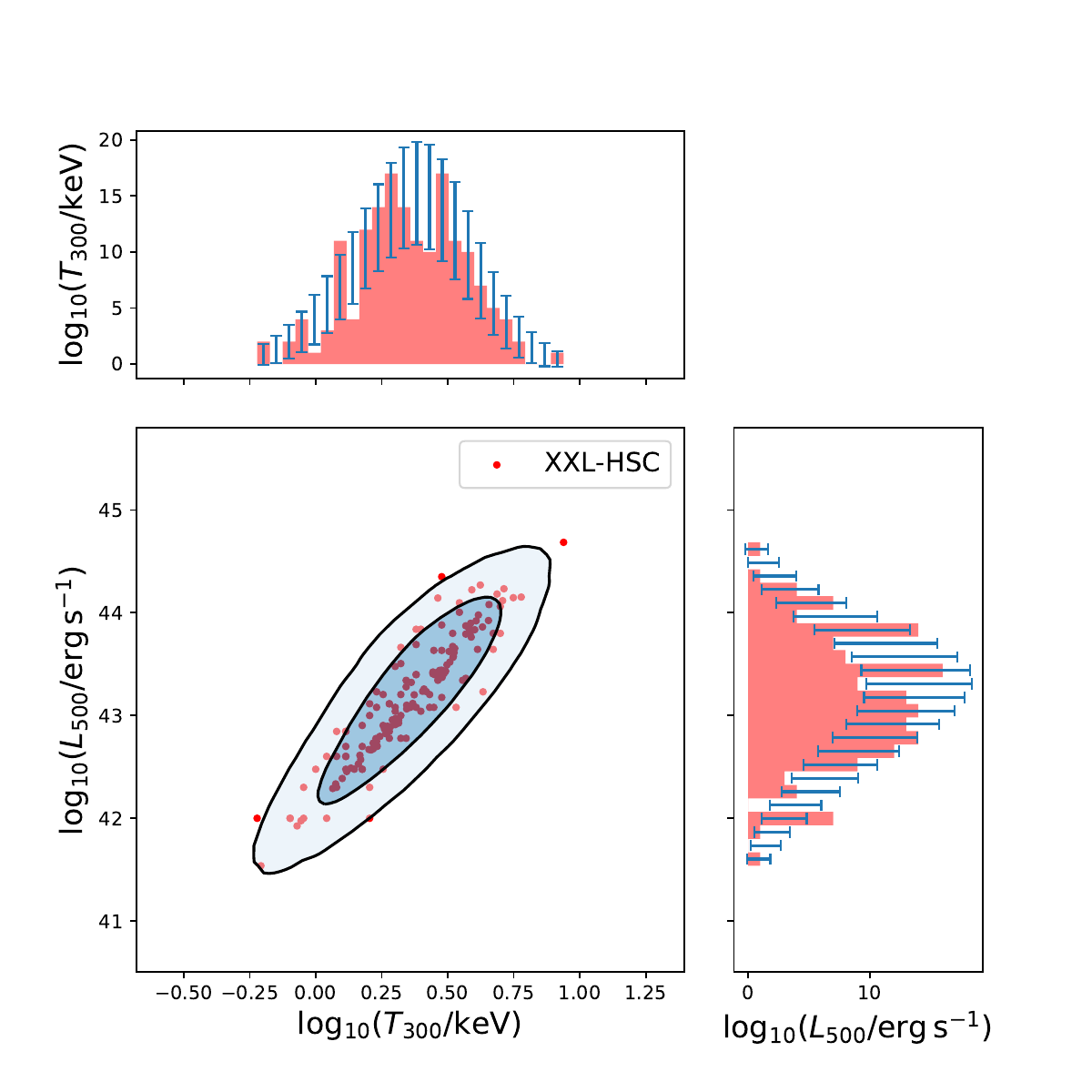}
\caption{The temperature--luminosity distribution of observed XXL C1 galaxy clusters (red points and histograms) compared with the distribution from synthetic cluster samples generated using approximately 5,000 posterior samples inferred via SBI. The error bars indicate the 68\% confidence intervals of the simulated distributions, while the shaded blue contours correspond to the 68\% and 95\% confidence levels of the posterior predictive distribution.}
 \label{fig:L_vs_T}
\end{figure}

\section {CLUSTER SAMPLE AND DATA} \label{sec:obs}
\subsection{XMM Observation} \label{sec:XMM}
In this study, we use the galaxy cluster sample drawn from the XXL second data release (DR2) catalog \citep{XXL_XX}, which contains 185 X-ray–selected systems classified as C1, including 98 clusters in the XXL-N region and 87 in XXL-S. 
To ensure high sample purity,
we exclude 10 systems that lack secure spectroscopic confirmation. The resulting final C1 sample comprises 175 confirmed clusters over a total survey area of 47.36 deg$^2$ spanning both the XXL-N and XXL-S fields. This selection strategy minimizes contamination from spurious detections and blended point sources, yielding a cluster catalog with well-controlled purity \citep{2006MNRAS.372..578P}.

Out of the full sample, 171 clusters satisfy the redshift selection criterion of \( z < 1 \) adopted in this study, with the minimum redshift at $z=0.044$. Among these, 162 clusters have measured X-ray temperatures $T_{300\,\rm{kpc}}$ measured within a fixed aperture of 300 kpc, along with X-ray soft-band luminosity $L_{500}$ estimated within the cluster radius of $R_{500}$. The temperature values span from 0.6 keV to 8.6 keV, while the luminosities range from \( 3.45 \times 10^{41} \) erg/s to \( 4.84 \times 10^{44} \) erg/s. Measurement uncertainties in both \( T_{300\,\mathrm{kpc}} \) and \( L_{500} \) are explicitly incorporated into the cluster forward modeling (see Section~\ref{sec:X-ray}).

For this study, we use the 171 C1 clusters with \( z < 1 \) for cluster abundance analysis (Figure~\ref{fig:Num_vs_z}). A subset of 162 clusters with available X-ray observable measurements is employed to characterize the distributions of temperature and luminosity, summarized through their quantile values. For cluster mass estimation, which requires external calibration through weak-lensing measurements, we utilize only those XXL-N clusters that lie within the HSC-SSP survey footprint, as detailed in Section~\ref{sec:HSC}. Table~\ref{tab:summary_stats} summaries the number of clusters we used for different summary statistics.

\begin{table}[ht]
\centering
\caption{Number of XXL C1 Clusters Used in Each Summary Statistic}
\begin{tabular}{lc}
\hline
Summary Statistic & Number of Clusters\\
\hline
Cluster number counts        & 171 \\
Weak lensing shear analysis    & 86  \\
$T_{300\,\rm{kpc}}$ and $L_{500}$ distribution & 162 \\
\hline
\end{tabular}

\label{tab:summary_stats}
\end{table}

\subsection{Subaru HSC Survey} \label{sec:HSC}
For the weak-lensing mass calibration, we utilize the three-year shear catalog (S19A) from the HSC-SSP, with full details provided in \citet{2022PASJ...74..421L}. To summarize, HSC obersevations consist of \textit{grizy} imaging. The shape catalog comprises galaxies selected based on basic quality cuts and a magnitude cut of \( i < 24.5\,\mathrm{mag} \). Galaxy shapes are measured from co-added \( i \)-band images using the re-Gaussianization method \citep{2003MNRAS.343..459H}.

For this study, we adopt the Direct Empirical Photometric (DEmP) code \citep[DEmP;][]{demp} to estimate photo-$z$ probability distributions $P(z)$ for individual galaxies. To securely identify background sources, we use the full \( P(z) \) information for each galaxy—an approach referred to as the P-cut method \citep{2014MNRAS.444..147O,2018PASJ...70...30M}. In this method, background galaxies are selected based on the following criteria:

\begin{equation}
\int_{z_l+\Delta z}^\infty P(z)\,dz > P_{\rm cut} \quad \text{and} \quad z_{mc} < z_{\rm max},
\end{equation}

where \( P_{\rm cut} \) is a constant threshold set to 98\%, $z_l$ is the lens redshift, \( z_{mc} \) is a redshift point estimator randomly drawn from \( P(z) \), and \( z_{\rm max} \) denotes the maximum redshift considered. Following \citet{2018PASJ...70...30M}, we adopt \( z_{\rm max} = 2.5 \) and \( \Delta z = 0.2 \) to minimize contamination from foreground and cluster member galaxies.

The three-year shape catalog spans approximately 430 deg$^2$, with the HSC--\XMM field overlapping the XXL-N survey area by 21.4 deg$^2$, with number density of source galaxies $n_{\rm{gal}}\approx22.9\,\rm{galaxies\,/armin}^2$ \citep{2022PASJ...74..421L}. For the weak-lensing analysis, we restrict our sample to XXL clusters that lie within this overlapping area. By selecting clusters with a projected comoving transverse separation from the HSC-SSP footprint of \( R_{\mathrm{min}} = 0.3\,h^{-1}\,\mathrm{Mpc} \), we identify 86 C1 clusters within the redshift range \( 0.044 \le z < 1.0 \), all of which have measured X-ray temperatures. To extract more detailed information from the sample, we divide the HSC-XXL overlapping cluster sample into three sub-samples based on their temperature, and measured the stacked lensing signal for each temperature bin. This approach avoids smoothing out important mass-dependent trends, as described in Section~\ref{sec:modeling}.

\section{Observational Measurement} \label{sec:obs_measure}

\subsection{Basics of Cluster Weak Lensing}
Weak gravitational lensing results from the deflection of light by mass fluctuations, such as galaxy clusters, inducing small but coherent distortions in the observed shapes of background galaxies \citep[see][for review]{Umetsu2020rev}. These distortions are characterized by two main quantities: the convergence (\( \kappa \)), which quantifies the isotropic magnification, and the complex shear (\( \gamma \)), which represents the anisotropic stretching of galaxy images.
The convergence is directly related to the projected surface mass density \( \Sigma \) of the lensing mass distribution, normalized by the critical surface mass density \( \Sigma_{\mathrm{crit}} \), expressed as 

\begin{equation}
\kappa = \Sigma/\Sigma_\mathrm{crit},
\label{eq:kappa}
\end{equation}
where \( \Sigma_{\mathrm{crit}} \) depends on the geometric configuration of the observer, lens, and source as 

\begin{equation}
\label{eq:cr}
\Sigma_\mathrm{crit}(z_l, z_s) =\frac{c^2 D_s(z_s)} {4\pi G (1+z_l)^2 D_l(z_l) D_{ls}(z_l,z_s)},
\end{equation}
where
$D_l(z_l)$, $D_s(z_s)$, $D_{ls}(z_l,z_s)$ are the angular diameter distances from the observer to the lens, from the observer to the source, and from the lens to the source, respectively. 

The factor of \( (1+z_l)^2 \) in Equation~(\ref{eq:cr}) arises due to our use of comoving surface mass densities ($\Sigma$ and $\Sigma_\mathrm{crit}$). We set $D_{ls}(z_l,z_s)/D_s(z_s)=0$ for unlensed sources with $z_s\le z_l$.

The complex shear field, $\gamma = \gamma_1 + i\gamma_2$, can be decomposed into the tangential component, $\gamma_{+}$, and the $45^\circ$-rotated cross component, $\gamma_{\times}$, defined with respect to a chosen reference point. The azimuthally averaged radial profile of the tangential component, $\gamma_{+}(R)$, measured around the selected center, is directly related to the excess surface mass density, $\Delta\Sigma(R)$, defined as
\begin{equation}
    \Delta\Sigma(R) \equiv \Sigma(<R) - \Sigma(R) = \Sigma_\mathrm{crit}\,
    \gamma_{+}(R),
    \label{eq:gmmat}
\end{equation}
where $\Sigma(R)$ is the azimuthally averaged surface mass density at projected comoving radius $R$, and $\Sigma(<R)$ denotes the mean surface mass density enclosed within radius $R$. 
We emphasize that this relationship holds for any mass distribution described by a scalar gravitational potential, regardless of its symmetry.

Similarly, the azimuthally averaged cross-shear ($\times$) component vanishes identically for any such potential. Consequently, any non-zero signal in the measured cross-component arises solely from statistical noise or systematic effects.

In general, the observable quantity in weak-lensing shape analyses is the complex reduced shear:
\begin{equation}
    g :=g_1+ig_2 = \frac{\gamma}{1-\kappa},
\end{equation}
which can be directly estimated from the image ellipticities of background galaxies. The reduced tangential shear signal, $g_+(R)$, measured as a function of cluster-centric radius $R$ is related to $\Sigma(R)$ and $\Delta\Sigma(R)$ via
\begin{equation}
\label{eq:gt}
\Delta\Sigma_+(R)\equiv \Sigma_\mathrm{crit} g_+(R) = \frac{\Delta\Sigma(R)}{1-\Sigma_\mathrm{crit}^{-1}\Sigma(R)},
\end{equation}
where $\Delta\Sigma_{+}(R)$ represents the nonlinearly corrected excess surface mass density.

\subsection{Weak Lensing Shear Profile for XXL-N Clusters} \label{sec:WL}

We measure the azimuthally averaged excess surface mass density, \( \Delta\Sigma_+(R) \), for stacked subsamples of galaxy clusters centered on their respective X-ray peak positions. The measurements are performed in 10 logarithmically spaced comoving radial bins, ranging from \( R = 0.3\,h^{-1}\,\mathrm{Mpc} \) to \( 3\,h^{-1}\,\mathrm{Mpc} \) \citep{U2020}, using the estimator described in \citet{Miyatake2019}: 
\begin{equation}
\label{eq:delta_sigma}
\langle \Delta\Sigma_+\rangle(R_i)=\frac{1}{2\mathcal{R}(R_i)}\frac{\sum_{l,s}w_{l,s}e_{+,ls}\langle\Sigma_{{\rm crit},ls}^{-1}\rangle^{-1}}{[1+m(R_i)]\sum_{l,s}w_{l,s}},
\end{equation}
where the summation running over all lens--source pairs ($l, s$), $e_{+}$ is an estimate of the tangential ellipticity of the source galaxy, $\mathcal{R}(R_i)$ is the shear responsivity defined as
\begin{equation}
\mathcal{R}(R_i)=1-\frac{\sum_{l,s}w_{l,s}e_{rms,ls}^2}{\sum_{l,s}w_{l,s}}.
\end{equation}
The factor of $1+m(R_i)$ is the shear calibration factor which corrects for multiplicative shear bias
\begin{equation}
1+m(R_i)=\frac{\sum_{l,s}w_{l,s}(1+m_s)}{\sum_{l,s}w_{l,s}},
\end{equation}
where $m_s$ is the multiplicative bias of the source.
The weight for each lens--source pair is defined as 
\begin{equation}
\label{eq:weight}
w_{l,s}=\frac{\langle\Sigma_{\textrm {crit},ls}^{-1}\rangle^2}{\sigma_{e,s}^2+\sigma_{{\rm rms}, s}^2},
\end{equation}
where $\sigma_e,s$ is the shape measurement uncertainty per ellipticity component and $\sigma_{\rm rms}$ is the rms ellipticity per each component. The average inverse critical surface mass density for
each lens–source pair, $\langle\Sigma_{{\rm crit},ls}^{-1}\rangle$, is calculated by averaging with the photo-z probability distribution function $P_s(z)$ as
\begin{equation}
\label{eq:sigma_cr_pdfz}
\langle\Sigma_{{\rm crit},ls}^{-1}\rangle=\frac{\int_{z_l}^\infty P_s(z)\Sigma_{\rm crit}^{-1}(z_l, z_s)dz_s}{\int_0^\infty P_s(z)dz_s},
\end{equation}

We also correct for additive shear bias by subtracting the appropriate weighted term from Equation~(\ref{eq:delta_sigma}) \citep[e.g.][]{2018PASJ...70S..25M, 2022PASJ...74..421L}. Furthermore, we account for the impact of both multiplicative and additive selection bias resulting from the anisotropic selection of the galaxy ensemble as presented in \cite{2022PASJ...74..421L}.

We quantify the signal-to-noise ratio (SNR) of the stacked lensing profile, integrated over the full radial range, using the linear SNR estimator, defined as $\mathrm{SNR} = \langle S \rangle / \sigma_{\langle S \rangle}$ \citep{2017MNRAS.472.1946S,U2020}, where  
\begin{equation}
\label{eq:SNR}
\begin{aligned}
\langle S\rangle &=\frac{\sum_{i=1}^{N}\langle \Delta\Sigma_+\rangle(R_i)/\sigma_{\rm{shape}}^2(R_i)}{\sum_{i=1}^{N}1/\sigma_{\rm{shape}}^2(R_i)}, \\
\sigma_{\langle S\rangle} &=\frac{1}{\sqrt{\sum_{i=1}^{N}1/\sigma_{\rm{shape}}^2(R_i)}}.
\end{aligned}
\end{equation}
Here, $\sigma_{\rm{shape}}(R_i)$ denotes the statistical uncertainty  in $\langle\Delta\Sigma_+\rangle(R_i)$ due to shape noise, which can be estimated from the cross-component measurements as
\begin{equation}
\label{eq:Sigma_shape}
\sigma_{\rm{shape}}^2(R_i)=\frac{\sum_{l,s}w_{l,s}^2e_{\times,ls}^2\langle\Sigma_{{\rm crit},ls}^{-1}\rangle^{-2}}{4\mathcal{R}(R_i)^2[1+m(R_i)]^2\left(\sum_{l,s}w_{l,s}\right)^2}.
\end{equation}

\begin{figure}[htbp]
 \centering
 \includegraphics[width=0.5\textwidth,trim={0.0cm 2cm 0cm 1cm},clip]{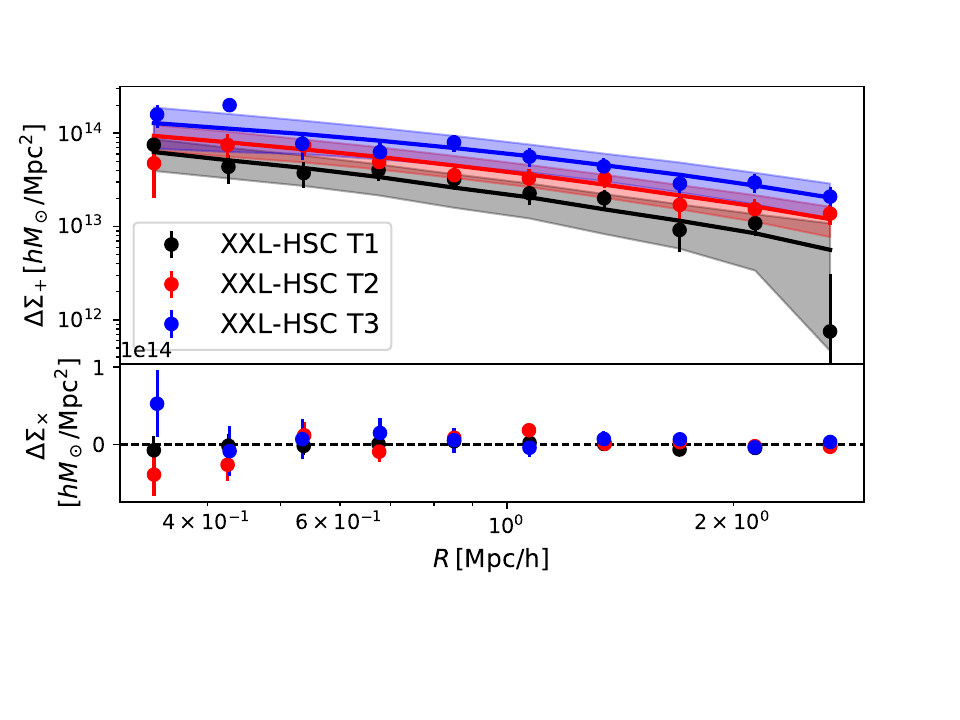}
 \caption{Azimuthally averaged stacked surface mass density profiles around X-ray-selected XXL-N C1 galaxy clusters, derived from the HSC S19A shape catalog. The upper panel shows the tangential shear component, $\langle\Delta\Sigma_{+}\rangle(R)$, while the lower panel displays the 45$^\circ$-rotated (cross) component, $\langle\Delta\Sigma_{\times}\rangle(R)$, both plotted as functions of the comoving cluster-centric radius $R$. Weak lensing signals for the three temperature bins are indicated using different colors. The shaded regions represent simulated signals based on the posterior obtained using the SBI approach.}
 \label{fig:XXLsel_gt}
\end{figure}

We adopt an empirical noise estimator based on the measured variance of the cross-component ($\times$-mode) shear, which provides a conservative noise budget for the forward simulations by capturing statistical noise, the scatter induced by halo non-circularity, and any residual systematics.
Relying on a single stacked lensing profile may obscure information about the underlying cluster population and limit sensitivity to cosmological signals, due to the averaging over heterogeneous systems. To preserve sensitivity to mass-dependent trends and intrinsic variations, we adopt a more refined approach by dividing the cluster sample into three distinct temperature bins:
$T_1\in[0, 1.9]$~keV, $T_2\in[1.9, 3.5]$~keV, and $T_3\in[3.5, 10]$~keV. We compute the stacked weak lensing profile  separately for each temperature bin. The characteristics of the temperature-binned subsamples are listed in Table~\ref{tab:Tx_bin}, and the corresponding stacked shear profiles are shown in Figure~\ref{fig:XXLsel_gt}.

We note that, when computing the observed shear profiles \( \Delta\Sigma_+(R) \), a fiducial cosmology with $\Omega_{m}=0.28$ is assumed to convert angular separations into comoving radii. As demonstrated in Appendix~\ref{App: shear}, this choice has a negligible impact on the inferred shear profiles.

\begin{deluxetable}{ccccc}
\tablecolumns{5}
\tablewidth{0pt}
\tabletypesize{\scriptsize}
\tablecaption{\label{tab:Tx_bin}
Summary of the $T_{300\,\mathrm{kpc}}$-binned Cluster Subsamples used in the HSC Weak-lensing Analysis}
\tablehead{
    \colhead{Bin} & 
    \colhead{$N_{\rm cl,T}$ \tablenotemark{a}} &
    \colhead{$\tilde{T}_{300\,\rm kpc}$(keV) \tablenotemark{b}} &
    \colhead{$\tilde{z}$ \tablenotemark{c}} &
    \colhead{SNR} 
}
\startdata
$T1\in[0, 1.9]\,\rm keV$  & 36 & 1.45 & 0.184 & 7.77  \\
$T2\in[1.9, 3.5]\,\rm keV$ & 36 & 2.5 & 0.336 & 10.27\\
$T3\in[3.5,10]\,\rm keV$ & 14 & 4.5 & 0.4285 & 11.08  \\
\enddata
\tablenotemark{a}{Number of clusters with measured X-ray temperatures.}
\tablenotetext{b}{Median X-ray temperature.}
\tablenotetext{c}{Median cluster redshift.}
\end{deluxetable}

\subsection{Blinding Analysis} \label{sec:blind}
To mitigate confirmation bias in the mass calibration analysis, we implement catalog-level blinding following the scheme outlined in \citet{2019PASJ...71...43H}. Specifically, the analysis is independently performed on three shape catalogs, each with different levels of blinding, achieved by perturbing the multiplicative shear bias as follows:
\begin{equation}
\label{eq:blinding}
m_{{\rm cat},i} = m_{{\rm true},i} + dm_{1,i} + dm_{2,i},
\end{equation}
where $m_{\rm true}$ denotes the true multiplicative bias in the HSC S19A catalog, and $dm_1$ and $dm_2$ are randomly generated offsets used for blinding. The subscript $i = 0,1,2$ indicates the three blinded catalogs.
The values of $dm_1$ and $dm_2$ are encrypted within the catalogs. The $dm_1$ term can only be decrypted and removed by the leader of analysis team prior to performing the main analysis, while $dm_2$ remains encrypted throughout and can only be decrypted by a designated 'blinder-in-chief' who does not participate in the analysis. Among the three catalogs, only one has $dm_2 = 0$, corresponding to the true shape catalogs. Final unblinding is performed only after the analysis for all three blinded catalogs is completed.

Following unblinding, no changes were made to the core cosmological inference pipeline or the data vector. However, we optimized the training strategy regarding temperature bins containing zero clusters. In the final analysis, we retain simulations corresponding to these empty bins by assigning them a zero lensing shear signal, rather than discarding the realizations. This approach substantially increases the effective training volume of the parameter space while preserving the information content from cluster abundance, X-ray distributions, and weak-lensing measurements in a self-consistent manner. Crucially, it ensures robust sampling across the full prior range without altering the underlying physical forward model.

\section{Modeling Procedure} \label{sec:modeling}

\begin{figure*}
 \centering
 \includegraphics[width=1\textwidth,trim={0cm 0cm 0cm 0cm},clip]{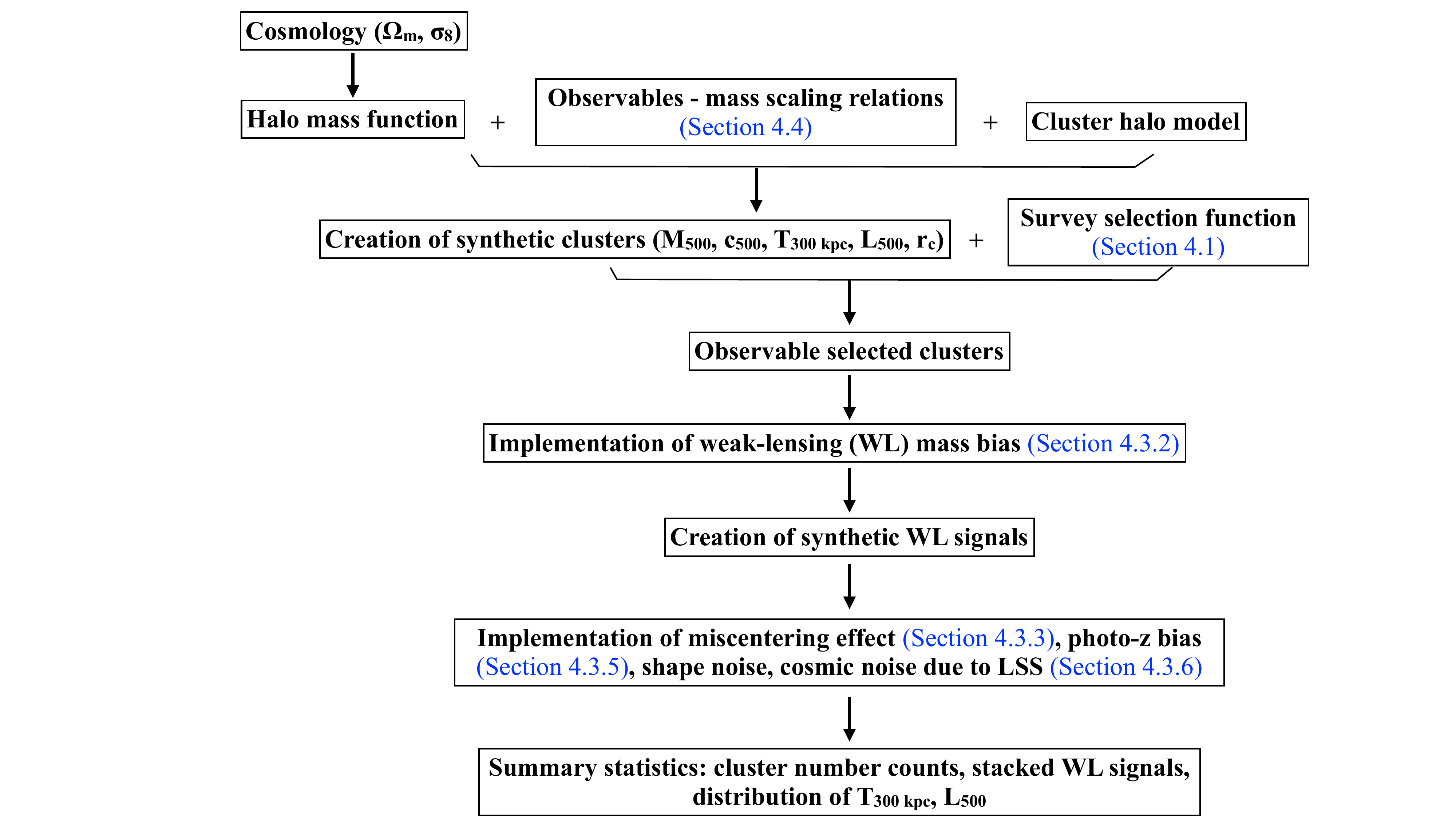}
 \caption{Schematic diagram illustrating the forward modelling procedure for the cosmological inference pipeline.}
 \label{fig:flowchart}
\end{figure*}

In this section, we describe the forward-modeling procedure adopted for our simulation-based cosmological analysis. A schematic overview of the full methodology is presented in Figure~\ref{fig:flowchart}.

\subsection{XXL Survey Selection} \label{sec:XXL_selection}

In this study, we utilize the well-characterized XXL C1 selection function derived from Monte Carlo simulations.  These simulations incorporate $\beta$-model sources with $\beta = 2/3$ and account for instrumental effects such as point spread function (PSF) distortion, vignetting, detector masks, background noise, and Poisson fluctuations across all three \textit{XMM-Newton} detectors \citep{2016A&A...592A...2P}. The XXL cluster detection pipeline is applied to these simulations to evaluate both completeness and purity levels of the survey.

\begin{figure}[htbp]
 \centering
 \includegraphics[width=0.5\textwidth,trim={0cm 1cm 0cm 2cm},clip]{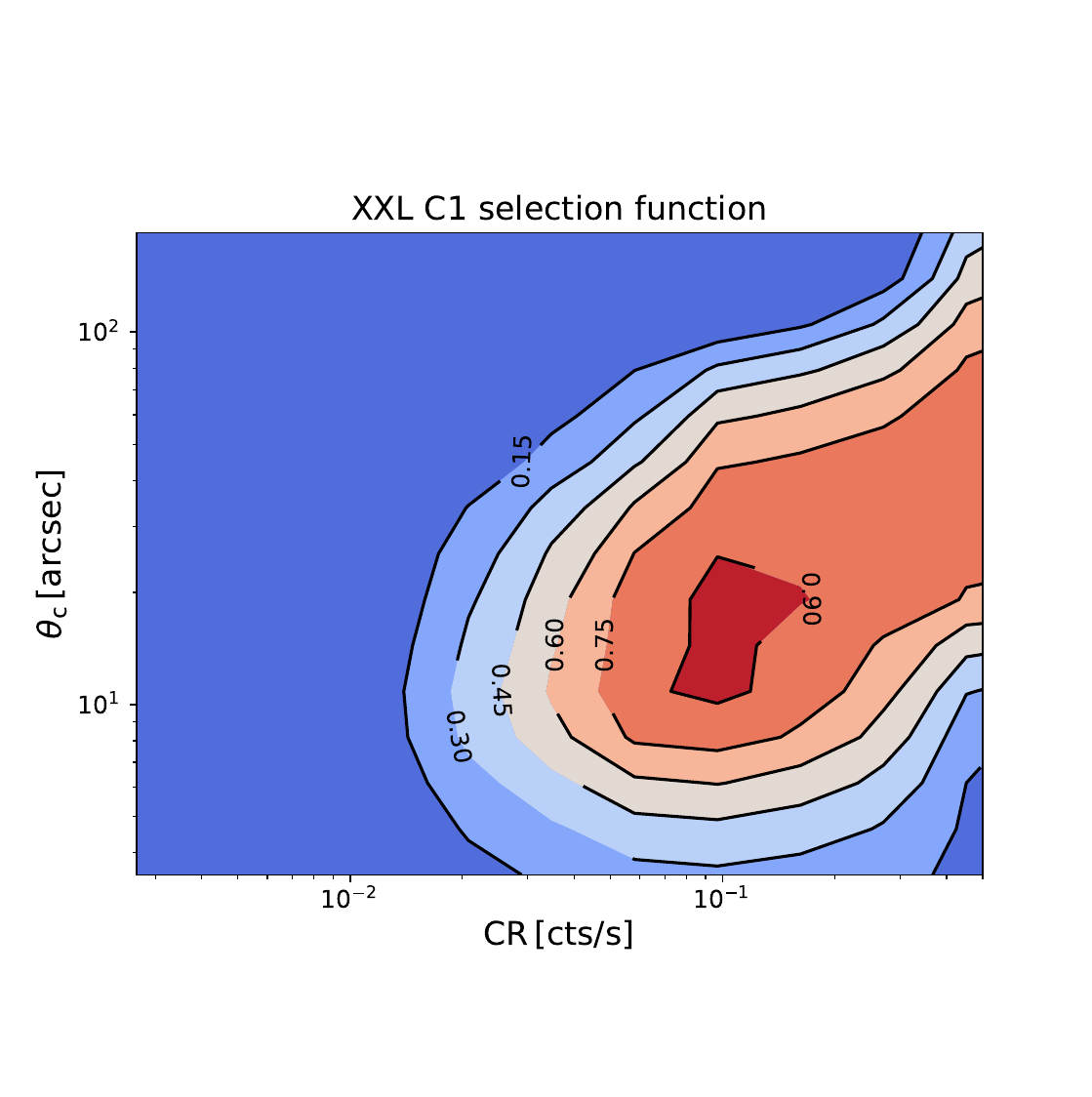}
 \caption{Selection function of X-ray galaxy clusters for the XXL survey. The color image shows the probability distribution of the XXL survey selection function, shown as a function of X-ray count rate and angular core radius.}
 \label{fig:XXLsel}
\end{figure}

Figure~\ref{fig:XXLsel} displays the selection function mapped onto the parameter space defined by the X-ray count rate (CR) and the apparent core radius ($\theta_c$). The CR refers to the number of X-ray counts per second in the [0.5–2]~keV energy band, normalized to the on-axis response. 

In our forward modeling framework, each synthetic cluster, defined by its halo mass and redshift, is assigned an X-ray temperature from the $T_{300\,\rm kpc}$--$M_{500}$ scaling relation and an X-ray luminosity from the $L_{500}$--$T_{300\,\rm kpc}$ relation (see Section~\ref{sec:X-ray}). Based on these X-ray properties, the source count rate, ${\rm CR}(T_{300\,\rm kpc}, L_{500}, z)$, is computed using the Astrophysical Plasma Emission Code (\textsc{APEC}), assuming a metallicity of $Z = 0.3\,Z_\odot$. The corresponding X-ray emission spectra are generated using the AtomDB database and folded through the \textit{XMM-Newton} EPIC response matrices to account for instrumental effects.

The core radius $r_c$ in the $\beta$-model is linked to the overdensity radius $R_{500}$, which is computed from $M_{500}$ sampled from the halo mass function. The distribution of scaled core radii, $r_c/R_{500}$, was derived from a collection of 322 individual surface brightness profiles from the sample presented in \citet{Comparat:2020}, each fitted with a $\beta$-model (fixing $\beta=2/3$). As shown in \citet{2025arXiv250605457R}, this distribution is well described by a log-normal function with a mean $r_c = 0.143\,R_{500}$ and a scatter $\sigma_{\ln r_c} = 0.332$. For further details on the derivation of this scaling relation, we refer to Appendix~B of their work.

For a comprehensive discussion of the selection function, including survey completeness and instrumental effects, we refer the reader to \citet{2006MNRAS.372..578P} and \citet{2022A&A...663A...3G}.

\subsection{Cluster Abundance} 
\label{sec:CL}

For a given cosmological model and the survey selection function described in Section~\ref{sec:XXL_selection}, we generate a  synthetic catalog of detectable clusters over the XXL survey footprint (\( \Omega_{\mathrm{s}} = 47.36\,\mathrm{deg}^2 \)) and within the redshift range \( 0.04 < z < 1 \). Halos are sampled from the \citet{2016MNRAS.456.2486D} comoving mass function \( dn(M, z)/dM \), with halo masses defined at an overdensity of \( \Delta = 500 \). Although the choice of halo mass function may introduce systematic biases, \citet{2025arXiv250605457R} have shown that its impact on cosmological constraints is negligible for an XXL-like survey.

For each simulated halo, we compute its expected X-ray observables using the scaling relations described in Section~\ref{sec:X-ray}. 
Each simulated cluster sampled from mass function is projected onto the $(\mathrm{CR}, \theta_c)$ parameter space and selected according to its probability of detection $P(\mathrm{CR}, \theta_c)$, as defined by the selection function evaluated on a predefined grid. This procedure ensures that the synthetic catalog accurately reflects the observational characteristics and detection efficiency of the XXL survey. The resulting redshift distribution of the selected synthetic clusters serves as a set of summary statistics.

As described in Section~\ref{sec:XMM}, the incompleteness of reliable \( T_{300\,\mathrm{kpc}} \) measurements in the XXL C1 sample is approximately 5\%.  This level of incompleteness is not expected to significantly affect the quantile summary statistics of \( L_{500} \) or \( T_{300\,\mathrm{kpc}} \) when averaged over redshift. In the forward modeling, we therefore assume that the subset of clusters with available \( T_{300\,\mathrm{kpc}} \) and \( L_{500} \) measurements is representative of the full XXL C1 sample. This assumption neglects any potential coupling between temperature measurement completeness and the X-ray selection function.

\subsection{Weak Lensing}
\label{subsec:WL}

For our lens modeling, we restrict the analysis to 45\% of the full suvery-selected cluster sample. This limitation arises from the fact that only 21.4~deg$^2$ of the XXL survey overlaps with the HSC-SSP coverage. To construct this subset for HSC weak lensing, we randomly select 45\% of the clusters from the synthetic XXL cluster sample, yielding a subsample size of $N_\mathrm{cl, N} = N_\mathrm{cl} \times 45\%$. In order to probe potential mass-dependent trends in the weak-lensing signal, we further subdivide the synthetic XXL-N C1 cluster sample into the three temperature bins defined in Section~\ref{sec:WL}. 

In this subsection, we outline our modeling procedure for cluster weak-lensing measurements, incorporating the lensing signal,  key systematic effects, and statistical noise.

\subsubsection{Density Profile}
\label{sec:density_nfw}
We model the mass distribution of XXL clusters with a spherical Navarro--Frenk--White \citep[NFW;][]{NFW1, NFW2} profile given by
\begin{equation}
\rho(r)=\frac{\rho_\mathrm{s}}{(r/r_\mathrm{s})(1+(r/r_\mathrm{s}))^2},
\end{equation}
where $\rho_\mathrm{s}$ and $r_\mathrm{s}$ denote the characteristic density and scale radius, respectively. 
We parametrize the NFW profile using the halo mass, \( M_{\Delta} \), and the concentration parameter, \( c_{\Delta} = r_{\Delta} / r_\mathrm{s} \). In this study, we adopt an overdensity definition of \( \Delta = 500 \).

For each synthetic cluster, we draw a concentration value, $c_{500}$, from a log-normal distribution with a mean \( \ln c_{500} \) specified by the cosmology-dependent concentration--mass (\( c \)--\( M \)) relation, \( c_{500}(M_{500}, z\,|\,\Omega_{m}, \sigma_8) \), from \citet{Diemer19}, and an intrinsic log-normal scatter of \( \sigma_\mathrm{int}(\ln c_{500}) = 0.368 \). 
In our inference, this $c$--$M$ relation is evaluated for each sampled set of cosmological parameters, with $\Omega_{m}$ and $\sigma_8$ treated as free parameters.

We use analytic expressions from \citet{2000ApJ...534...34W} for the radial dependence of the projected NFW profiles, \( \Sigma^\mathrm{NFW}(R|z_l) \) and \( \Delta\Sigma^\mathrm{NFW}(R|z_l) \), which provide accurate approximations of the projected matter distribution around clusters. This radial behavior is supported by cosmological \( N \)-body simulations \citep[e.g.,][]{Oguri+Hamana2011} as well as by direct lensing measurements \citep[e.g.,][]{Okabe+Smith2016,Umetsu2016}. The contribution from the 2-halo term to \( \Delta\Sigma \) becomes significant only at scales beyond several virial radii \citep{Oguri+Hamana2011}, which lie beyond our outer fitting limit of \( R = 3\,h^{-1}\,\mathrm{Mpc} \).

\subsubsection{Mass Bias Relation}

Weak-lensing mass estimates, $M_{500}^\mathrm{WL}$, can systematically deviate from the true halo mass, $M_{500}$, due to projection effects and mismatches with the assumed NFW profile. The magnitude and nature of this bias depend on several factors, including the true halo mass, the assumed mass model (e.g., density profile and priors), the binning scheme and radial fitting range, and the choice of covariance matrix, which governs the relative weighting across radial bins \citep{U2020}.

Using synthetic shear data from the BAHAMAS dark matter-only cosmological simulations \citep{BAHAMAS17}, \citet{U2020} characterized the weak-lensing mass bias as a function of true halo mass for both $M_{500}$ and $M_{200}$. In this work, we adopt their $M_{500}^\mathrm{WL}$--$M_{500}$ relation, which covers the mass range relevant to the XXL sample. Their calibration was derived by fitting reduced tangential shear profiles of simulated halos using the same radial binning scheme and NFW profile assumption as our analysis. 

We express this \( M_{500}^{\rm WL}\)--\(M_{500} \) relation as
\begin{equation}
\ln M_{500}^\mathrm{WL} = \ln M_{500} + \ln f_\mathrm{WL}(M_{500}) \pm \sigma_\mathrm{int}\left(\ln M_{500}^\mathrm{WL}\right),
\end{equation}
where \( f_\mathrm{WL}(M_{500}) \) quantifies the average mass bias at fixed \( M_{500} \), and \( \sigma_\mathrm{int}\left(\ln M_{500}^\mathrm{WL}\right) \) denotes the intrinsic scatter in weak-lensing mass at fixed true mass. This bias relation is well described by the functional form proposed by \citet[][see their Figure~1]{2022PASJ...74..175A}:
\begin{equation}
f_\mathrm{WL}(M_{500}) = a\,\tanh\left(\frac{M_{500}}{10^{14}M_{\odot}}\frac{1}{b}\right) + c,
\end{equation}
with best-fit parameters \( a = 0.29 \), \( b = 1.57 \), \( c = 0.73 \), and \( \sigma_\mathrm{int}(\ln M_{500}^{\rm WL}) = 0.21 \). In this work, we adopt \( \sigma_\mathrm{int}(\ln M_{500}^{\rm WL}) = 0.2 \).

Importantly, \citet{U2020} found no evidence for a dependence of the weak-lensing mass bias on the signal-to-noise ratio (or the shot noise level quantified by the source galaxy number density, $n_\mathrm{g}$), implying that non-linear noise bias effects are negligible. This robustness arises because the weak-lensing shear signal scales approximately linearly with the underlying halo mass. In our forward modeling framework, we explicitly include the same noise components---shape noise and uncorrelated large-scale structure (see Section~\ref{sec:noise})---as in the synthetic analysis of \citet{U2020}, thereby ensuring consistency with the conditions under which the mass bias relation was characterized.

We incorporate this mass-dependent bias model and associated scatter into our forward model, parameterizing the NFW profile in terms of the weak-lensing mass \( M_{500}^\mathrm{WL} \) and halo concentration $c_{500}$.

\begin{figure}[htbp]
 \centering
 \includegraphics[width=0.5\textwidth,trim={0cm 0cm 1cm 0cm},clip]{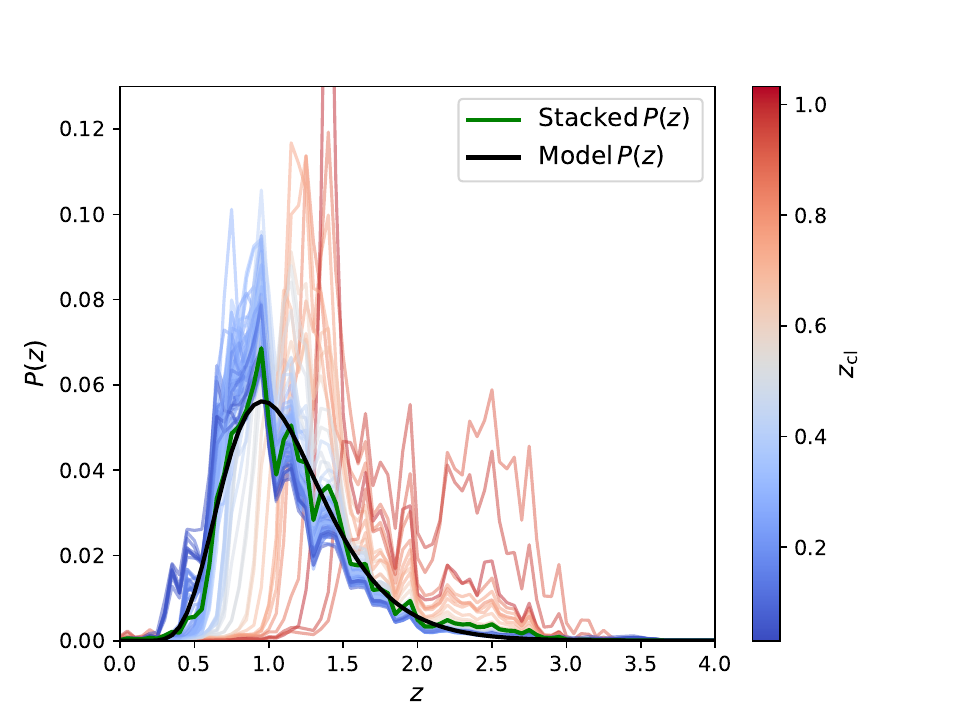}
 \caption{The $P(z)$ distribution of background sources. The stacked $P(z)$ over the source galaxies for each of XXL-C1 cluster are shown, with each curve color-coded according to the corresponding cluster redshift, as indicated by the colorbar. The  variations are primarily driven by shot noise due to the finite number of source galaxies. }The green distribution represents the stacked $P(z)$ over all the XXL-N C1 galaxy clusters. The black one represents the best-fit $P(z)$ model. 
 \label{fig:photoz}
\end{figure}

\subsubsection{Cluster Miscentering}
\label{sec:miscentering}

As described in Section~\ref{sec:WL}, we measure the stacked lensing profile around XXL-N subsamples, centered on their respective X-ray peak positions. However, these X-ray-defined centers do not always coincide with the true centers of the underlying dark matter halos---a discrepancy commonly referred to as miscentering. Such offsets can suppress the observed tangential shear signal, particularly at small cluster-centric distances, and thereby bias mass estimates. To ensure accurate mass calibration using forward simulations, it is essential to incorporate the effects of miscentering into our weak-lensing modeling.

To model cluster miscentering, we introduce a fraction ($f_{\rm mis}$) of the cluster sample that is miscentered from the true halo center, with a miscentering offset $R_{\rm mis}$ following a Rayleigh distribution:
\begin{equation}
P_{\rm mis}(x|\sigma_{\rm mis})=\frac{x}{\sigma_{\rm mis}^2}\exp{\left[-\frac{1}{2}\left(\frac{x}{\sigma_{\rm mis}}\right)^2\right]},
\end{equation}
where $x=R_{\rm mis}/r_s$ with $r_s$ the NFW scale radius. 
The surface mass density for these miscentered clusters is then computed following \citep{2006MNRAS.373.1159Y, 2007ApJ...656...27J} as
\begin{equation}
\begin{aligned}
\Delta\Sigma^{\rm mis}(R|R_{\rm mis})=&\frac{2}{R^2}\int_0^RR'\Sigma^{\rm mis}(R'|R_{\rm mis})dR'\\
&-\Sigma^{\rm mis}(R|R_{\rm mis}),\\
\end{aligned}
\end{equation}
where the azimuthally averaged miscentered surface mass density is given by
\begin{equation}
\begin{aligned}
\Sigma^{\rm mis}(R|R_{\rm mis})=&\int_0^{2\pi}\frac{d\phi}{2\pi}\Sigma^{\rm cen}(\sqrt{R^2+R_{\rm mis}^2+2RR_{\rm mis}\cos{\phi}}),
\end{aligned}
\end{equation}
and \( \Sigma^{\rm cen}(R) \) denotes the surface mass density centered on the true halo center, described by the NFW profile \( \Delta\Sigma^\mathrm{NFW}(R) \) (see Section~\ref{sec:density_nfw}).

For the remaining fraction of the cluster sample, \( 1 - f_{\rm mis} \), the excess surface mass density is assumed to be perfectly centered and is modeled using the NFW profile, such that \( \Delta\Sigma_{+}^{\rm cen} = \Delta\Sigma_{+}^{\rm NFW} \). To ensure unbiased parameter inference, our analysis marginalizes over the miscentering fraction $f_{\rm mis}$ and scale $\sigma_{\rm mis}$ using uninformative priors as specified in Table~\ref{tab:prior}.

\subsubsection{Photometric-redshift Distribution}
\label{sec:photoz}

Another critical component for realistic modeling of the weak-lensing signal is a reliable characterization of the photo-$z$ distribution, \( P(z) \), of background galaxies selected for the analysis. We compute the stacked distribution  $\langle P\rangle(z)$ for the $P$-cut-selected source galaxies across the 86 XXL clusters within the HSC-SSP footprint, as shown in Figure~\ref{fig:photoz}. We find that the stacked distribution $\langle P\rangle(z)$ is well described by the following parametric form:
\begin{equation}
\begin{aligned}
P_{\rm model}(z) &= \frac{A\beta}{z_1}\left(\frac{z}{z_1}\right)^{\alpha} \exp\left[-\left(\frac{z}{z_1}\right)^\beta\right], \\
A &= \frac{1}{\Gamma\left(\frac{1+\alpha}{\beta}\right)}, \\
z_1 &= z_0 \frac{\Gamma\left(\frac{1+\alpha}{\beta}\right)}{\left(\frac{2+\alpha}{\beta}\right)},
\end{aligned}
\end{equation}
where $\alpha = 32$, $\beta = 0.32$, and $z_0 = 1.13$. 

We adopt the best-fit model, \( P_{\rm model}(z) \), as the effective source redshift distribution for computing the source-averaged excess surface mass density, \( \Delta\Sigma(R|z_l) \), for synthetic clusters. For simplicity, we assume this common \( P_{\rm model}(z) \) distribution applies to all clusters in the XXL-N subsample.

The source-averaged reduced tangential shear profile for each synthetic cluster is then computed using the underlying surface mass densities around the cluster, $\Sigma(R|z_l)$ and $\Delta\Sigma(R|z_l)$ (see also Equation~(\ref{eq:gt})), as \citep{Umetsu2020rev}
\begin{equation}\label{eq:source-ave-deltasigma}
\Delta\Sigma_+(R|z_l) = \frac{\Delta\Sigma(R|z_l)}{1 - \langle \Sigma_{\rm crit}^{-1} \rangle(z_l)\, \Sigma(R|z_l)},
\end{equation}
where the source-averaged inverse critical surface density is defined as
\begin{equation}\label{eq:sigma_cr}
\langle \Sigma_{\rm crit}^{-1} \rangle(z_l) = \frac{\int_{z_l}^{\infty} dz\, P_{\rm model}(z)\, \Sigma_{\rm crit}^{-1}(z_l, z)}{\int_{0}^{\infty} dz\, P_{\rm model}(z)}.
\end{equation}

We emphasize that the excess surface mass density (see Equation~\ref{eq:source-ave-deltasigma}) is evaluated for each synthetic cluster using its corresponding weak-lensing mass, \( M_{500}^\mathrm{WL} \), which includes intrinsic scatter and mass-dependent bias relative to the true halo mass, \( M_{500} \).

\begin{deluxetable}{cccc}
\tablecolumns{4}
\tablewidth{0pt}
\tabletypesize{\scriptsize}
\tablecaption{\label{tab:prior}
Summary of Parameters used in the Forward Simulator and their Prior Distributions}
\tablehead{
    \colhead{Parameter} & 
    \colhead{Prior}    & 
    \colhead{Fudical Values\tablenotemark{a}}  &
    \colhead{Best-fit Values\tablenotemark{b}}
    } 
\startdata
    \multicolumn{4}{c}{Cosmology} \\
    \hline
$\Omega_{m}$  & $\mathcal{U}(0.15, 0.55)$ & 0.28 & $0.366\pm0.091$ \\
$\sigma_8$ & $\mathcal{U}(0.55, 1.0)$  &0.82 & $0.800\pm 0.100$\\
\hline
\multicolumn{4}{c}{Scaling relations} \\
\hline
$\alpha_{T\text{--}M}$ & $\mathcal{U}(1, 5)$ &2.46 & $2.34\pm0.228$\\
$\beta_{T\text{--}M}$ & $\mathcal{U}(0.3, 2)$ & 0.85 & $0.634\pm0.100$\\
$\gamma_{T\text{--}M}$ & $\mathcal{U}(-1.5, 2)$ &0.32 & $0.213\pm0.493$\\
$\alpha_{L\text{--}T}$ & $\mathcal{U}(2, 25)$ &20.9 & $10.9\pm 2.54$\\
$\beta_{L\text{--}T}$ & $\mathcal{U}(2, 3.5)$ &2.63 & $2.99\pm0.285$\\
$\gamma_{L\text{--}T}$ & $\mathcal{U}(-1, 3.5)$ &2.17 & $1.61\pm1.21$\\
$\sigma_{T\text{--}M}$\tablenotemark{c} & $\mathcal{U}(0.01, 0.35)$ & 0.13 & $0.168\pm0.088$\\
$\sigma_{L\text{--}T}$\tablenotemark{c} & $\mathcal{U}(0.01, 0.55)$ &$0.38$ & $0.286\pm0.157$\\
$\delta_{\textrm{ln }T}$ & \multicolumn{3}{c}{Fixed to $0.137$} \\
$\delta_{\textrm{ln }L}$ & \multicolumn{3}{c}{Fixed to $0.154$} \\
\hline
\multicolumn{4}{c}{Core Radius} \\
\hline
$r_c$ & \multicolumn{3}{c}{Fixed to $0.143 R_{500}$} \\
$\sigma(\ln{r_c})$ & \multicolumn{3}{c}{Fixed to 0.332} \\
\hline
\multicolumn{4}{c}{Miscentering} \\
\hline
$f_{\rm mis}$ & $\mathcal{U}(0, 1.0)$ &0.6 & $0.381\pm0.246$\\
$\sigma_{\rm mis}$ & $\mathcal{U}(0, 2.0)$& 0.5 & $1.04\pm0.543$ \\
\hline
\multicolumn{4}{c}{WL mass bias} \\
\hline
$a$ & \multicolumn{3}{c}{Fixed to 0.29}  \\
$b$ & \multicolumn{3}{c}{Fixed to 1.57}  \\
$c$ & \multicolumn{3}{c}{Fixed to 0.73} \\
$\sigma_\mathrm{int}(\ln M_{500}^\mathrm{WL})$ & \multicolumn{3}{c}{Fixed to 0.2} \\
\hline
\multicolumn{4}{c}{Photo-z bias} \\
\hline
$b_z$ & \multicolumn{3}{c}{Fixed to $0.68\%$}  \\
\enddata
\tablenotetext{a}{Fiducial values used to generate synthetic cluster survey simulations in Section~\ref{sec:mock}.}
\tablenotetext{b}{Marginalized posterior constraints (or best-fit values) computed using the biweight location and scale estimators applied to the marginalized posterior distributions obtained from the SBI analysis.}
\tablenotetext{c}{Prior ranges for the intrinsic scatters of the cluster scaling relations are specified in natural logarithmic units.}
\end{deluxetable}

\subsubsection{Photometric-redshift Bias} 
\label{sec:photozbias}

Accurate estimation of photo-$z$s for source galaxies is essential in cluster weak-lensing analyses. Any discrepancies in photo-$z$ estimates can introduce systematic biases in the calculation of the source-averaged critical surface mass density, thereby affecting cluster mass calibration. 

The level of this bias can be quantified using the methodology outlined in \citet{Miyatake2019}. Specifically, the bias in the weak-lensing signal for a cluster at redshift $z_l$ can be estimated as
\begin{equation}
\frac{\Delta\Sigma_+}{\Delta\Sigma_+^{\rm true}} = 1 + b_z(z_l) = \frac{\sum_s w_{ls} \langle\Sigma_{\textrm{crit}, ls}^{-1}\rangle^{-1} [\Sigma_{\textrm{crit}, ls}^{\rm true}]^{-1}}{\sum_s w_{ls}},
\end{equation}
where the summation over $s$ includes all source galaxies, and quantities with the ``true'' superscript are derived from spectroscopic redshifts. Although the HSC shape catalog overlaps with public spectroscopic datasets such as GAMA and VVDS, the number of galaxies with reliable spectroscopic redshifts remains insufficient to fully characterize the HSC source sample.

To address this, \citet{Miyatake2019} introduced a method to estimate photo-\( z \) bias using the COSMOS-30 photo-\( z \) catalog \citep{2009ApJ...690.1236I}. For further methodological details, we refer the reader to their study. Applying this approach, \citet{U2020} estimated the weighted average photo-\( z \) bias \( \langle b_z \rangle \) for a sample of 136 XXL C1+C2 clusters, finding a negligible bias of approximately \( \langle b_z \rangle = 0.68\% \), which was shown to be largely insensitive to the choice of weighting scheme. We adopt this photo-\( z \) bias of \( b_z = 0.68\% \) in our weak-lensing modeling.

\subsubsection{Noise Implementation}
\label{sec:noise}

In the SBI framework, the forward simulator incorporates realistic realizations of statistical noise. To account for uncertainties in the weak-lensing signal, we model the covariance matrix of the \( \Delta\Sigma_+ \) profile as the sum of two components:
\begin{equation}
C = C^{\rm shape} + C^{\rm lss},
\end{equation}
where \( C^{\rm shape}_{ij} = \sigma_{\rm shape}^2(R_i)\, \delta_{ij} \) represents the diagonal shape noise covariance, with \( \delta_{ij} \) denoting the Kronecker delta, and \( C^{\rm lss}_{ij} \) accounts for cosmic noise due to uncorrelated large-scale structure \citep[LSS;][]{Schneider1998, Hoekstra2003}. The shape noise variance, \( \sigma_{\rm shape}^2(R_i) \), is defined in Equation~(\ref{eq:Sigma_shape}).

Shape noise originates from the intrinsic ellipticities of background galaxies, combined with measurement uncertainties in their shape estimation. We adopt a shape dispersion of $\sigma_{g}=0.25$ per galaxy per shear component, consistent with the average shape noise level estimated from the cross-component signals in the HSC shape catalog. 
In our forward simulator, we implement this by adding random Gaussian noise with zero mean and a standard deviation of \( \sigma_{g, \mathrm{eff}} = \sigma_{g} / \sqrt{N_\mathrm{gal}} \) to the shear signal in each radial bin for every cluster. 
The effective shape noise of excess surface mass density, $\Delta\Sigma_+$, is then calculated as \( \sigma_{g, \mathrm{eff}} \, \langle\Sigma_{{\rm crit},ls}^{-1}\rangle(z_l)^{-1}\),
where \(\langle\Sigma_{{\rm crit},ls}^{-1}\rangle(z_l)\) is the source-averaged inverse critical surface density defined in Equation~(\ref{eq:sigma_cr}). Here, $N_\mathrm{gal}$ represents the expected number of source galaxies within each radial bin $[R_i,R_{i+1}]$, calculated as $N_\mathrm{g}=\pi n_\mathrm{g}(R^2_{i+1}-R^2_{i})/D_l^2$, where $n_\mathrm{g}$ denotes the mean surface number density of background galaxies, set as $n_\mathrm{g}=15$~galaxies~arcmin$^{-2}$. This value reflects the effective source density after applying the $P(z)$ cut in HSC weak lensing analyses, also consistent with findings in \citet{2018PASJ...70...30M}.

The cosmic noise contribution introduces correlated uncertainties between radial bins and becomes increasingly significant at larger projected radii, where the cluster lensing signal diminishes \citep{Miyatake2019}. We compute the elements of the LSS covariance matrix following the formalism outlined in Appendix A of \citet{Miyatake2019}, given by
\begin{equation}
    C^{\rm lss}_{i,j}=\langle\Sigma_{\textrm{crit,}ls}^{-1}\rangle^{-2}\int \frac{\ell  d\ell }{2\pi}P_k(\ell)J_2\left(\frac{\ell R_i}{D_l}\right)J_2\left(\frac{\ell R_j}{D_l}\right),
\end{equation}
where $\langle\Sigma_{\textrm{crit,}ls}^{-1}\rangle$ is the source-averaged inverse critical mass density given in Equation~(\ref{eq:sigma_cr}), $J_2(x)$ is the second-order Bessel function of the first kind, 
and $P_k(\ell)$ is the two-dimensional weak-lensing power spectrum as a function of angular multiple $\ell$. 

Given the LSS covariance matrix \( C^{\rm lss} \), we generate correlated noise realizations for \( \Delta\Sigma_+(R_i\,|\,z_l) \) across 10 radial bins for each synthetic cluster using Cholesky decomposition \citep[e.g.,][their Section~6.1.2]{Umetsu2010}.
This procedure is performed independently for each cluster, assuming a fiducial flat $\Lambda$CDM model based on the WMAP 9-year cosmology \citep{Hinshaw2013}, characterized by $\Omega_{m}=0.286$ and $\sigma_8=0.82$.

While the LSS covariance \( C^{\rm lss} \) does depend on cosmological parameters, we fix it to the fiducial cosmology in our forward simulations to reduce computational cost. This simplification is justified as the dominant source of uncertainty in \( \Delta\Sigma_+ \) across our radial range arises from shape noise, making the impact of cosmology-dependent variations in \( C^{\rm lss} \) subdominant in the current analysis.

\subsubsection{Data Summary: Stacked Lensing Profile}

Accounting for the miscentering effect (Section~\ref{sec:miscentering}), source population averaging (Section~\ref{sec:photoz}), photo-$z$ bias (Section~\ref{sec:photozbias}), and statistical uncertainties including shape noise and cosmic noise (Section~\ref{sec:noise}), we compute the stacked \( \langle \Delta\Sigma_+ \rangle(R) \) profile for each temperature-binned subsample of synthetic XXL-N C1 clusters as
\begin{equation}
\begin{aligned}
\label{eq:stackeq}
\langle\Delta\Sigma_{+,j}\rangle(R_i)
=\frac{1}{N_{{\rm cl},j}}\left(\sum_l\Delta\Sigma_{+,l}^{\rm cen}(R_i)+\sum_k\Delta\Sigma_{+,k}^{\rm mis}(R_i)\right)
\end{aligned}
\end{equation}
 
where the indices \( l \) and \( k \) run over the centered and miscentered components, respectively, of clusters in the \( j \)-th temperature bin, and \( N_{{\rm cl},j} \) denotes the total number of clusters in that bin.

\subsection{X-ray Scaling Relations} 
\label{sec:X-ray}

In this study, the X-ray observables  are linked to cluster mass through two fundamental scaling relations: the temperature–mass ($ T_{300\,\mathrm{kpc}}$--$M_{500})$ and the luminosity–temperature ($ L_{500}$--$T_{300\,\mathrm{kpc}}$) scaling relations. 
These relations encode the physical link between the thermodynamic state of the intracluster medium and the underlying dark matter halo mass at each redshift, and are parameterized as power laws of the form:

\begin{equation}
\frac{T_{300\,\mathrm{kpc}}}{T_{\mathrm{piv}}} = \alpha_{T\text{--}M}
\left( \frac{M_{500}}{M_{\mathrm{piv}}} \right)^{\beta_{T\text{--}M}}
\left( \frac{E(z)}{E(z=0.3)} \right)^{\gamma_{T\text{--}M}},
\end{equation}

\begin{equation}
\frac{L_{500}}{L_{\mathrm{piv}}} = \alpha_{L\text{--}T} 
\left( \frac{T_{300\,\mathrm{kpc}}}{T_{\mathrm{piv}}} \right)^{\beta_{L\text{--}T}}
\left( \frac{E(z)}{E(z=0.3)} \right)^{\gamma_{L\text{--}T}},
\end{equation}
where $M_{\mathrm{piv}} = 10^{14}\,h^{-1}\,M_\odot$, $T_{\mathrm{piv}} = 1\,\mathrm{keV}$, and $L_{\mathrm{piv}} = 10^{41}\,\mathrm{erg\,s^{-1}}$ are the adopted pivot values. The quantity $E(z) \equiv H(z)/H_0$ denotes the dimensionless Hubble function, which accounts for the redshift dependence of the scaling relations. The intrinsic scatter in these relations is modeled as a log-normal distribution with standard deviations $\sigma_{T\text{--}M}$ and $\sigma_{L\text{--}T}$, respectively, expressed in natural logarithmic units.

Measurement uncertainties in the X-ray temperature and luminosity are incorporated into the forward modeling. These uncertainties, adopted from the XXL survey data products, are modeled as fractional uncertainties in natural logarithmic units, with values of \( \delta_{\ln T} = 0.137 \) and \( \delta_{\ln L} = 0.154 \), respectively.

For the full synthetic XXL C1 population, the distributions of $T_{300\,\mathrm{kpc}}$ and $L_{500}$ are summarized by their respective quantiles, evaluated at cumulative probability levels of (0.050, 0.169, 0.288, 0.406, 0.525, 0.644, 0.763, 0.881). These quantiles constitute the final set of summary statistics used in our SBI analysis. We note that these marginal distributions of temperature and luminosity discard the cluster-by-cluster pairing information, effectively making the inference insensitive to the correlation coefficient between the intrinsic scatters. Constraining this correlation would require joint summary statistics—such as two-dimensional density estimates in the $L-T$ plane—which would substantially increase the dimensionality of the data vector and the complexity of the inference. We therefore treat the intrinsic scatters as independent and acknowledge this as a limitation of the current analysis.

The normalization, slope, and redshift-evolution parameters, along with the intrinsic scatter, of both scaling relations---$(\alpha, \beta, \gamma, \sigma)_{T\text{--}M}$ and $(\alpha, \beta, \gamma, \sigma)_{L\text{--}T}$---are treated as free parameters and are jointly inferred using the SBI framework, as described in Section~\ref{sec:SBI}.

\subsection{Model Parameters}
\label{sec:model_parameter}

Our forward simulator for cosmological weak-lensing mass calibration is governed by a parameter vector $\boldsymbol{p}$ consisting of:
\begin{itemize}
    \item The cosmological parameters $(\Omega_{m}, \sigma_8)$;
    \item The scaling relation parameters $(\alpha, \beta, \gamma, \sigma)_{T\text{--}M}$ and $(\alpha, \beta, \gamma, \sigma)_{L\text{--}T}$;
    \item The parameters characterizing the miscentering effect $(f_{\rm mis}, \sigma_{\rm mis})$.
\end{itemize}

All relevant prior distributions, assumed to be uninformative, are summarized in Table~\ref{tab:prior}. This modeling approach allows for a self-consistent calibration of X-ray observables in terms of cluster mass, while simultaneously incorporating cosmological uncertainties and systematic effects.

\section{Simulation-based Inference} \label{sec:SBI}

SBI encompasses a class of statistical methods designed to estimate posterior distributions when the likelihood function is computationally intractable. Instead of relying on explicit likelihood evaluations, SBI leverages forward-modelled simulations to compare theoretical predictions with observations. Prominent techniques within this framework include Approximate Bayesian Computation \citep[ABC;][]{ABC}, Sequential Monte Carlo \citep[SMC;][]{SMC}, and Density Estimation Likelihood-Free Inference \citep[DELFI;][]{2012arXiv1212.1479F,2016arXiv160506376P,2017arXiv171101861L,2018arXiv180507226P,2018arXiv180509294L,2018MNRAS.477.2874A}.

In our previous work \citep{Tam2022}, we evaluated the performance of both ABC and DELFI in the context of cluster cosmology. While both approaches produced unbiased parameter constraints, DELFI proved to be significantly more efficient, achieving comparable precision with an order of magnitude fewer simulations. Based on these findings, we adopt DELFI for the present analysis of the XXL and HSC datasets.

DELFI employs neural density estimators to model the conditional probability distribution $ p(\boldsymbol{t}|\boldsymbol{p}) $, where \( \boldsymbol{t} \) denotes summary statistics derived from the data, and \( \boldsymbol{p} \) are the model parameters defined in Section~\ref{sec:model_parameter}. Once trained, the likelihood for a given observed data vector \( \boldsymbol{t}_o \) is approximated as \( p(\boldsymbol{t}_o|\boldsymbol{p}) \), from which the posterior distribution is obtained via Bayes’ theorem:  
\[
p(\boldsymbol{p}|\boldsymbol{t}_o) \propto p(\boldsymbol{t}_o|\boldsymbol{p}) \, p(\boldsymbol{p}).
\]

We implement DELFI using the publicly available \textsc{pydelfi} package\footnote{\url{https://github.com/justinalsing/pydelfi}} \citep{2019MNRAS.488.4440A}, which combines adaptive simulation acquisition with flexible neural density estimators. We integrated DELFI with our forward model described in Section~\ref{sec:modeling} with the workflow as follows: The initial set of model parameters $\boldsymbol{p}$ are sampled from the prior distributions specified in Table~\ref{tab:prior}, ensuring broad coverage of the parameter space. In subsequent training, parameters are drawn adaptively using the active acquisition strategy implemented in \textsc{pydelfi}, which concentrates simulations in regions of high posterior density inferred from previous training iterations. This sequential sampling improves computational efficiency while preserving accuracy in the inferred posterior distributions. For each sampled parameter set, we generate synthetic data summaries \( \boldsymbol{t}(\boldsymbol{p}) \) using the forward model. The summary statistics include redshift-binned cluster number counts, stacked weak-lensing profiles in three temperature bins, and quartile values describing the distribution of X-ray luminosity and temperature.  The simulations are used to train the neural density estimator, and the procedure is iteratively repeated until convergence of the posterior distribution is reached.

We construct the neural network architecture using an ensemble of eight neural density estimators (NDEs) to improve robustness and reduce sensitivity to individual network configurations. The ensemble consists of four masked autoregressive flows (MAFs), each composed of five masked autoencoders for density estimation (MADEs), with each MADE incorporating two hidden layers containing 50 units per layer. In addition, the ensemble includes four mixture density networks (MDNs), each using five Gaussian components, and each MDN configured with two hidden layers of 50 units. All networks utilize the $\tanh$ activation function.

To mitigate overfitting, we reserve 10\% of the simulation data as a validation set during training.
To summarize the posterior constraints, we employ the biweight estimators of \citet{1990AJ....100...32B} for the central location and scale of the marginalized posterior distributions. We adopt these metrics as our best-fit parameter estimates and uncertainties, respectively, throughout this study.

\section{Results of Cosmological Inference} \label{sec:result}

\subsection{Pipeline validation using Synthetic Catalogs} \label{sec:mock}
Before presenting the main cosmological inference results based on the XXL+HSC dataset, we validate the performance and reliability of our SBI pipeline using synthetic XXL-N surveys generated from our forward modeling framework.

To construct these synthetic surveys, we adopt a fiducial parameter model 
$F$, with 12 fixed parameters listed in Table~\ref{tab:prior}. We employ the neural network architecture described in Section~\ref{sec:SBI}, and perform SBI using the \textsc{pydelfi} package.
The procedure begins with 1,200 initial simulations, followed by iterative batches of 800 simulations for each subsequent round of training.

We first assess the robustness of our inference pipeline by performing a parameter recovery test. We ran our SBI pipeline on 12 independent realizations of a synthetic XXL+HSC dataset generated from the fiducial parameter model (Model $F$), each incorporating realistic observational noise. Figure~\ref{fig:mock} displays the posterior distributions obtained via DELFI. The vertical dashed lines indicate the true input values of the synthetic data. While the cosmological parameters and the $M_{500}$--$T_{300\,\rm kpc}$ relation are well constrained, the $L_{500}$--$T_{300\,\rm kpc}$ relation is weakly constrained in its redshift evolution and exhibits strong covariance between normalization and slope. This limitation primarily reflects our choice of summary statistics: by characterizing the luminosity and temperature distributions using global quantiles without explicit redshift binning, the inference has limited sensitivity to the evolution parameter $\gamma_{L-T}$. Incorporating redshift-binned X-ray summaries could improve constraints on the $L_{500}$--$T_{300\,\rm kpc}$ evolution in future work.

Nuisance parameters---including the intrinsic scatters ($\sigma_{T\text{--}M}$, $\sigma_{L\text{--}T}$) and the miscentering parameters ($\sigma_{\rm mis}$, $f_{\rm mis}$)---are marginalized over wide prior ranges to ensure unbiased inference. The observed deviations of the posterior means from the fiducial values arise from stochastic variations and observational noise propagated through the forward model. Using a consistent neural density estimator architecture across all realizations, we find the median of the 12 posterior samples for the cosmological parameters to be $\Omega_{m} = 0.311 \pm 0.042$ and $\sigma_8 = 0.790 \pm 0.039$. The fiducial values lie within approximately $0.75\sigma$ of these ensemble means for both parameters, indicating that the inference is statistically consistent and unbiased under realistic data variability.

We further evaluate the sensitivity of our results to network initialization and architecture by varying the number of layers and mixture components in both MAF and MDN models, using a fixed synthetic dataset. Across four independent trainings of different neural density estimators, we observe systematic shifts of $\sim13\%$ in the posterior mean of $\Omega_{m}$ and $\sim9\%$ in $\sigma_8$. We account for this effect by incorporating it as an additional systematic uncertainty in our final cosmological parameter estimates. Notably, the derived parameter combination $S_8$ remains robust, with systematic shifts staying below $3\%$. This indicates that neural network variability primarily affects the solution along the $\Omega_{m}$--$\sigma_8$ degeneracy direction rather than the overall signal amplitude constrained by the data. Beyond this specific degeneracy, the parameters of the two scaling relations show consistent and converged posterior distributions across all network configurations.

\begin{figure*}[htbp]
 \centering
 \includegraphics[width=1\textwidth,trim={0.2cm 0cm 0cm 0cm},clip]{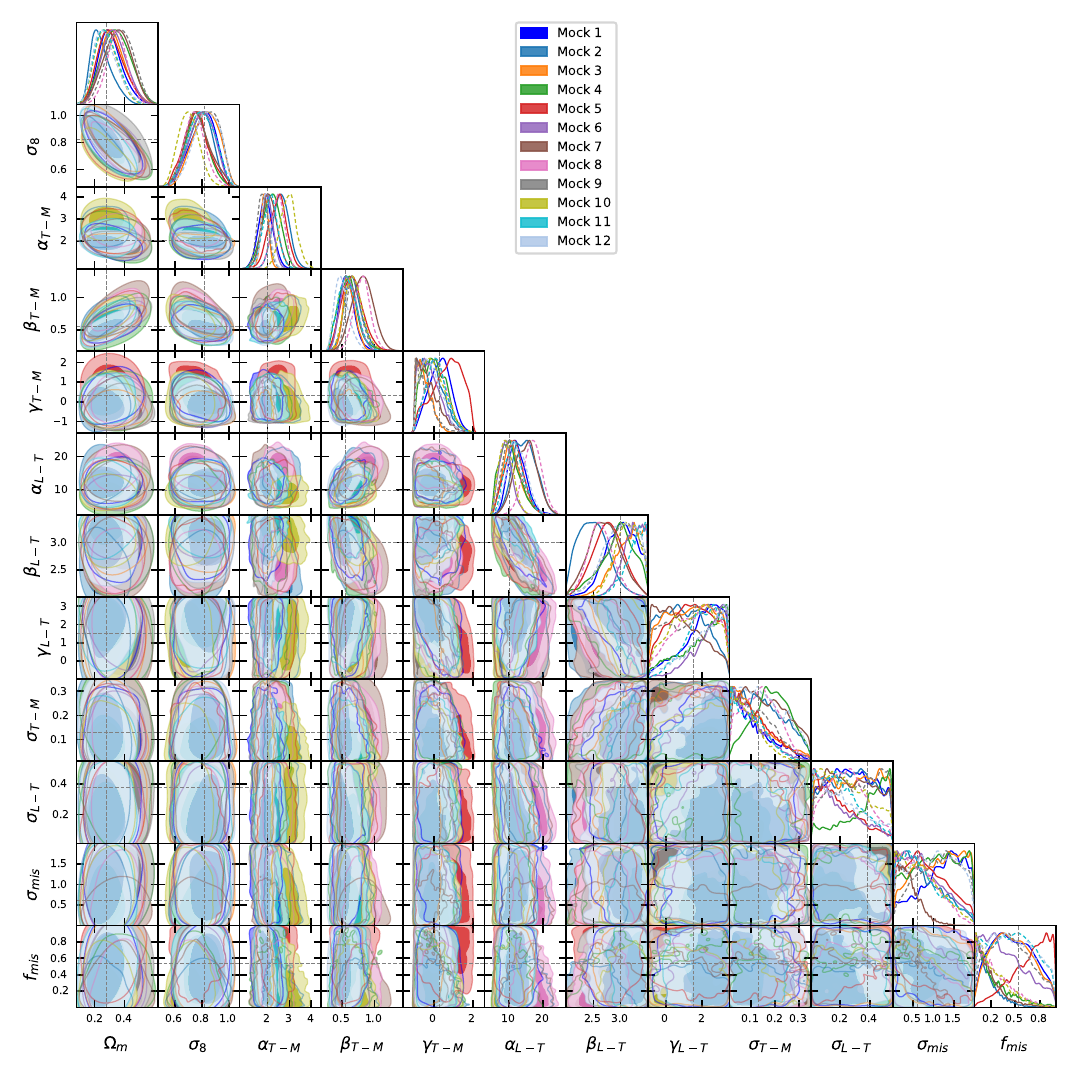}
 \caption{Parameter constraints showing the marginalized one-dimensional (histograms) and two-dimensional (68\% and 95\% confidence level contours) posterior distributions for 12 independent random realizations of synthetic XXL-HSC observations, inferred using \textsc{pydelfi}. The black dashed lines indicate the fiducial values used to generate the synthetic surveys.}
 \label{fig:mock}
\end{figure*}

\begin{figure*}[htbp]
 \centering
 \includegraphics[width=1\textwidth,trim={0.2cm 0cm 0cm 0cm},clip]{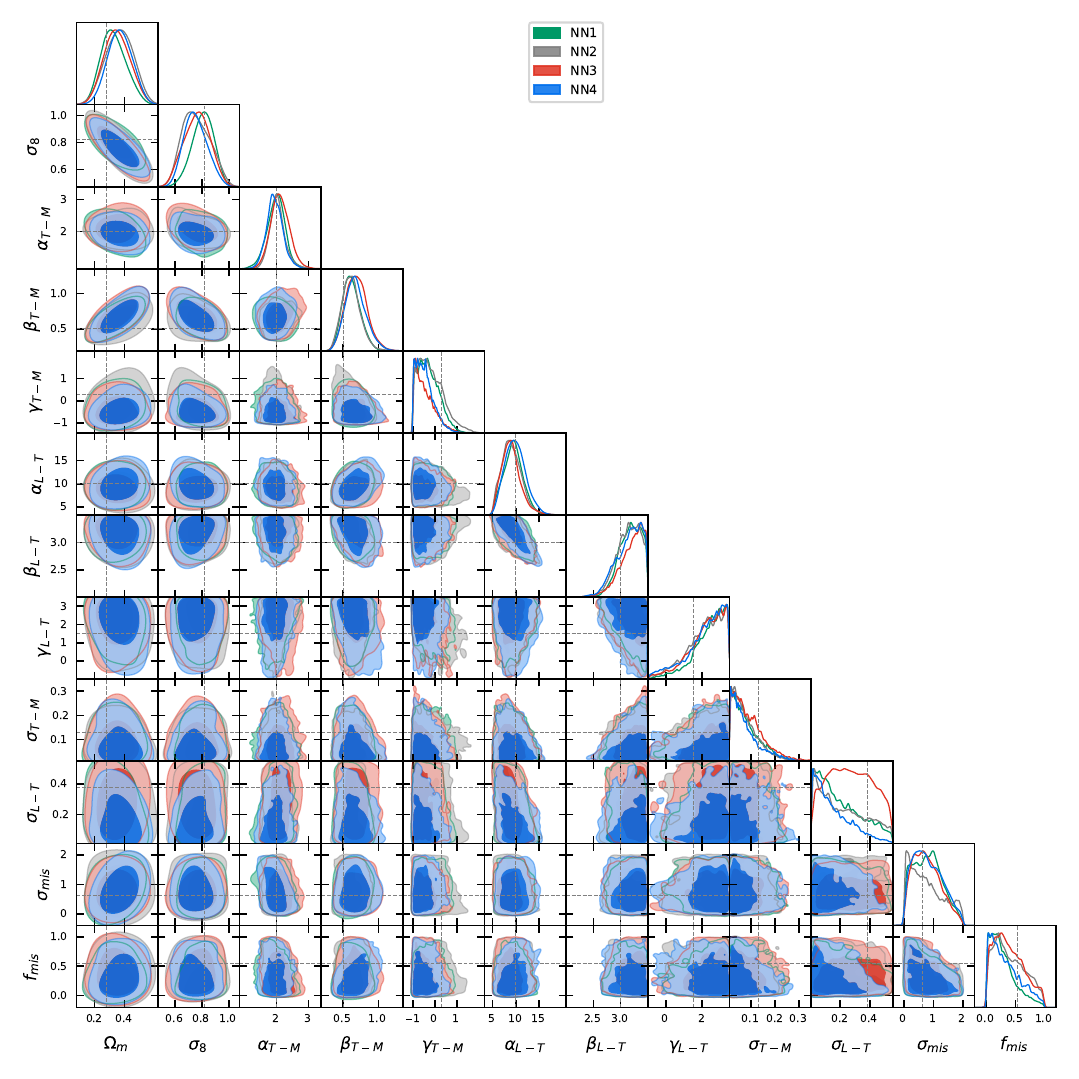}
 \caption{Parameter constraints showing the marginalized one-dimensional (histograms) and two-dimensional (68\% and 95\% confidence level contours) posterior distributions from four independent \textsc{pydelfi} runs, each using a different neural-network architecture, applied to the same synthetic XXL--HSC observations. The black dashed lines indicate the fiducial values used to generate the synthetic survey.}
 \label{fig:mock}
\end{figure*}

\begin{figure*}[htbp]
 \centering
 \includegraphics[width=1\textwidth,trim={0.2cm 0.2cm 0cm 0cm},clip]{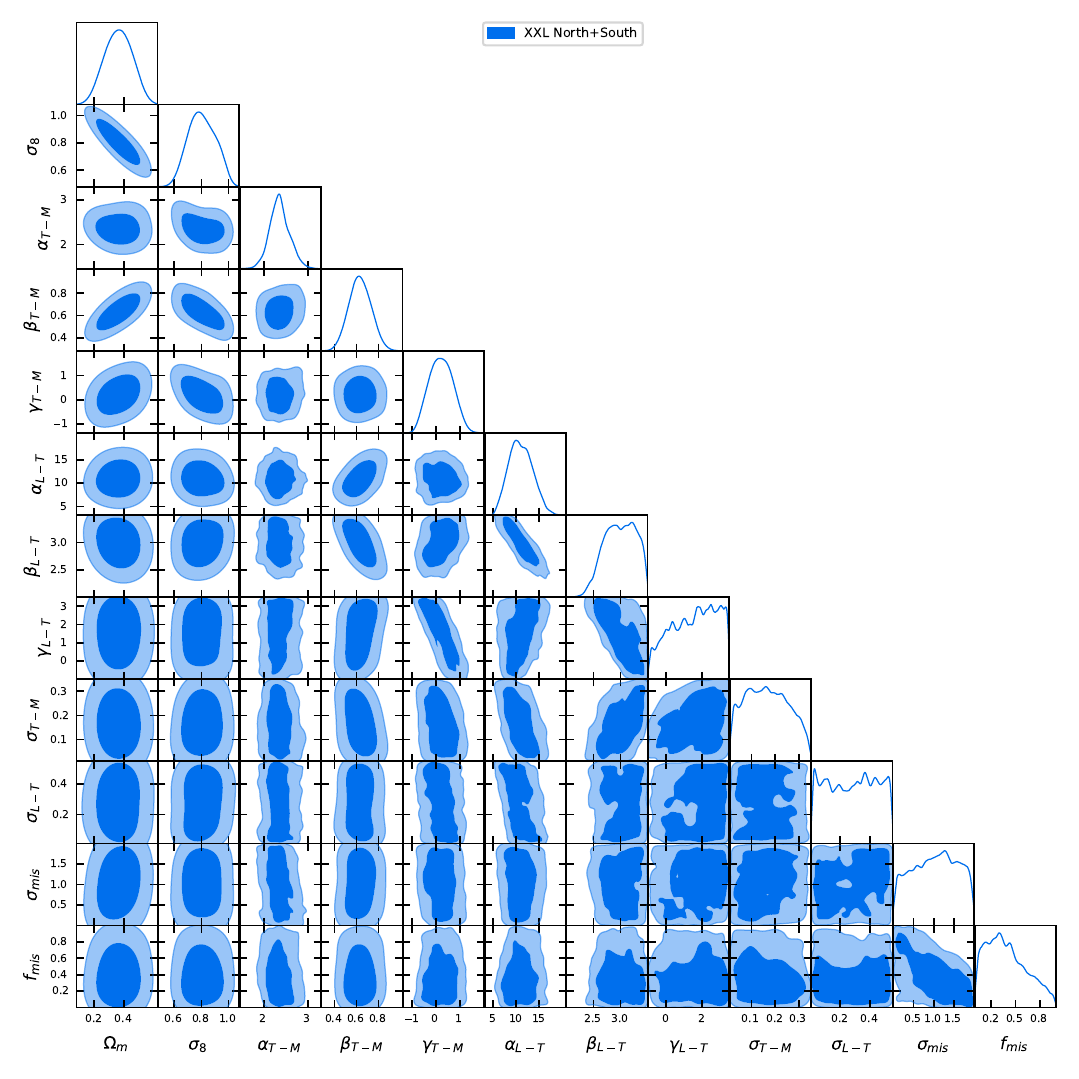}
 \caption{Parameter constraints showing the marginalized one-dimensional (histograms) and two-dimensional (68\% and 95\% confidence level contours) posterior distributions derived from the observed joint XXL--HSC dataset, inferred using \textsc{pydelfi}.}
 \label{fig:real}
\end{figure*}

\begin{figure}[htbp]
 \centering
 \includegraphics[width=0.5\textwidth,trim={0.5cm 0cm 0cm 1cm},clip]{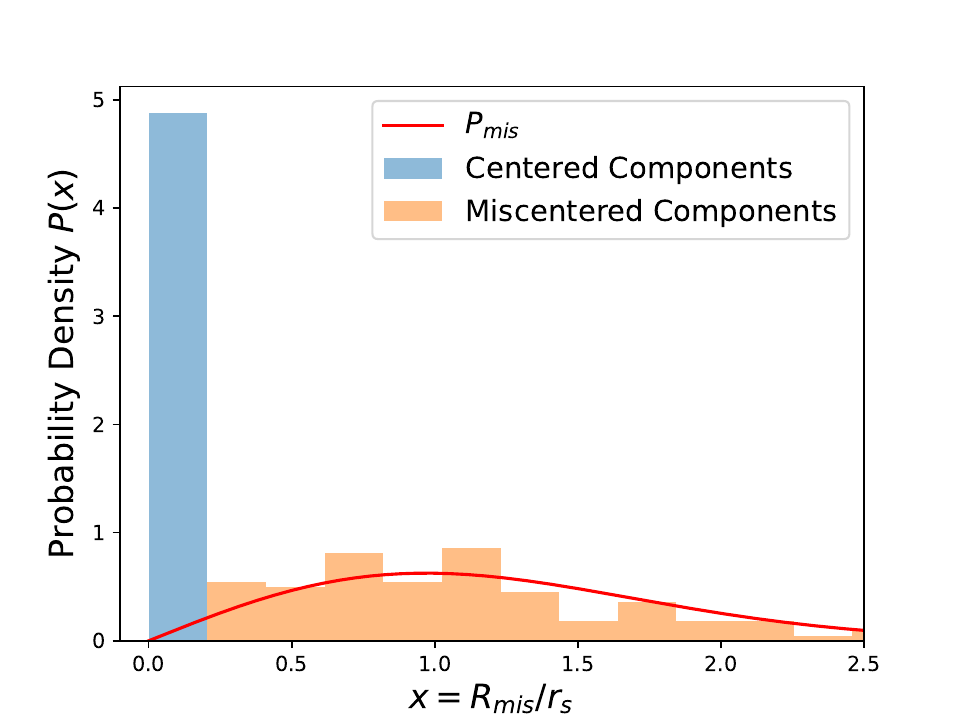}
 \caption{The probability density distribution of the miscentering offset, $x = R_{\mathrm{mis}}/r_s$, derived from the best-fit values of the miscentering fraction $f_{\mathrm{mis}}$ and the miscentering scatter $\sigma_{\mathrm{mis}}$. The blue bar corresponds to the fraction $(1-f_{\mathrm{mis}})$ of clusters that are well-centered ($x = 0$). The orange histogram represents the remaining $f_{\mathrm{mis}}$ fraction of clusters, whose miscentering offsets follow the Rayleigh distribution shown by the red solid line.}
 \label{fig:Prob_mis}
\end{figure}

\begin{figure}[htbp]
 \centering
\includegraphics[width=0.45\textwidth,trim={0.1cm 0cm 0.3cm 0cm},clip]{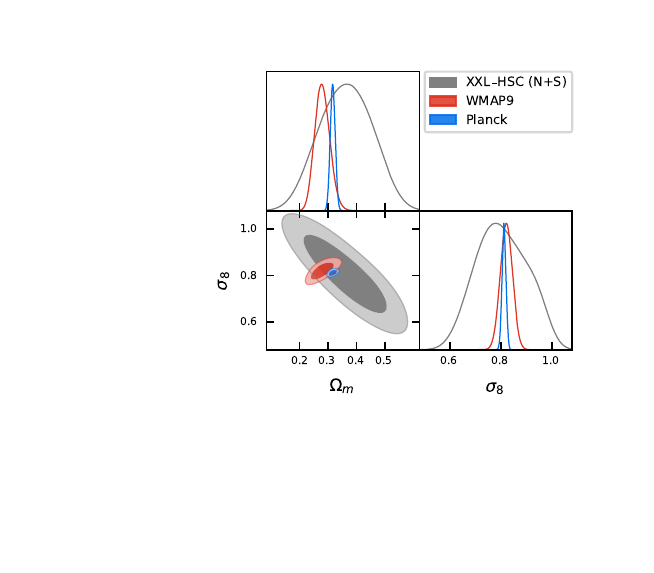}
 \caption{Comparison of the $(\Omega_{m}, \sigma_8)$ constraints from our XXL-HSC cluster-based analysis with the CMB results from \textit{WMAP9} and \textit{Planck}.}
 \label{fig:compare_with_CMB}
\end{figure}

\begin{figure*}[htbp]
 \centering
 \includegraphics[width=0.8\textwidth,trim={0.5cm 0cm 0cm 0cm},clip]{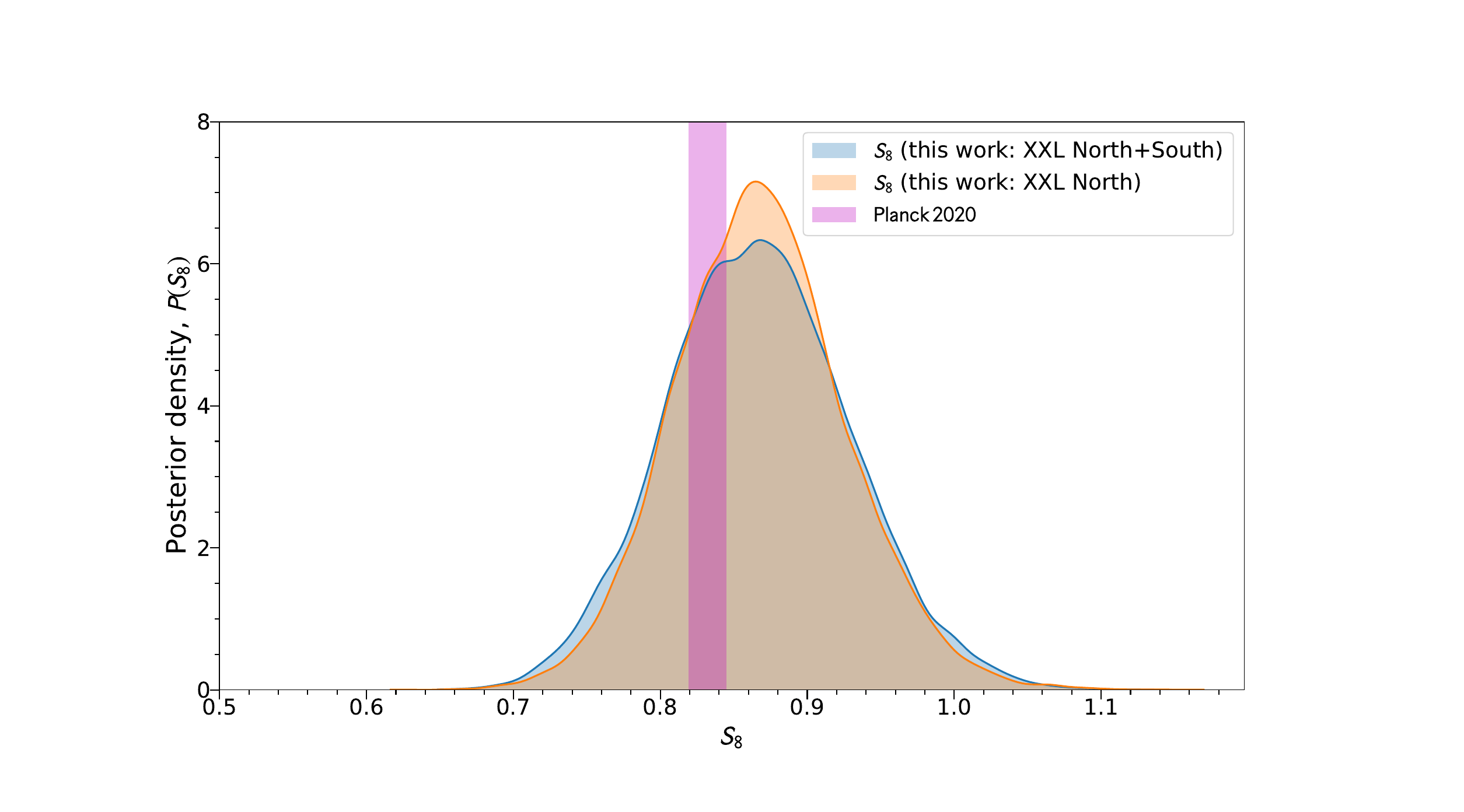}
 \caption{Marginalized posterior distribution of $S_8=\sigma_8(\Omega_{m}/0.3)^{0.5}$ derived from the joint XXL-HSC survey data using \textsc{pydelfi}. The blue distribution shows the posterior derived from the full XXL C1 sample (XXL-N+S), yielding a best-fit value of $S_8=0.867\pm0.063$. The orange distribution corresponds to the posterior from the XXL-N C1 subsample.The pink shaded region indicates the \Planck CMB measurement of $S_8$
for comparison.}
 \label{fig:S8}
\end{figure*}

\begin{figure}
\centering
 \includegraphics[width=0.5\textwidth,trim={0.0cm 0cm 0cm 0cm},clip]{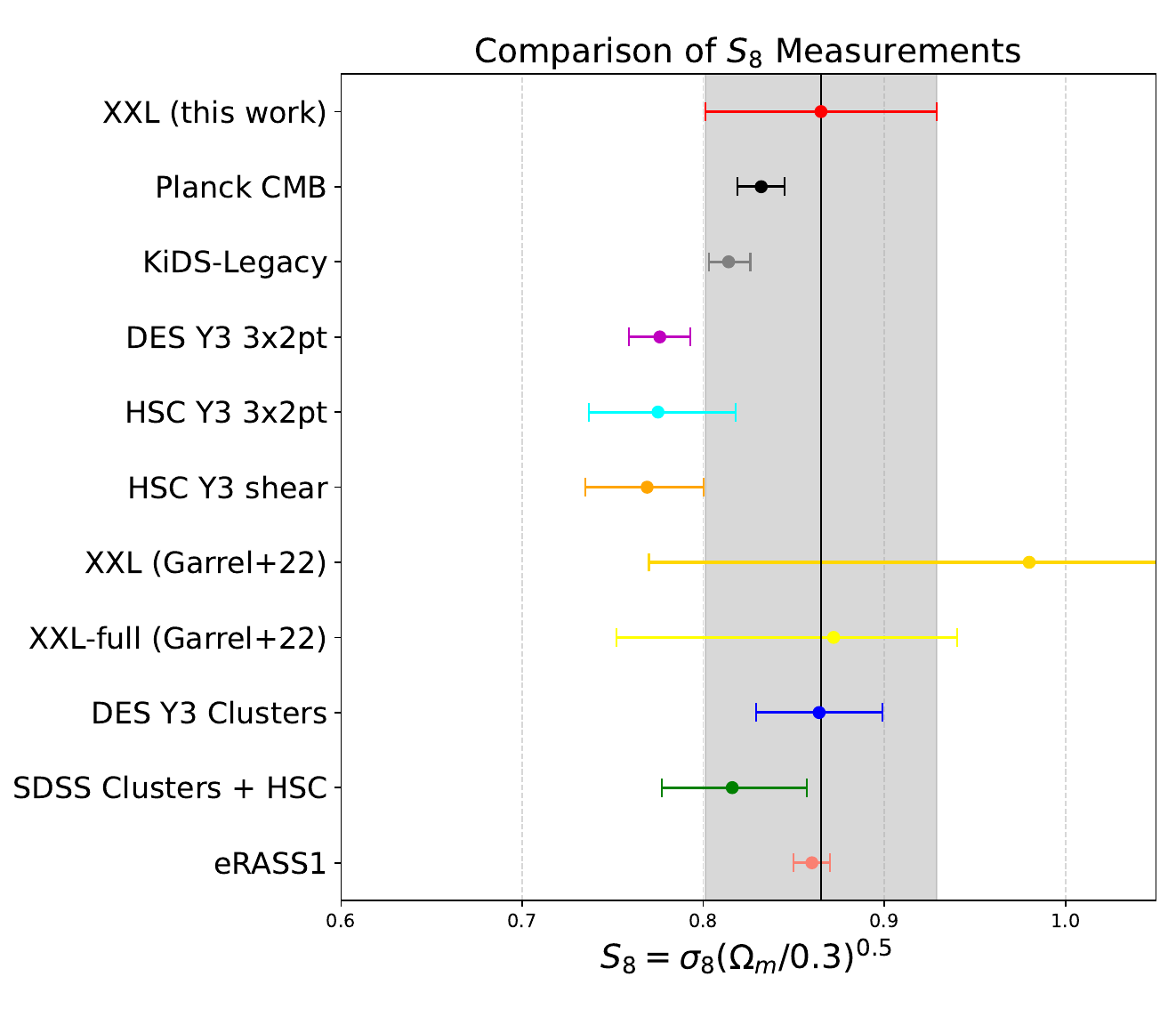}
 \caption{Comparison of our $S_8$ estimation with various cosmological probes, including cosmic shear analyses, cluster-based measurements, and the \Planck CMB result. }
 \label{fig:S8_hist}
\end{figure}

\subsection{Results with XXL C1 clusters }
In this section, we perform a cosmological weak-lensing mass calibration utilizing \textsc{pydelfi} on XXL-HSC observational data. We employ the neural network configuration described in Section~\ref{sec:mock}. The resulting posterior distributions of all free parameters are presented in Figure~\ref{fig:real} and their corresponding best-fit values computed from the posteriors are listed in Table~\ref{tab:prior}. 

To assess the validity of the posterior samples in accurately describing the observed data, we input the posterior parameter set into our forward model to generate a large ensemble of random realizations. Figure~\ref{fig:XXLsel_gt} presents the observed stacked shear profile for the XXL C1 sample alongside the simulated lensing signals derived from $\sim$ 5000 posterior samples. These simulated signals inherently incorporate all noise characteristics, effectively replicating the statistical and systematic uncertainties present in real observational measurements. The comparison shows that both the magnitude and shape of the simulated lensing profiles exhibit good agreement with the observed data.

In addition to the lensing signals, Figure~\ref{fig:Num_vs_z} and Figure~\ref{fig:L_vs_T} present three additional simulated summary statistics: the cluster number abundance, and the distributions of X-ray temperature and luminosity. These simulated observables are generated from posterior samples obtained through the forward-modeling pipeline and are consistent with the corresponding distributions observed in the XXL C1 cluster sample. We note that the intrinsic scatter in the observed XXL data appears visually tighter than that of the simulated distributions. This difference arises from two main effects. 

First, in the XXL catalog, the X-ray luminosity $L_{500}$ is measured within $R_{500}$, which is itself derived from the $M_{500}$--$T_{300\,\rm kpc}$ relation of \citet{2016A&A...592A...4L}. This procedure induces a positive correlation between the measurement uncertainties of $T_{300\,\rm kpc}$ and $L_{500}$. In contrast, within our SBI forward model, $T_{300\,\rm kpc}$ and $L_{500}$are generated from their respective scaling relations, and we do not explicitly model this aperture-induced covariance in the measurement errors.

Second, the shaded blue contours in Figure~\ref{fig:L_vs_T} represent the posterior predictive distribution, obtained by stacking approximately 5,000 realizations drawn from the posterior chain. These contours therefore incorporate both intrinsic scatter and parameter uncertainty, and are consequently broader than the scatter observed in any single realization of the data. However, statistically, the XXL data points fall well within the 2$\sigma$ and 3$\sigma$ contours of the prediction, indicating that our inferred cosmology and scaling relations are consistent with the observations despite this visual difference.

As part of the same posterior predictive framework, we also infer the distribution of miscentering offsets that impact the weak-lensing signal. Figure~\ref{fig:Prob_mis} presents the miscentering-offset distribution for the synthetic XXL C1 clusters based on the best-fit parameters, 
$f_{\mathrm mis}$ and $\sigma_{\rm mis}$.
It is worth noting that a physical degeneracy exists between cluster miscentering and baryonic effects in the inner halo. Miscentering suppresses the central weak-lensing signal by offsetting the assumed cluster center, while strong non-thermal energy injection mechanisms, such as AGN feedback, can similarly reduce the inner mass density relative to a dark-matter-only NFW profile. Since our forward model does not explicitly include baryonic modifications, part of the inferred miscentering signal may effectively absorb these unmodeled baryonic effects. This limits the extent to which the miscentering parameters can be interpreted purely as geometric offsets.

In addition to the full XXL C1 sample, we perform a consistency check in Appendix~\ref{sec:C1-N} using only the C1 clusters located in the XXL-N region, which overlaps with the HSC survey footprint. The posterior distributions inferred from this subsample are consistent with those obtained from the full C1 analysis that combines the XXL-N and XXL-S regions, as shown in Figure~\ref{fig:post_all}. This agreement demonstrates that our results are not driven by a specific sky region and highlights the robustness of the SBI framework in modeling cluster observables across different subsamples.

\subsection{Cosmological constraints}

We obtain marginalized constraints on the cosmological parameters, $\Omega_{m} = 0.366 \pm 0.091$ and $\sigma_8 = 0.80 \pm 0.10$.
These quoted uncertainties represent the combined budget of statistical noise and physical systematics propagated through the forward model, excluding the additional methodological uncertainty from neural network stochasticity (see Section~\ref{sec:mock}).
Figure~\ref{fig:compare_with_CMB} compares the resulting $\Omega_{m}\textrm{--}\sigma_8$ posterior with the CMB constraints from \textit{WMAP9} \citep{WMAP9} and \textit{Planck} \citep{Planck2018VI} measurement. Our results are consistent with the CMB measurements within the 2$\sigma$ level. Figure~\ref{fig:S8} shows the marginalized posterior distribution of $S_8=\sigma_8(\Omega_{m}/0.3)^{0.5}$. 

From the XXL cluster sample, we obtain a constraint of $S_8=0.867\pm0.063$, corresponding to a precision of approximately $7\%$. This result is also consistent with the \textit{Planck} CMB measurement \citep{Planck2018VI}. A comparison with recent literature constraints on $S_8$ is shown in Figure~\ref{fig:S8_hist}.

As discussed in Section~\ref{sec:mock}, we estimate an additional systematic uncertainty arising from stochastic variability of the neural density estimators, corresponding to approximately $13\%$ in  $\Omega_{m}$, $9\%$ in $\sigma_8$, and $3\%$ in $S_8$. These contributions reflect the stability of the inference framework rather than physical uncertainty in the data and are therefore quoted separately and not incorporated into the posterior contours shown here.

\citet{2022A&A...663A...3G} performed a forward cosmological analysis of spectroscopically confirmed XXL C1 galaxy clusters using the XXL DR1 catalog. Using the XXL C1 sample alone, they constrained the cosmological parameters to be $\Omega_{m} = 0.378 ^{+0.068}_{-0.13}$, $\sigma_8 = 0.89^{+0.12}_{-0.28}$, and $S_8 = 0.97^{+0.067}_{-0.21}$, where the inferred value of $S_8$ being higher than that derived in this work. By further combining their analysis with the two-point correlation function of XXL clusters and baryon acoustic oscillation (BAO) data, they achieved a tighter constraint of $S_8 = 0.872^{+0.068}_{-0.12}$ which is consistent with our findings. 

Although our analysis and \citet{2022A&A...663A...3G} use the same dataset of XXL C1 clusters, our inference yields tighter constraints on cosmological parameters. The improved precision and the shift in the inferred value of $S_8$ likely arise from several key methodological differences. First, our forward model incorporates population modeling of halo structure through a cosmology-dependent concentration–mass ($c$--$M$) relation. This allows additional cosmological information to be extracted from the internal structure of dark matter halos and helps to break parameter degeneracies. In contrast, \citet{2022A&A...663A...3G} relied on the mass calibration of \citet{U2020}, who treated halo concentration as a free parameter with broad priors to ensure unbiased mass estimation. While this approach minimizes model dependency, it effectively decouples the halo concentration from the background cosmology, meaning that the constraining power of the $c$--$M$ relation was not utilized in their analysis.

Secondly, our SBI framework integrates the weak-lensing mass calibration and cosmological inference into a single, simultaneous forward-modeling loop. This ensures that the covariance between scaling-relation parameters and cosmological parameters is fully captured. In contrast, \citet{2022A&A...663A...3G} adopted a step-wise approach, using the mass calibration constraints derived by \citet{U2020} as inputs for their cluster abundance analysis. While both analyses account for the survey selection function, our joint inference avoids the potential loss of information associated with separating the mass calibration from the cosmological fit.

Furthermore, our forward model employs self-consistent geometry, with cosmological parameters directly determining angular diameter distances and critical surface densities at every step of the simulation. In comparison, \citet{2022A&A...663A...3G} employed an approximate rescaling of distances relative to a fixed \textit{WMAP9} reference cosmology, which may not fully capture the cosmology dependence of the weak-lensing observables.

We also compare our results with constraints from other cluster-based measurements. \citet{Sunayama2024} report \( S_8 = 0.816^{+0.041}_{-0.039} \) based on optically selected clusters from the Sloan Digital Sky Survey (SDSS), using weak lensing data from the HSC survey. \citet{DES2025} find \( S_8 = 0.864 \pm 0.035 \) based on optically selected redMaPPer clusters from DES Year 3 data. Recent constraints from weak-lensing shear-selected clusters based on the HSC survey reports \( S_8 = 0.993^{+0.084}_{-0.126}\) \citep{Chiu2024}. These results are  statistically consistent with our $S_8$ constraint. In contrast, a joint analysis of cluster abundance and weak-lensing measurements of AMICO galaxy clusters in the KiDS-1000 survey finds a significantly lower value of $S_8=0.74\pm0.03$, compared with our measurement.

Our result shows a higher value of $S_8$ compared to the cosmic shear measurement from HSC survey \citep[e.g.][]{HSC_cosmicshear, HSC_cosmicshear_twopoint, HSC_cosmicshear3X2} and DES \citep{DESYear3cosmicshear} with mild discrepancy at
levels of $<$ 2$\sigma$. Nevertheless, our result remains consistent with the $S_8$ constraint from the KiDS-Legacy cosmic shear analysis \citep{KiDS2025}.

\subsection{X-ray Observables -- Cluster Mass Scaling Relations} \label{sec:scaling relations}
The inferred X-ray observable scaling relations are expressed as follows:

\begin{align}\label{eq:bf-T_M}
\frac{T_{300\,\mathrm{kpc}}}{T_{\mathrm{piv}}} &= (2.34 \pm 0.228) 
\left( \frac{M_{500}}{M_{\mathrm{piv}}} \right)^{0.634 \pm 0.100} \nonumber \\
&\quad \times \left( \frac{E(z)}{E(z=0.3)} \right)^{0.213 \pm 0.493},
\end{align}

\begin{align}\label{eq:bf-L_T}
\frac{L_{500}}{L_{\mathrm{piv}}} &= (10.9 \pm 2.54) 
\left( \frac{T_{300\,\mathrm{kpc}}}{T_{\mathrm{piv}}} \right)^{2.99 \pm 0.285} \nonumber \\
&\quad \times \left( \frac{E(z)}{E(z=0.3)} \right)^{1.61 \pm 1.21},
\end{align}

where the pivotal quantities are defined in Section~\ref{sec:X-ray}. The log-normal scatters, $\sigma_{T\text{--}M}=0.168\pm0.088$ and $\sigma_{L\text{--}T}=0.286\pm0.157$, remain weakly constrained.

We compare the constraints on the scaling relations with predictions from the self-similar gravitational collapse model. For  $T_{300\,\rm kpc}$--$M_{500}$ relation, self-similar predicts the relation $T_{X} \propto E^{2/3}(z) M_{\Delta}^{2/3}$ by assuming that the system is in virial equilibrium, where $G M_{\Delta} / R_{\Delta} \sim T_{X}$, and that mass scales with radius as $M_{\Delta} \propto R_{\Delta}^3 \rho_c$. 

Our results yield a mass trend for the $T_{300\,\mathrm{kpc}}$-- $M_{500}$ relation of \( \beta_{T\text{--}M} = 0.634 \pm 0.100 \), nearly in agreement with the self-similar prediction of \( \beta_{ T\text{--}M,\mathrm{ss}} = 2/3 \).  The redshift evolution parameter we infer is \( \gamma_{T\text{--}M} = 0.213 \pm 0.493 \), which is shallower than the self-similar expectation but remains consistent within the uncertainty.

\citet{2016A&A...592A...4L} derived the $T_{300\,\mathrm{kpc}}$--$M_{500}^{\mathrm{WL}}$ relation using the XXL Data Release 1 (DR1) and the weak-lensing shear catalog from the Canada-France-Hawaii Telescope Lensing Survey \citep[CFHTLenS;][]{2012MNRAS.427..146H,2013MNRAS.433.2545E}, focusing on a subsample of 38 XXL-N clusters at \( z \leq 0.6 \). They reported a best-fit mass slope of \( \sim 0.56^{+0.12}_{-0.10} \). By combining this sample with 48 massive clusters from the Canadian Cluster Comparison Project \citep[CCCP;][]{CCCP} and 10 X-ray selected groups from COSMOS \citep{COSMOS}, they obtained a revised slope of \( \sim 0.60^{+0.04}_{-0.05} \) which is shallower than our result but consistent within uncertainties.
\citet{U2020} used S16A HSC lensing data to measure the $T_{300\,\mathrm{kpc}}$--$M_{500}$ relation for 136 systems at \( 0.031 \leq z \leq 1.033 \) detected in the 25 deg\(^2\) XXL-N region from the \XMM-XXL survey. They obtained the mass slope as $0.85 \pm 0.31$ with redshift trend $0.18\pm 0.66$. When redshift scaling fixed to the self-similar
prediction, they mass trend become as $0.75 \pm 0.27$. \cite{2020MNRAS.492.4528S} studied the scaling relations between X-ray properties and weak lensing mass for for XXL Survey systems and  find \( \beta_{T\text{--}M_{wl}} = 0.78 \pm 0.43 \). \citet{2022A&A...663A...3G} performed a forward cosmological analysis of the XXL C1 clusters. They reported a mass slope of  $0.85 \pm 0.39$
for the $T_{300\,\rm kpc}$--$M_{500}$ relation, with a redshift evolution trend of $0.32 \pm 0.75$. These previous measurements of the $T_{300\,\mathrm{kpc}}$--$M_{500}$ relation for XXL systems exhibit higher values of the mass trend parameter, but remain broadly consistent with our results within the uncertainties.

\citet{2022A&A...661A..11C} performed weak-lensing mass calibration and X-ray scaling relation analysis for systems selected in the eROSITA Final Equatorial Depth Survey (eFEDS). Their \( T_{300\,\rm kpc}\text{--}M_{500} \) relation yields a mass slope of \( \beta_{T\text{--}M} = 0.65 \pm 0.11 \), along with a similarly shallow redshift evolution trend, also consistent with our findings.

For the $L_{500}$--$T_{300\,\mathrm{kpc}}$ scaling relation, the self-similar model predicts a soft-band dependence of the form $L_{X} \propto E(z) T_{X}^{3/2}$. Our results indicate a significantly steeper trend, with $\beta_{L\text{--}T} = 2.99 \pm 0.29$. This behavior is consistent with previous studies \citep[e.g.,][]{1998ApJ...504...27M,2012MNRAS.421.1583M,2016MNRAS.463..820Z}, which have shown that the $L_{500}$--$T_{300\,\rm kpc}$ relation is systematically steeper than the self-similar prediction. This deviation is commonly attributed to radiative cooling and feedback mechanisms. In practice, the thermodynamic state of the intracluster medium is not determined solely by gravitational collapse; various forms of feedback---including star formation and active galactic nucleus (AGN) activity---inject non-thermal energy, introducing additional scatter and steepening the scaling relation.
The redshift evolution of the $L_{500}$--$T_{300\,\mathrm{kpc}}$ relation is weakly constrained in our analysis, consistent with the validation results presented in Section~\ref{sec:mock}. In future work, incorporating redshift-binned temperature and luminosity distributions could improve sensitivity to redshift-dependent trends, thereby tightening constraints on the evolution parameter $\gamma_{L\text{--}T}$.

Our measured $L_{\rm 500}$--$T_{300\,\rm kpc}$ relation aligns with the findings of \citet{2016A&A...592A...3G}, who analyzed the $L_{500}$--$T_{300\,\rm kpc}$ relation for 100 brightest clusters in the XXL Survey and obtained a slope of $3.08 \pm 0.15$ with a redshift evolution factor of $E(z)^{1.64\pm0.77}$.
\citet{2022A&A...663A...3G} measured mass slope of  
$\beta_{L\text{--}T} = 2.63 \pm 0.34$, with a redshift evolution trend of  
$\gamma_{L\text{--}T} = 2.17 \pm 0.94$ for XXL C1  clusters, which exhibits a slightly steeper redshift dependence than ours.

For the log-normal intrinsic scatters of $T_{300\,\rm kpc}$--$M_{500}$ and $L_{\rm 500}$--$T_{300\,\rm kpc}$ relations, we found a smaller scatter of $T_{300\,\rm kpc}$ with $\sigma_{T\text{--}M}=0.168\pm0.088$, compared with $\sigma_{L\text{--}T}=0.286\pm0.157$,  both expressed in natural logarithmic units.
This result indicates that temperature is more tightly correlated with mass than luminosity is with temperature, consistent with previous findings \citep[e.g.,][]{2019ApJ...871...50B, 2022A&A...661A..11C, 2022A&A...663A...3G}.

In summary, our $T_{300\,\rm kpc}$--$M_{500}$ relation is consistent with self-similar predictions in both slope and redshift evolution within uncertainties. The $L_{\rm 500}$--$T_{300\,\rm kpc}$ relation exhibits a steeper slope than expected from self-similarity, in agreement with previous findings, while its redshift evolution remains weakly constrained. We also observe a strong covariance between the slope and normalization of the $L_{\rm 500}$--$T_{300\,\rm kpc}$ relation in the posterior distribution. Overall, our constraints on the scaling relations are in good agreement with earlier analyses of XXL data.

\begin{figure}[htbp]
 \centering
 \includegraphics[width=0.495\textwidth,trim={0cm 0cm 0cm 0cm},clip]{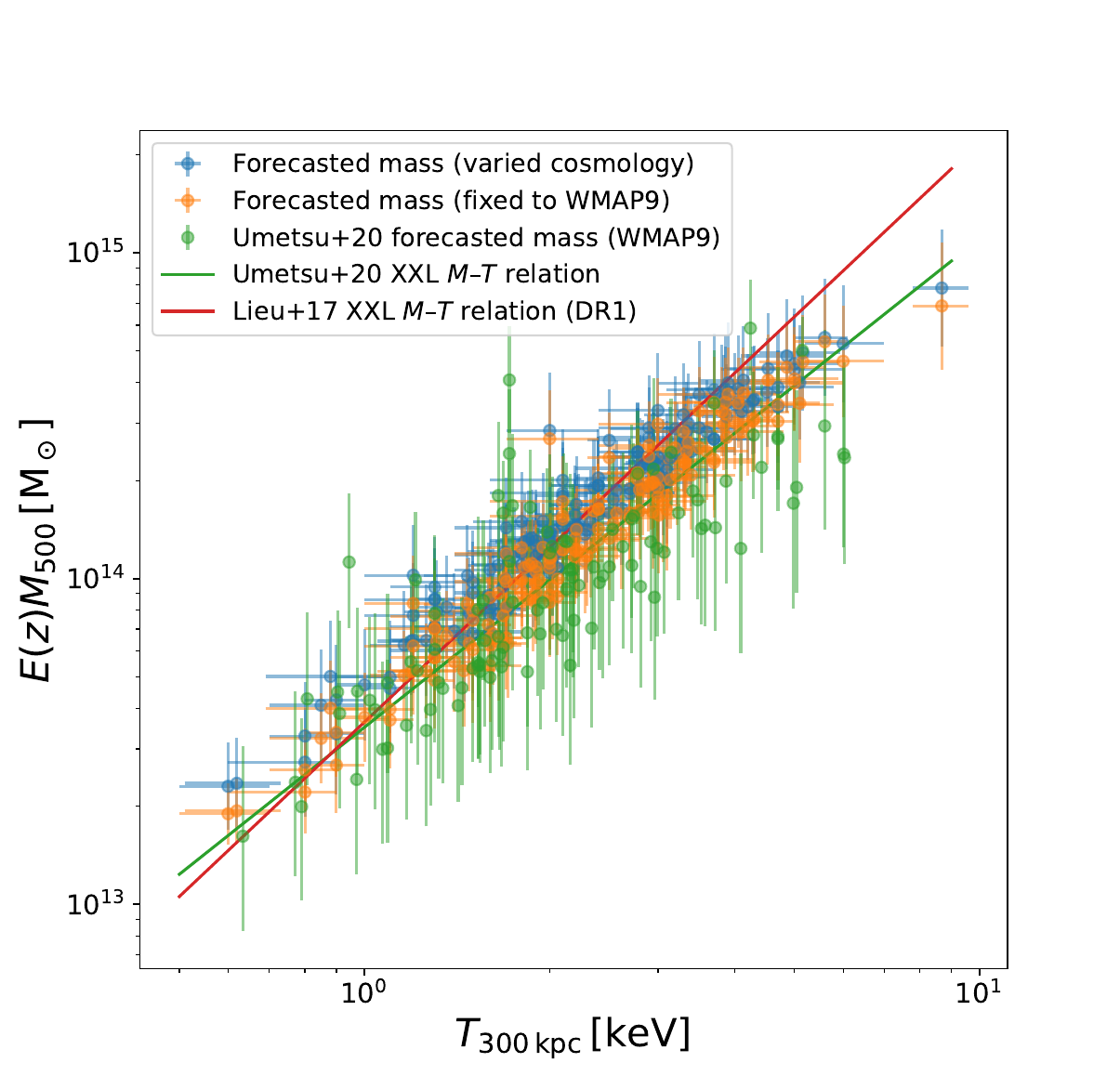}
 \caption{Mass forecasting based on X-ray temperature measurements. The blue data points show the forecasted masses for the XXL DR2 C1 sample derived from the posterior distribution of our SBI analysis, marginalizing over all parameters including cosmology. The orange data points represent the masses derived from the posterior distribution under a fixed \textit{WMAP9} cosmology. For comparison, the green data points indicate the XXL DR2 weak-lensing masses from \citet{U2020}. The green solid line shows the best-fit scaling relation reported in \citet{U2020}, while the red solid line corresponds to the earlier XXL DR1 calibration from \citet{2016A&A...592A...4L}.} 
  \label{fig:mass_T}
\end{figure}

\begin{figure}[htbp]
 \centering
 \includegraphics[width=0.52\textwidth,trim={0cm 0cm 0cm 0cm},clip]{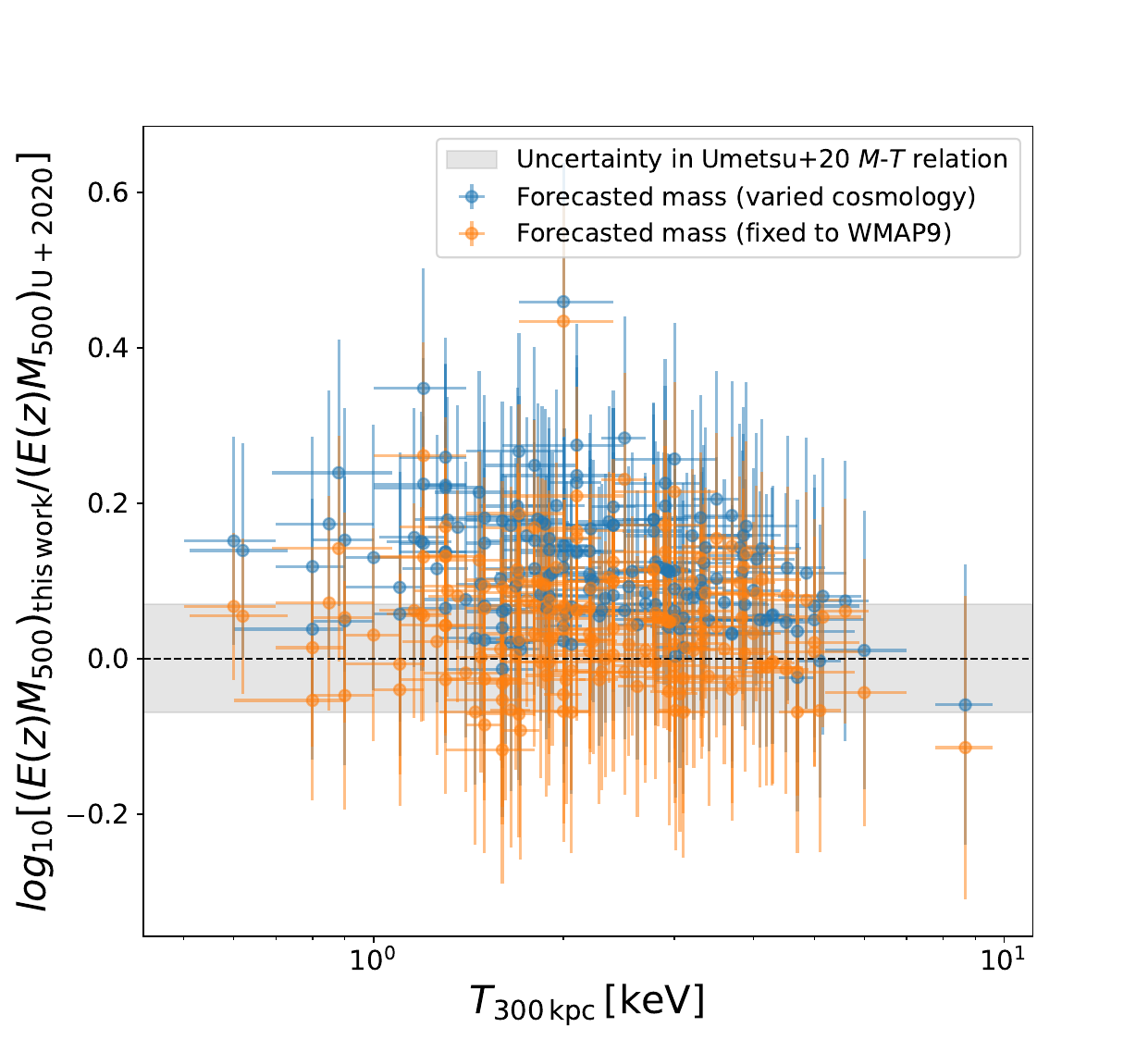}
\caption{Comparison of cluster mass estimates derived in this work with those from \citet{U2020}. Blue points show the forecasted masses inferred when cosmological parameters are allowed to vary, while orange points correspond to the analysis with cosmology fixed to the \textit{WMAP9} values. The grey shaded region indicates the uncertainty of the $M$–$T$ relation reported by \citet{U2020}.}
 \label{fig:mass_T_2}
\end{figure}

\begin{figure}[htbp]
 \centering
\includegraphics[width=0.47\textwidth,trim={0.4cm 0cm 0cm 0cm},clip]{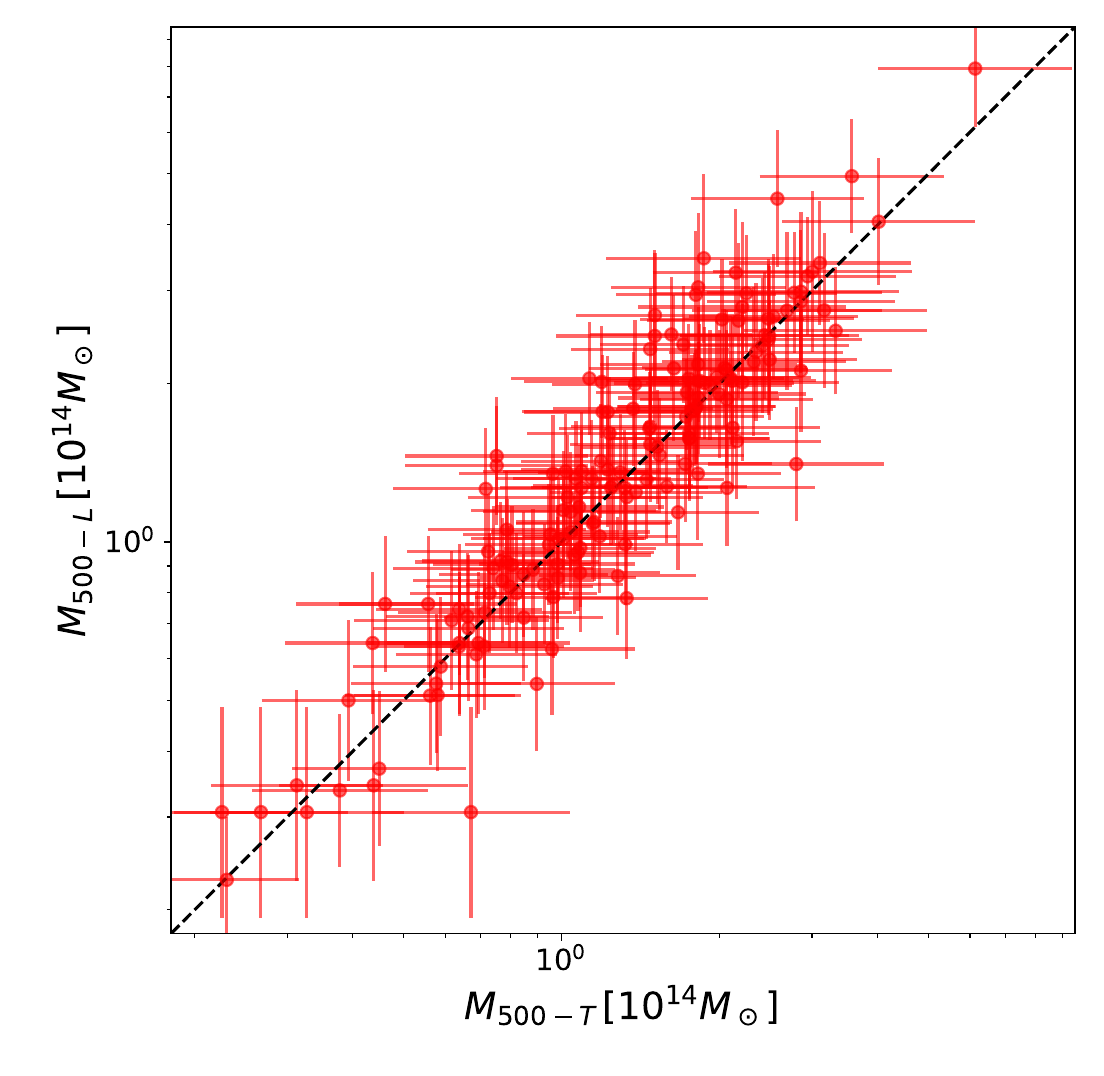}
\caption{Comparison of forecasted cluster masses derived from X-ray luminosity (y-axis) and temperature (x-axis). The dashed line indicates the one-to-one relation.}
 \label{fig:mass_comparison_LT}
\end{figure}

\subsection{Mass Estimates for Individual XXL Galaxy Clusters} 
\label{sec:Masscalibration}

In this section, we derive forecasted cluster mass estimates for individual galaxy clusters in the XXL survey using the posterior samples obtained from our SBI analysis described in Section~\ref{sec:scaling relations}. We refer to this predictive procedure as mass forecasting. We begin by drawing 5000 parameter sets from the posterior distribution, denoted as $(\boldsymbol{p}_{{\rm post},i})$ with $i=1,2,\dots, 5000$. For each realization, we apply our forward model $F(\boldsymbol{p}_{{\rm post},i})$ to generate a predictive simulated cluster catalog characterized by its physical properties, including redshift ($z$), true halo mass ($M_{500}$), concentration ($c_{500}$), X-ray temperature ($T_{300\,\mathrm{kpc}}$), and luminosity ($L_{500}$).

We then identify simulated counterparts whose redshifts and X-ray observables fall within specified uncertainty ranges: specifically, a redshift interval of $z_{\mathrm{cl}} \pm 0.03$ and a temperature (or luminosity) range defined by the individual measurement uncertainties of each XXL cluster.

Using this selected subset of simulated clusters, we compute the mean and standard deviation of $\log(M_{500}/M_\odot)$ to derive the forecasted mass estimate for the specific cluster. This approach ensures that the inferred cluster masses are directly derived from the posterior distribution, thereby marginalizing over all cosmological and systematic uncertainties. All cluster mass estimates inferred from the XXL observed X-ray temperature or luminosity are presented in Table~\ref{tab:WL_mass}. The XXL C1 sample covers a mass range from $2.30\times10^{13}\,M_{\odot}$ to $6.14\times10^{14}\,M_{\odot}$, with a typical uncertainty of approximately $5\times10^{13}\,M_{\odot}$ for estimates inferred from the $T_{300\,\textrm{kpc}}$ measurements.

We compare our forecasted cluster mass estimates for the XXL sample with previous studies. \citet{2016A&A...592A...4L} analyzed the XXL DR1 sample using the weak-lensing shear catalog from CFHTLenS to derive the $M^{\rm WL}_{500}$--$T_{300\,\mathrm{kpc}}$ relation for a subsample of 38 XXL-N clusters at $z \leq 0.6$. Their resulting $M^{\rm WL}_{500}$--$T_{300\,\mathrm{kpc}}$ relation is shown in Figure~\ref{fig:mass_T}. Compared to our results, the DR1-based calibration yields higher masses for clusters with higher temperatures ($T_{300\,\textrm{kpc}}$) and lower masses for clusters with lower temperatures. This systematic difference is driven by the shallower mass slope of the $T_{300\,\rm kpc}$--$M_{500}$ relation reported in their analysis.

\citet{U2020} analyzed 136 C1 and C2 clusters in the XXL-N field using S16A HSC weak-lensing data and derived bias-corrected mass estimates predicted from X-ray temperature measurements assuming a \textit{WMAP9} cosmology. As shown in Figure~\ref{fig:mass_T}, our mass estimates are systematically higher than theirs. We attribute this difference primarily to the higher $S_8$ value, and particularly the higher value of $\Omega_{m}$, inferred from our analysis. In a flat $\Lambda$CDM model, a higher value of $\Omega_{m}$ implies a lower value of $\Omega_\Lambda$, leading to smaller angular diameter distances and thus lower geometric lensing efficiencies (higher critical surface density, $\Sigma_\mathrm{crit}$). Consequently, higher cluster masses are required to reproduce the same observed lensing signal.

To assess the cosmology dependence, we performed an additional inference in which cosmology was fixed to \textit{WMAP9}. The resulting posteriors and scaling relations are presented in Appendix~\ref{sc:fixed cosmo}. The corresponding forecasted cluster masses are listed in Table~\ref{tab:WL_mass} and shown in Figure~\ref{fig:mass_T}, exhibiting improved consistency with the results of \citet{U2020}. Figure~\ref{fig:mass_T_2} further illustrates the logarithmic differences between the mass estimates. These comparisons suggest that the discrepancies in forecasted cluster masses are primarily driven by differences in the assumed cosmology.

Overall, our forecasted mass estimates inferred from $T_{300\,\rm kpc}$ for high-temperature clusters fall between those of \citet{2016A&A...592A...4L} and \citet{U2020}, while systematically exceeding both at the low-temperature end. A distinguishing feature of our approach is the explicit marginalization over cosmological parameters, which enables an internally consistent inference of cluster masses, in contrast to previous analyses that relied on a fixed cosmology.

Finally, we also derived forecasted cluster masses from X-ray luminosity $L_{500}$ using the same methodology. These mass estimates are listed in Table~\ref{tab:WL_mass}. As shown in Figure~\ref{fig:mass_comparison_LT}, the mass estimates derived from $L_{500}$ are generally in good agreement with those based on $T_{300\,\rm kpc}$, closely following the one-to-one relation.

\section{Discussion and Systematics}
\label{sc:Discussion}

Our forward model, described in Section~\ref{sec:modeling}, is built from semi-analytic components, including the halo mass function, selection function, NFW lensing profiles, and X-ray scaling relations. Although these ingredients are individually well defined, the inference is performed on a set of summary statistics---such as redshift-binned cluster counts and stacked weak-lensing profiles---whose joint likelihood is non-trivial to construct analytically. The mapping from model parameters to these summaries is highly non-linear and involves correlated noise, selection-induced couplings between observables, and intrinsic scatter in halo properties.

In particular, the stacked weak-lensing signal depends non-linearly on halo concentration, which follows a cosmology-dependent concentration--mass relation with log-normal intrinsic scatter, resulting in a non-Gaussian distribution of excess surface mass density even at fixed halo mass. Moreover, the cluster selection process couples number counts and lensing observables, as stochastic fluctuations affect both the detected abundance and the stacked lensing amplitude. Simulation-based inference naturally captures these effects by forward-modeling the full data vector from a single stochastic realization, without requiring explicit modeling of the joint likelihood.

The reliability of SBI nevertheless depends on the completeness of the generative model. Our framework explicitly accounts for several key sources of systematic uncertainty, including cluster miscentering, photometric-redshift bias, weak-lensing mass bias, and the cosmology dependence of the concentration--mass relation. Residual systematics may persist if relevant physical or observational effects are not explicitly modeled, such as calibration uncertainties in the XXL detection pipeline or correlated intrinsic scatter between mass proxies.

Finally, we view this analysis as a stepping stone toward future SBI applications based on higher-fidelity forward models. Incorporating hydrodynamical simulations would enable a more realistic treatment of baryonic effects and correlated scatter, for which analytic likelihood-based approaches would become prohibitively complex.

\section{Summary and Conclusions} 
\label{sec:conclusions}

We have carried out a cosmological analysis of the X-ray-selected XXL survey cluster sample using an SBI framework to jointly constrain cosmological parameters and X-ray scaling relations. Our approach employed forward modeling of cluster number counts, X-ray observables, and HSC-SSP weak-lensing measurements. We focused on XXL C1 clusters at $z \leq 1$ in the XXL-N and XXL-S regions. Specifically, we utilized the full sample of 171 clusters for the number-count analysis, a subset of 162 clusters for X-ray scaling relations, and 86 clusters in the XXL-N region---where the XXL footprint overlaps with HSC-SSP---for weak-lensing calibration.

This work advances cluster mass calibration by integrating cosmological forward modeling with likelihood-free Bayesian inference, enabling simultaneous treatment of systematic uncertainties and cosmological dependence. We adopted the DELFI algorithm to learn the conditional distribution of summary statistics given model parameters via an ensemble of neural density estimators. Our forward model generates synthetic XXL cluster samples that self-consistently incorporate the XXL-C1 selection function, the X-ray scaling relations, weak-lensing modeling, and systematic effects, including weak-lensing mass bias, cluster miscentering, and photo-$z$ calibration bias. We validated our SBI cosmological analysis using synthetic cluster samples, demonstrating that the posterior distributions converge robustly and remain stable when evaluated against the fiducial input parameters. The SBI framework facilitates full marginalization over nuisance parameters and cosmological priors, providing statistically rigorous constraints without relying on analytic likelihood approximations.

From the joint analysis, we obtain cosmological constraints of $\Omega_{m} = 0.366 \pm 0.091$, $\sigma_8 = 0.80 \pm 0.10$, and $S_8 = 0.867 \pm 0.063$, with an additional systematic uncertainty of approximately 13\% in $\Omega_{m}$ ($\delta\Omega_m=0.048$), 9\% in $\sigma_8$ ($\delta\sigma_8=0.072$), and 3\% in $S_8$ ($\delta S_8=0.026$), arising from stochastic variability of the neural density estimators. 
These results are consistent with other cluster-based cosmological studies and with \Planck measurements, showing a mild ($<2\sigma$) tension with recent cosmic-shear measurements from HSC-SSP and DES, while remaining compatible with KiDS-Legacy constraints.

The derived scaling relations, $T_{300\,\mathrm{kpc}}$--$M_{500}$ and $L_{500}$--$T_{300\,\mathrm{kpc}}$, are summarized in Equations~(\ref{eq:bf-T_M}) and~(\ref{eq:bf-L_T}), with intrinsic scatters of $\sigma_{T\text{--}M} = 0.168 \pm 0.088$ and $\sigma_{L\text{--}T} = 0.286 \pm 0.157$, respectively. The mass dependence of the $T_{300\,\mathrm{kpc}}$--$M_{500}$ relation is consistent with the self-similar prediction, although its inferred redshift evolution is marginally shallower. The $L_{500}$--$T_{300\,\mathrm{kpc}}$ relation is steeper than the self-similar expectation, in agreement with previous studies \citep[e.g.,][]{2016A&A...592A...3G, 2022A&A...663A...3G, 2022A&A...661A..11C}.

By sampling from the posterior predictive distribution, we provide forecasted mass estimates for all individual XXL C1 clusters with measured X-ray temperatures or luminosities, marginalized over cosmological and population model parameters. These results will be a valuable resource for future multi-probe cosmological analyses of the XXL sample, including those incorporating cluster clustering \citep[e.g.,][]{2022A&A...663A...3G}.

This work highlights the strength and flexibility of simulation-based inference in complex cosmological applications, particularly for self-consistent forward modeling of multi-probe cluster datasets that combine optical, X-ray, and SZE surveys with weak-lensing mass calibration. The forward-modeling pipeline developed here is highly adaptable and will be essential for interpreting forthcoming multi-wavelength cluster surveys \citep[e.g.][]{CHEX-MATE}. In particular, future applications of the SBI approach to joint weak-lensing and redshift survey data \citep[e.g.][]{Geller2005,Umetsu2025} are especially promising in the current era of wide-field imaging and spectroscopic surveys, including DESI \citep{DESI2016}, Subaru PFS \citep{Takada2014}, and DESI-II \citep{Schlegel2022}.

\begin{acknowledgments}
We thank Fox Davidson, Tomomi Sunayama and SkyPy collaboration for fruitful discussions and providing helpful suggestions. This work is supported by the National Science and Technology Council (grants NSTC 113-2112-M-A49-041-MY3). K.U. acknowledges support from the National Science and Technology Council, Taiwan (grant NSTC 112-2112-M-001-027-MY3) and the Academia Sinica Investigator award (grant AS-IA-112-M04). M.S. acknowledges financial contributions from contract INAF mainstream project 1.05.01.86.10, INAF Theory Grant 2023: Gravitational lensing detection of matter distribution at galaxy cluster boundaries and beyond (1.05.23.06.17), INAF Guest Observer Grant 2024: Towards anchoring the mass scale of galaxy clusters with galaxy kinematics (1.05.24.02.15), and contract Prin-MUR 2022 supported by Next Generation EU (n.20227RNLY3, The concordance cosmological model: stress-tests with galaxy clusters).

XXL is an international project based on an XMM Very Large Program surveying two 25\,deg$^2$ extragalactic fields at a depth of  $\sim 6\times10^{-15}$\,erg\,s$^{-1}$\,cm$^{-2}$ in the
0.5--2\,keV band. The XXL website is \href{http://irfu.cea.fr/xxl}{http://irfu.cea.fr/xxl}. Multiband information and spectroscopic follow-up of the X-ray sources are obtained through a number of survey programs, summarized at \href{http://xxlmultiwave.pbworks.com/}{http://xxlmultiwave.pbworks.com/}.

The HSC Collaboration includes the astronomical communities of Japan and Taiwan, as well as Princeton University. The HSC instrumentation and software were developed by the National Astronomical Observatory of Japan (NAOJ), the Kavli Institute for the Physics and Mathematics of the Universe (Kavli IPMU), the University of Tokyo, the High Energy Accelerator Research Organization (KEK), the Academia Sinica Institute for Astronomy and Astrophysics in Taiwan (ASIAA), and Princeton University. Funding was contributed by the FIRST program from the Japanese Cabinet Office, the Ministry of Education, Culture, Sports, Science and Technology (MEXT), the Japan Society for the Promotion of Science (JSPS), the Japan Science and Technology Agency (JST), the Toray Science Foundation, NAOJ, Kavli IPMU, KEK, ASIAA, and Princeton University.



This work is based on data collected at the Subaru Telescope and retrieved from the HSC data archive system, which is operated by the Subaru Telescope and Astronomy Data Center, National Astronomical Observatory of Japan. 

This work is based on observations obtained with \textit{XMM-Newton}, an ESA science mission with instruments and contributions directly funded by ESA Member States and NASA. 
\end{acknowledgments}
\newpage
\software{Astropy \citep{2018AJ....156..123A}, Colossus \citep{2018ApJS..239...35D}, emcee \citep{emcee}, GetDist \citep{getdist}, matplotlib \citep{Matplotlib}, NumPy \citep{numpy}, pathos \citep{2012arXiv1202.1056M, pythos}, PyDelfi \citep{2019MNRAS.488.4440A}, pygtc \citep{pygtc}, Python \citep{python3}, Scipy \citep{scipy}, TensorFlow \citep{tensorflow2015-whitepaper} }


\appendix

\section{Inference with the XXL-N C1 sample}
\label{sec:C1-N}

In this appendix, we present the posterior distributions derived from our SBI analysis using only the C1 cluster sample located in the XXL-N region, where the X-ray footprint overlaps with the HSC-SSP weak-lensing data.

Figure~\ref{fig:post_all} compares the resulting posterior with that of the full-sample analysis (XXL-N+S). The results are in excellent agreement, demonstrating the robustness of our SBI framework with respect to sample selection. The constraints on cosmological parameters and the $T_{300\mathrm{kpc}}$--$M_{500}$ scaling relation remain virtually unchanged relative to the full-sample result. This is expected, as the constraining power is dominated by the XXL-N clusters which provide the weak-lensing mass calibration. In contrast, the $L_{500}$--$T_{300\mathrm{kpc}}$ relation shows more sensitivity to the sample definition. Including the additional X-ray measurements from the XXL-S region leads to a modest shift in the normalization of this relation. As indicated by the validation tests in Section~\ref{sec:mock}, the $L_{500}$--$T_{300\mathrm{kpc}}$ relation is less tightly constrained by the current summary statistics, making it more susceptible to sample variance.

We note that our current forward-modeling framework simplifies the analysis by assuming identical X-ray selection functions for the XXL-N and XXL-S regions. In reality, the X-ray coverage in XXL-N is slightly deeper than in XXL-S, potentially leading to differences in detection probability of up to $\sim 0.1$ between the two fields. Nevertheless, the consistency between the posteriors from the combined XXL-N+S analysis and the XXL-N–only analysis suggests that this approximation has a negligible impact on our results. The error budget is currently dominated by statistical uncertainties arising from the limited sample size and weak-lensing shape noise, rendering the effects of selection inhomogeneity sub-dominant. Incorporating region-specific selection functions remains a natural refinement for future work to further improve the fidelity of the forward modeling.

\section{Mass Calibration with Fixed Cosmology}
\label{sc:fixed cosmo}

\begin{figure*}[htbp]
 \centering
 \includegraphics[width=1\textwidth,trim={0.2cm 0.2cm 0cm 0cm},clip]{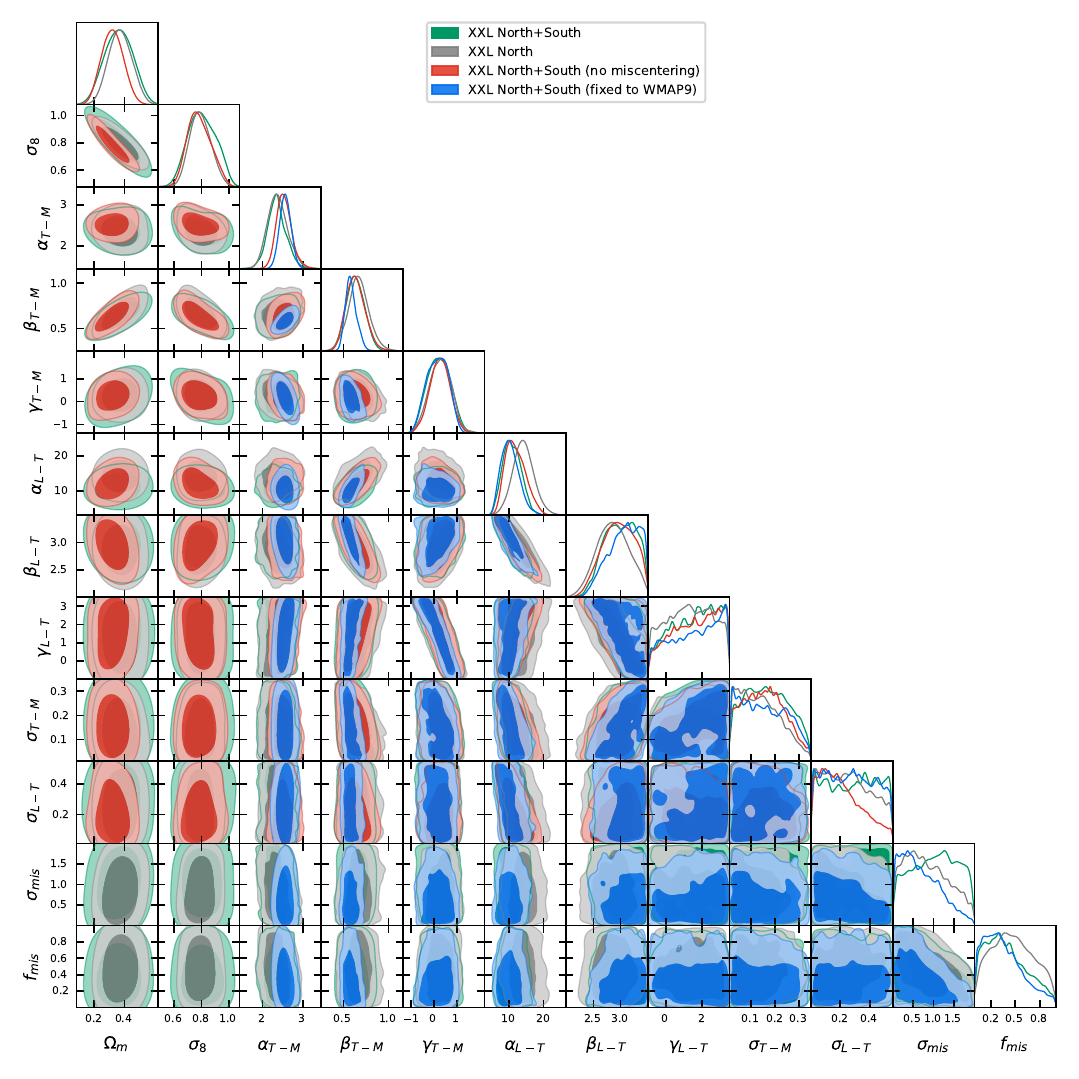}
\caption{Same as Figure~\ref{fig:real}, but comparing posterior distributions inferred under different model assumptions. The green contours show the fiducial posterior obtained using C1 clusters in both the XXL-N and XXL-S regions. The grey contours show the posterior derived using only the C1 clusters from XXL-N. The red contours represent the posterior inferred assuming perfect cluster centering (no miscentering). The blue contours correspond to the posterior obtained when cosmological parameters are fixed to the \textit{WMAP9} values.} 
 \label{fig:post_all}
\end{figure*}

\begin{figure}
 \centering
 \includegraphics[width=0.5\textwidth,trim={0.2cm 0.2cm 0cm 0cm},clip]{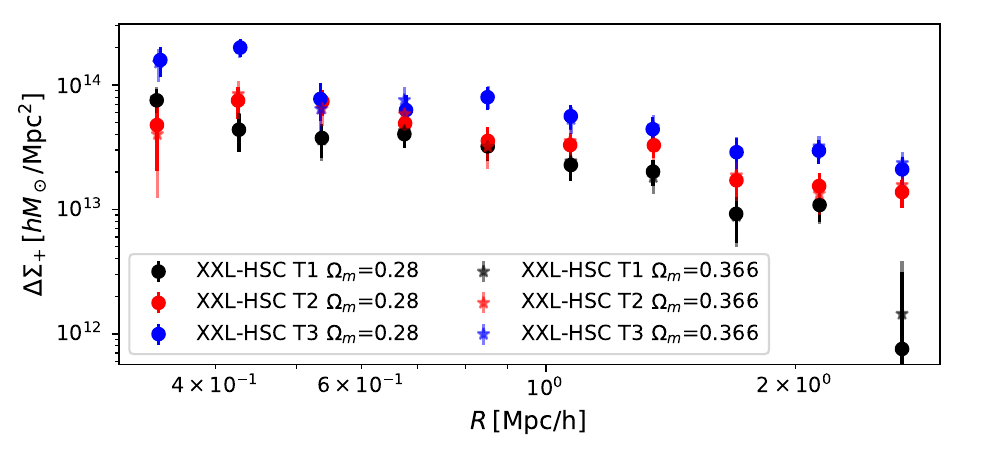}
\caption{Comparison of azimuthally averaged stacked tangential shear profiles around XXL-N C1 galaxy clusters, derived from the HSC S19A shape catalog. The profiles were computed assuming two different reference cosmologies for the geometric conversion of angular separations to comoving radii: the fiducial \textit{WMAP9} model ($\Omega_{m}=0.28$, blue) and the best-fit SBI model ($\Omega_{m}=0.366$, orange).}  
 \label{fig:compare_HSC_Om}
\end{figure}

To assess the impact of implicit cosmological assumptions in the X-ray measurements, we note that the $L_{500}$ values reported in the XXL catalog are derived by extrapolating the aperture luminosity measured within 300 kpc, $L_{300\,\mathrm{kpc}}$, to $R_{500,\mathrm{MT}}$. This extrapolation relies on the $M_{500}$--$T_{300\,\rm kpc}$ relation from \cite{2016A&A...592A...4L}, which was calibrated under a fixed \textit{WMAP9} cosmology \citep{WMAP9}. Consequently, the published XXL $L_{500}$ measurements are intrinsically tied to these assumptions and carry an implicit cosmology dependence.

However, this dependence is expected to be weak for the assumed $\beta$-profile. Given the statistical uncertainties of the current sample, the residual cosmology dependence of $L_{500}$ is therefore expected to have a negligible impact on the inferred cosmological constraints.

Nevertheless, to explicitly verify this, we performed a validation test by fixing the cosmology in our forward modeling to \textit{WMAP9}, while leaving the scaling relation and nuisance parameters free:
\begin{align*}
\boldsymbol{p} = \{& \alpha_{T\text{--}M}, \beta_{T\text{--}M}, \gamma_{T\text{--}M},
\alpha_{L\text{--}T}, \beta_{L\text{--}T}, \gamma_{L\text{--}T}, \\ &\sigma_{T\text{--}M}, \sigma_{L\text{--}T}, 
                   f_{\rm mis}, \sigma_{\rm mis}\}.
\end{align*}

The posterior distribution obtained under the fixed cosmology assumption is shown in Figure~\ref{fig:post_all}. The $T_{300\,\mathrm{kpc}}$--$M_{500}$ relation is more tightly constrained under fixed cosmology, with the normalization parameter increasing to $\alpha_{T\text{--}M} = 2.55 \pm 0.14$ compared to $2.34 \pm 0.23$ in the fiducial cosmology-varying scenario. The slope parameter decreases slightly from $\beta_{T\text{--}M} = 0.63 \pm 0.10$ to $0.58 \pm 0.06$. The lower value of $\beta_{T\text{--}M}$ is expected given the positive degeneracy between $\beta_{T\text{--}M}$ and $\Omega_{m}$ evident in Figure~\ref{fig:real}. The resulting best-fit scaling relation under the \textit{WMAP9} cosmology is given by:

\begin{align}
\frac{T_{300\,\mathrm{kpc}}}{T_{\mathrm{piv}}} &= (2.55 \pm 0.136) 
\left( \frac{M_{500}}{M_{\mathrm{piv}}} \right)^{0.58\pm0.06} \nonumber \\
&\quad \times \left( \frac{E(z)}{E(z=0.3)} \right)^{0.18 \pm 0.45},
\end{align}

\begin{align}
\frac{L_{500}}{L_{\mathrm{piv}}} &= (10.3 \pm 2.43) 
\left( \frac{T_{300\,\mathrm{kpc}}}{T_{\mathrm{piv}}} \right)^{3.07 \pm 0.27} \nonumber \\
&\quad \times \left( \frac{E(z)}{E(z=0.3)} \right)^{1.56 \pm 1.27},
\end{align}

We derive forecasted cluster masses using the method described in Section~\ref{sec:Masscalibration}. The resulting masses are listed in Table~\ref{tab:WL_mass}. Figure~\ref{fig:mass_T} compares these estimates with those obtained in our fiducial (cosmology-varying) analysis and with previous studies. Fixing the cosmology to \textit{WMAP9} yields systematically lower mass estimates, bringing them into closer agreement with the weak-lensing results of \citet{U2020}, as illustrated in Figure~\ref{fig:mass_T_2}.

These findings confirm that the higher estimated masses and different scaling relations found in our main analysis are primarily driven by the cosmological dependence of the weak lensing geometric scaling. This highlights the importance of self-consistently propagating cosmological dependence in cluster mass calibration.

\section{Testing without Miscentering Effect}
\label{sec:no_miscenter}

In this section, we investigate the impact of cluster miscentering on our parameter inference. To isolate this effect, we perform a validation test in which the miscentering component is removed from the forward model, effectively assuming that all clusters are perfectly centered on their X-ray peaks. We then repeat the full inference procedure under this assumption to assess the resulting shifts in the parameter constraints.

Figure~\ref{fig:post_all} shows the resulting posterior distributions. While largely consistent with our fiducial analysis, the contours exhibit a systematic shift toward a lower matter density, $\Omega_{m} = 0.32 \pm 0.071$, and a correspondingly lower value of $S_8 = 0.80 \pm 0.039$. This shift allows for a clear physical interpretation. By excluding the miscentering treatment, the forward model assumes that the weak-lensing signal is perfectly centered. In reality, miscentering suppresses the signal amplitude and flattens the density profile at small scales.

To fit this suppressed, flattened signal using a perfectly centered model, the inference is forced to favor systematically lower halo concentrations and masses. This bias propagates to the cosmological parameters in two ways. First, lower concentrations physically correspond to later formation times and lower fluctuation amplitudes, favoring a lower $\sigma_8$ \citep{DiemerKravtsov2015}. Second, assigning lower masses to the observed clusters creates a potential overprediction of the cluster abundance, as low-mass halos are far more numerous. To reconcile these low inferred masses with the observed number counts, the inference must suppress the predicted halo population, driving the solution toward lower values of $\Omega_{m}$ and $\sigma_8$.

This test demonstrates that miscentering is a non-negligible systematic in lensing-based mass calibration and must be explicitly modeled to avoid biased cosmological inferences.

\section{Cosmology Dependence in Shear Profile Measurement}
\label{App: shear}

As detailed in Section~\ref{sec:WL}, we assume a fiducial \textit{WMAP9} cosmology ($\Omega_{m}=0.28$) when converting angular separations into physical radial bins ($R$ in Mpc/$h$) and when computing the critical surface density, $\Sigma_{\rm{crit}}$, for the shear measurements. In principle, this geometric assumption could introduce bias if the true cosmology deviates significantly from the fiducial value used for data reduction.

To assess the impact of this assumption, we performed a validation test in which the observed shear profiles were re-computed using the best-fit cosmology inferred from our SBI analysis ($\Omega_{m} = 0.366$). As shown in Figure~\ref{fig:compare_HSC_Om}, the resulting shear profiles are statistically indistinguishable from the fiducial measurements, with differences that are negligible compared to the statistical uncertainties dominated by weak-lensing shape noise.

This robustness stems from the fact that both the angular diameter distance $D_A(z)$ and the lensing efficiency $\Sigma_{\rm crit}^{-1}$ vary only weakly with $\Omega_{m}$ at the typical redshifts of our cluster sample ($z \sim 0.3$). Furthermore, since the cluster lensing signal follows a relatively shallow radial dependence (approximately $\propto R^{-1}$), the small shifts in radial bin edges induced by changes in the distance scale result in minimal corrections to the binned shear signal. Consequently, our SBI inference results are robust against the choice of fiducial cosmology used in constructing the observed shear profiles.

\section{Calibrated Cluster Masses for XXL-C1 Clusters}
\label{app:mass_table}

In this appendix, we present a table summarizing the key properties of the XXL-C1 galaxy clusters. We explicitly list the calibrated cluster masses inferred from the X-ray observables, derived using the SBI approach detailed in Section~\ref{sec:Masscalibration} and Appendix~\ref{sc:fixed cosmo}.

\begin{longtable*}{cccccccccc}
\caption{Cluster Properties and Weak-lensing Measurements} 
\label{tab:WL_mass} \\
\hline
ID & R.A. (deg) & Decl. (deg) & $z$ & $T_{300\,\rm{kpc}}$ & $L_{500}$ & $M_{500, \textrm {T}}$ & $M_{500, \textrm{L}}$ & $M_{500, \textrm {T},\textrm {WMAP9}}$ & $M_{500, \textrm{L} ,\textrm {WMAP9}}$ \\
\hline
\endfirsthead

\multicolumn{10}{c}{{\tablename\ \thetable{} -- (Continued)}} \\
\hline
ID & R.A. (deg) & Decl. (deg) & $z$ & $T_{300\,\rm{kpc}}$  & $L_{500}$  & $M_{500, \textrm {T}}$  & $M_{500, \rm L}$  & $M_{500, \textrm {T},\textrm {WMAP9}}$ & $M_{500, \rm L,\textrm {WMAP9}}$\\
\hline
\endhead
\hline\\
\multicolumn{10}{l}{
  \parbox{\textwidth}{
    \textbf{Note.} This table summarizes the key properties of each XXL-C1 galaxy cluster, including the XLSSC identifier (1--499: XXL-N; 500--999: XXL-S), X-ray coordinates in R.A./Decl.\ (J2000.0), redshift, $T_{300\,\mathrm{kpc}}$ (keV), and $L_{500}$ ($10^{42}\,\mathrm{erg\,s^{-1}}$). We also list the forecasted cluster masses $M_{500,\mathrm{T}}$ and $M_{500,\mathrm{L}}$, inferred from the X-ray temperature and luminosity while marginalizing over cosmology, as well as the corresponding masses $M_{500,\mathrm{T,WMAP9}}$ and $M_{500,\mathrm{L,WMAP9}}$ estimated under a fixed \textit{WMAP9} cosmology. All mass estimates are reported in units of $10^{14}\,M_\odot$.
  }
}
\endfoot
2 & 36.384 & -3.920 & 0.771 & $2.5^{+0.2}_{-0.2}$ & $69.00\pm5.00$ & $1.64^{+0.74}_{-0.51}$ & $2.14^{+0.80}_{-0.58}$ & $1.56^{+0.58}_{-0.42}$ & $2.08^{+0.50}_{-0.41}$ \\
3 & 36.909 & -3.300 & 0.836 & $3.5^{+0.3}_{-0.3}$ & $125.00\pm9.00$ & $2.17^{+1.07}_{-0.72}$ & $2.63^{+1.07}_{-0.76}$ & $2.07^{+0.76}_{-0.56}$ & $2.71^{+0.67}_{-0.54}$ \\
5 & 36.788 & -4.301 & 1.058 & $2.1^{+0.3}_{-0.4}$ & $52.00\pm8.00$ & -- & -- & -- & -- \\
6 & 35.439 & -3.772 & 0.429 & $4.2^{+0.5}_{-0.7}$ & $186.00\pm16.00$ & $2.58^{+1.19}_{-0.81}$ & $4.49^{+1.57}_{-1.16}$ & $2.35^{+0.99}_{-0.69}$ & $4.28^{+1.03}_{-0.83}$ \\
8 & 36.336 & -3.801 & 0.299 & $1.6^{+0.2}_{-0.2}$ & $2.00\pm1.00$ & $0.90^{+0.37}_{-0.26}$ & $0.54^{+0.18}_{-0.14}$ & $0.76^{+0.28}_{-0.21}$ & $0.45^{+0.11}_{-0.09}$ \\
10 & 36.843 & -3.362 & 0.33 & $2.4^{+0.3}_{-0.4}$ & $25.00\pm3.00$ & $1.37^{+0.57}_{-0.40}$ & $1.79^{+0.51}_{-0.39}$ & $1.20^{+0.47}_{-0.34}$ & $1.57^{+0.37}_{-0.30}$ \\
11 & 36.540 & -4.969 & 0.054 & $1.6^{+0.2}_{-0.3}$ & $1.00\pm1.00$ & $0.67^{+0.37}_{-0.24}$ & $0.31^{+0.18}_{-0.11}$ & $0.53^{+0.26}_{-0.17}$ & $0.26^{+0.12}_{-0.08}$ \\
13 & 36.858 & -4.538 & 0.308 & $1.3^{+0.2}_{-0.3}$ & $7.00\pm1.00$ & $0.72^{+0.31}_{-0.22}$ & $0.96^{+0.28}_{-0.21}$ & $0.61^{+0.23}_{-0.17}$ & $0.81^{+0.19}_{-0.15}$ \\
18 & 36.008 & -5.091 & 0.324 & $1.7^{+0.2}_{-0.3}$ & $5.00\pm1.00$ & $0.96^{+0.40}_{-0.28}$ & $0.79^{+0.24}_{-0.18}$ & $0.83^{+0.31}_{-0.23}$ & $0.67^{+0.16}_{-0.13}$ \\
21 & 36.233 & -5.134 & 0.085 & $0.8^{+0.1}_{-0.1}$ & $1.00\pm1.00$ & $0.31^{+0.14}_{-0.10}$ & $0.34^{+0.18}_{-0.12}$ & $0.25^{+0.08}_{-0.06}$ & $0.28^{+0.11}_{-0.08}$ \\
22 & 36.917 & -4.858 & 0.293 & $2.0^{+0.2}_{-0.2}$ & $30.00\pm2.00$ & $1.13^{+0.46}_{-0.33}$ & $2.04^{+0.57}_{-0.45}$ & $0.97^{+0.36}_{-0.27}$ & $1.79^{+0.44}_{-0.35}$ \\
25 & 36.353 & -4.680 & 0.265 & $2.2^{+0.2}_{-0.3}$ & $22.00\pm2.00$ & $1.20^{+0.51}_{-0.36}$ & $1.77^{+0.49}_{-0.39}$ & $1.03^{+0.41}_{-0.29}$ & $1.54^{+0.36}_{-0.29}$ \\
27 & 37.012 & -4.851 & 0.295 & $2.7^{+0.5}_{-0.5}$ & $16.00\pm2.00$ & $1.54^{+0.69}_{-0.48}$ & $1.46^{+0.42}_{-0.33}$ & $1.34^{+0.58}_{-0.40}$ & $1.27^{+0.29}_{-0.24}$ \\
28 & 35.984 & -3.098 & 0.297 & $1.5^{+0.2}_{-0.3}$ & $5.00\pm1.00$ & $0.82^{+0.35}_{-0.24}$ & $0.80^{+0.24}_{-0.18}$ & $0.69^{+0.26}_{-0.19}$ & $0.67^{+0.16}_{-0.13}$ \\
29 & 36.017 & -4.225 & 1.05 & $3.2^{+0.5}_{-0.7}$ & $192.00\pm21.00$ & -- & -- & -- & -- \\
35 & 35.949 & -2.858 & 0.174 & $1.3^{+0.1}_{-0.1}$ & $4.00\pm1.00$ & $0.64^{+0.28}_{-0.19}$ & $0.74^{+0.25}_{-0.19}$ & $0.52^{+0.19}_{-0.14}$ & $0.62^{+0.15}_{-0.12}$ \\
36 & 35.527 & -3.054 & 0.492 & $3.5^{+0.5}_{-0.6}$ & $101.00\pm10.00$ & $2.15^{+0.94}_{-0.66}$ & $3.25^{+1.04}_{-0.79}$ & $1.96^{+0.76}_{-0.55}$ & $2.99^{+0.71}_{-0.58}$ \\
40 & 35.523 & -4.547 & 0.32 & $1.9^{+0.3}_{-0.3}$ & $6.00\pm1.00$ & $1.08^{+0.45}_{-0.32}$ & $0.87^{+0.26}_{-0.20}$ & $0.93^{+0.36}_{-0.26}$ & $0.74^{+0.17}_{-0.14}$ \\
41 & 36.378 & -4.239 & 0.142 & $1.7^{+0.1}_{-0.3}$ & $12.00\pm2.00$ & $0.75^{+0.37}_{-0.25}$ & $1.40^{+0.41}_{-0.32}$ & $0.62^{+0.27}_{-0.19}$ & $1.17^{+0.29}_{-0.23}$ \\
44 & 36.141 & -4.236 & 0.263 & $1.2^{+0.2}_{-0.2}$ & $4.00\pm1.00$ & $0.66^{+0.29}_{-0.20}$ & $0.72^{+0.23}_{-0.18}$ & $0.55^{+0.21}_{-0.15}$ & $0.61^{+0.15}_{-0.12}$ \\
49 & 35.988 & -4.588 & 0.494 & $2.1^{+0.2}_{-0.2}$ & $10.00\pm1.00$ & $1.32^{+0.53}_{-0.38}$ & $0.99^{+0.30}_{-0.23}$ & $1.18^{+0.43}_{-0.32}$ & $0.89^{+0.20}_{-0.17}$ \\
50 & 36.421 & -3.189 & 0.14 & $3.1^{+0.3}_{-0.3}$ & $27.00\pm2.00$ & $1.82^{+0.98}_{-0.64}$ & $2.17^{+0.62}_{-0.48}$ & $1.53^{+0.82}_{-0.53}$ & $1.88^{+0.46}_{-0.37}$ \\
51 & 36.498 & -2.825 & 0.279 & $1.3^{+0.1}_{-0.1}$ & $5.00\pm1.00$ & $0.73^{+0.30}_{-0.21}$ & $0.80^{+0.24}_{-0.18}$ & $0.61^{+0.22}_{-0.16}$ & $0.67^{+0.16}_{-0.13}$ \\
52 & 36.567 & -2.666 & 0.056 & $0.6^{+0.1}_{-0.1}$ & $1.00\pm1.00$ & $0.23^{+0.08}_{-0.06}$ & $0.31^{+0.18}_{-0.11}$ & $0.19^{+0.05}_{-0.04}$ & $0.26^{+0.12}_{-0.08}$ \\
54 & 36.319 & -5.887 & 0.054 & $1.5^{+0.1}_{-0.1}$ & $3.00\pm1.00$ & $0.66^{+0.35}_{-0.23}$ & $0.69^{+0.26}_{-0.19}$ & $0.52^{+0.24}_{-0.16}$ & $0.55^{+0.14}_{-0.11}$ \\
55 & 36.454 & -5.896 & 0.232 & $3.1^{+0.6}_{-0.4}$ & $27.00\pm2.00$ & $1.87^{+0.91}_{-0.61}$ & $2.00^{+0.56}_{-0.44}$ & $1.61^{+0.79}_{-0.53}$ & $1.74^{+0.41}_{-0.33}$ \\
56 & 33.871 & -4.682 & 0.348 & $3.0^{+0.4}_{-0.6}$ & $43.00\pm4.00$ & $1.71^{+0.76}_{-0.52}$ & $2.37^{+0.69}_{-0.54}$ & $1.51^{+0.63}_{-0.45}$ & $2.12^{+0.50}_{-0.41}$ \\
57 & 34.051 & -4.242 & 0.153 & $2.0^{+0.2}_{-0.3}$ & $11.00\pm1.00$ & $0.96^{+0.49}_{-0.33}$ & $1.35^{+0.39}_{-0.31}$ & $0.79^{+0.38}_{-0.26}$ & $1.13^{+0.27}_{-0.22}$ \\
58 & 34.935 & -4.889 & 0.332 & $2.2^{+0.3}_{-0.3}$ & $6.00\pm1.00$ & $1.28^{+0.53}_{-0.37}$ & $0.86^{+0.25}_{-0.19}$ & $1.12^{+0.43}_{-0.31}$ & $0.74^{+0.17}_{-0.14}$ \\
59 & 34.397 & -5.223 & 0.645 & $2.9^{+0.4}_{-0.5}$ & $22.00\pm3.00$ & $1.82^{+0.80}_{-0.56}$ & $1.35^{+0.45}_{-0.34}$ & $1.69^{+0.62}_{-0.45}$ & $1.25^{+0.30}_{-0.24}$ \\
60 & 33.668 & -4.553 & 0.139 & $4.7^{+0.3}_{-0.3}$ & $63.00\pm2.00$ & $3.10^{+1.52}_{-1.02}$ & $3.38^{+1.05}_{-0.80}$ & $2.85^{+1.48}_{-0.97}$ & $2.98^{+0.81}_{-0.64}$ \\
61 & 35.485 & -5.758 & 0.259 & $1.9^{+0.3}_{-0.3}$ & $13.00\pm2.00$ & $1.02^{+0.44}_{-0.31}$ & $1.36^{+0.39}_{-0.30}$ & $0.86^{+0.34}_{-0.25}$ & $1.16^{+0.28}_{-0.23}$ \\
62 & 36.061 & -2.721 & 0.059 & $0.8^{+0.1}_{-0.2}$ & $1.00\pm1.00$ & $0.27^{+0.12}_{-0.09}$ & $0.31^{+0.18}_{-0.11}$ & $0.22^{+0.07}_{-0.06}$ & $0.26^{+0.12}_{-0.08}$ \\
64 & 34.632 & -5.017 & 0.874 & -- & -- & -- & -- & -- & -- \\
67 & 34.681 & -5.549 & 0.382 & $1.2^{+0.2}_{-0.2}$ & $7.00\pm1.00$ & $0.80^{+0.33}_{-0.24}$ & $0.90^{+0.26}_{-0.20}$ & $0.69^{+0.28}_{-0.20}$ & $0.78^{+0.18}_{-0.15}$ \\
72 & 33.850 & -3.726 & 1.002 & $2.0^{+0.4}_{-0.3}$ & $139.00\pm19.00$ & -- & -- & -- & -- \\
75 & 35.834 & -5.454 & 0.211 & -- & -- & -- & -- & -- & -- \\
76 & 33.682 & -3.823 & 0.75 & -- & -- & -- & -- & -- & -- \\
82 & 32.714 & -6.173 & 0.427 & $3.6^{+0.6}_{-0.7}$ & $22.00\pm3.00$ & $2.16^{+0.96}_{-0.66}$ & $1.55^{+0.45}_{-0.35}$ & $1.97^{+0.80}_{-0.57}$ & $1.39^{+0.31}_{-0.26}$ \\
85 & 32.870 & -6.196 & 0.428 & $4.1^{+0.7}_{-0.8}$ & $44.00\pm5.00$ & $2.49^{+1.16}_{-0.79}$ & $2.22^{+0.67}_{-0.51}$ & $2.26^{+0.97}_{-0.68}$ & $2.01^{+0.47}_{-0.38}$ \\
86 & 32.809 & -6.162 & 0.424 & $2.8^{+0.5}_{-0.6}$ & $12.00\pm2.00$ & $1.67^{+0.71}_{-0.50}$ & $1.14^{+0.33}_{-0.25}$ & $1.50^{+0.59}_{-0.42}$ & $1.01^{+0.23}_{-0.19}$ \\
87 & 37.720 & -4.348 & 0.141 & $1.6^{+0.2}_{-0.2}$ & $13.00\pm2.00$ & $0.75^{+0.37}_{-0.25}$ & $1.45^{+0.43}_{-0.33}$ & $0.62^{+0.27}_{-0.19}$ & $1.23^{+0.31}_{-0.24}$ \\
88 & 37.611 & -4.581 & 0.295 & $1.9^{+0.3}_{-0.3}$ & $13.00\pm2.00$ & $1.06^{+0.45}_{-0.32}$ & $1.32^{+0.38}_{-0.29}$ & $0.91^{+0.35}_{-0.25}$ & $1.14^{+0.26}_{-0.21}$ \\
89 & 37.127 & -4.733 & 0.609 & $2.1^{+0.4}_{-0.5}$ & $46.00\pm9.00$ & $1.38^{+0.60}_{-0.42}$ & $1.99^{+0.65}_{-0.49}$ & $1.27^{+0.48}_{-0.35}$ & $1.86^{+0.44}_{-0.36}$ \\
90 & 37.121 & -4.857 & 0.141 & $1.1^{+0.1}_{-0.2}$ & $4.00\pm1.00$ & $0.46^{+0.22}_{-0.15}$ & $0.76^{+0.26}_{-0.20}$ & $0.37^{+0.14}_{-0.10}$ & $0.63^{+0.16}_{-0.13}$ \\
91 & 37.926 & -4.881 & 0.186 & $5.1^{+0.4}_{-0.4}$ & $131.00\pm4.00$ & $3.57^{+1.79}_{-1.19}$ & $4.95^{+1.40}_{-1.09}$ & $3.16^{+1.64}_{-1.08}$ & $4.25^{+1.06}_{-0.85}$ \\
92 & 32.071 & -7.276 & 0.432 & $2.4^{+0.4}_{-0.4}$ & $25.00\pm4.00$ & $1.47^{+0.61}_{-0.43}$ & $1.65^{+0.49}_{-0.38}$ & $1.31^{+0.50}_{-0.36}$ & $1.49^{+0.35}_{-0.28}$ \\
93 & 31.699 & -6.948 & 0.429 & $3.3^{+0.5}_{-0.5}$ & $63.00\pm7.00$ & $2.02^{+0.87}_{-0.61}$ & $2.65^{+0.80}_{-0.62}$ & $1.84^{+0.72}_{-0.52}$ & $2.42^{+0.56}_{-0.45}$ \\
94 & 30.648 & -6.732 & 0.886 & $3.0^{+0.5}_{-0.6}$ & $224.00\pm32.00$ & $1.87^{+0.99}_{-0.65}$ & $3.46^{+1.55}_{-1.07}$ & $1.84^{+0.70}_{-0.51}$ & $3.40^{+0.82}_{-0.66}$ \\
95 & 31.962 & -5.206 & 0.138 & $0.9^{+0.1}_{-0.1}$ & $2.00\pm1.00$ & $0.39^{+0.18}_{-0.12}$ & $0.50^{+0.21}_{-0.15}$ & $0.32^{+0.12}_{-0.08}$ & $0.41^{+0.13}_{-0.10}$ \\
96 & 30.973 & -5.027 & 0.52 & $5.0^{+0.9}_{-0.5}$ & $63.00\pm8.00$ & $3.33^{+1.64}_{-1.10}$ & $2.52^{+0.81}_{-0.61}$ & $3.14^{+1.38}_{-0.96}$ & $2.31^{+0.53}_{-0.43}$ \\
97 & 33.342 & -6.098 & 0.697 & $5.0^{+1.0}_{-1.2}$ & $115.00\pm19.00$ & $2.85^{+1.46}_{-0.96}$ & $2.86^{+1.06}_{-0.77}$ & $2.76^{+1.13}_{-0.80}$ & $2.80^{+0.66}_{-0.54}$ \\
98 & 33.115 & -6.076 & 0.297 & $3.0^{+0.6}_{-0.6}$ & $15.00\pm3.00$ & $1.72^{+0.80}_{-0.54}$ & $1.41^{+0.41}_{-0.32}$ & $1.50^{+0.68}_{-0.47}$ & $1.22^{+0.29}_{-0.23}$ \\
99 & 33.220 & -6.202 & 0.391 & $3.7^{+0.6}_{-0.9}$ & $23.00\pm4.00$ & $2.12^{+0.98}_{-0.67}$ & $1.65^{+0.48}_{-0.37}$ & $1.91^{+0.83}_{-0.58}$ & $1.46^{+0.35}_{-0.28}$ \\
100 & 31.549 & -6.193 & 0.915 & $5.6^{+0.5}_{-0.6}$ & $140.00\pm13.00$ & $3.16^{+1.80}_{-1.15}$ & $2.75^{+1.11}_{-0.79}$ & $3.29^{+1.30}_{-0.93}$ & $2.75^{+0.68}_{-0.54}$ \\
101 & 32.193 & -4.436 & 0.756 & $2.9^{+0.4}_{-0.5}$ & $139.00\pm16.00$ & $1.82^{+0.85}_{-0.58}$ & $3.04^{+1.16}_{-0.84}$ & $1.72^{+0.63}_{-0.46}$ & $3.01^{+0.72}_{-0.58}$ \\
102 & 31.322 & -4.652 & 0.969 & $3.9^{+0.8}_{-0.9}$ & $167.00\pm25.00$ & $2.21^{+1.28}_{-0.81}$ & $2.79^{+1.27}_{-0.87}$ & $2.20^{+0.86}_{-0.62}$ & $2.87^{+0.71}_{-0.57}$ \\
103 & 36.886 & -5.961 & 0.233 & $2.5^{+0.4}_{-0.4}$ & $11.00\pm2.00$ & $1.38^{+0.63}_{-0.43}$ & $1.24^{+0.36}_{-0.28}$ & $1.19^{+0.52}_{-0.36}$ & $1.06^{+0.26}_{-0.21}$ \\
104 & 37.324 & -5.895 & 0.294 & -- & -- & -- & -- & -- & -- \\
105 & 38.411 & -5.506 & 0.432 & $6.0^{+1.0}_{-0.8}$ & $142.00\pm14.00$ & $4.02^{+2.12}_{-1.39}$ & $4.06^{+1.29}_{-0.98}$ & $3.72^{+1.80}_{-1.21}$ & $3.72^{+0.91}_{-0.73}$ \\
106 & 31.351 & -5.732 & 0.3 & $2.8^{+0.2}_{-0.3}$ & $43.00\pm3.00$ & $1.62^{+0.69}_{-0.48}$ & $2.48^{+0.70}_{-0.55}$ & $1.41^{+0.58}_{-0.41}$ & $2.19^{+0.51}_{-0.42}$ \\
107 & 31.354 & -7.594 & 0.436 & $2.4^{+0.4}_{-0.4}$ & $49.00\pm6.00$ & $1.47^{+0.61}_{-0.43}$ & $2.32^{+0.71}_{-0.54}$ & $1.31^{+0.50}_{-0.36}$ & $2.13^{+0.49}_{-0.40}$ \\
108 & 31.832 & -4.827 & 0.254 & $2.3^{+0.3}_{-0.4}$ & $21.00\pm2.00$ & $1.22^{+0.54}_{-0.38}$ & $1.76^{+0.49}_{-0.38}$ & $1.05^{+0.44}_{-0.31}$ & $1.52^{+0.36}_{-0.29}$ \\
109 & 32.296 & -6.346 & 0.491 & $2.4^{+0.2}_{-0.2}$ & $69.00\pm5.00$ & $1.51^{+0.62}_{-0.44}$ & $2.70^{+0.85}_{-0.65}$ & $1.35^{+0.48}_{-0.35}$ & $2.49^{+0.57}_{-0.47}$ \\
110 & 33.537 & -5.585 & 0.445 & $1.7^{+0.3}_{-0.3}$ & $17.00\pm3.00$ & $1.09^{+0.45}_{-0.32}$ & $1.36^{+0.40}_{-0.31}$ & $0.95^{+0.36}_{-0.26}$ & $1.21^{+0.28}_{-0.23}$ \\
111 & 33.111 & -5.627 & 0.3 & $3.7^{+0.6}_{-0.6}$ & $62.00\pm5.00$ & $2.25^{+1.09}_{-0.73}$ & $2.96^{+0.87}_{-0.67}$ & $1.98^{+0.94}_{-0.64}$ & $2.65^{+0.64}_{-0.52}$ \\
112 & 32.514 & -5.462 & 0.139 & $1.0^{+0.1}_{-0.1}$ & $3.00\pm1.00$ & $0.44^{+0.21}_{-0.14}$ & $0.64^{+0.23}_{-0.17}$ & $0.35^{+0.13}_{-0.10}$ & $0.53^{+0.14}_{-0.11}$ \\
117 & 33.121 & -5.528 & 0.298 & $3.4^{+0.6}_{-0.5}$ & $12.00\pm2.00$ & $2.07^{+0.98}_{-0.66}$ & $1.27^{+0.36}_{-0.28}$ & $1.81^{+0.85}_{-0.58}$ & $1.09^{+0.26}_{-0.21}$ \\
122 & 34.433 & -3.759 & 1.99 & -- & -- & -- & -- & -- & -- \\
123 & 36.487 & -5.643 & 0.194 & -- & -- & -- & -- & -- & -- \\
124 & 34.425 & -4.863 & 0.516 & $2.1^{+0.4}_{-0.5}$ & $6.00\pm1.00$ & $1.33^{+0.57}_{-0.40}$ & $0.78^{+0.24}_{-0.18}$ & $1.19^{+0.46}_{-0.33}$ & $0.69^{+0.15}_{-0.12}$ \\
146 & 37.462 & -4.150 & 0.254 & $1.8^{+0.3}_{-0.3}$ & $3.00\pm1.00$ & $0.96^{+0.42}_{-0.29}$ & $0.63^{+0.21}_{-0.16}$ & $0.81^{+0.32}_{-0.23}$ & $0.52^{+0.13}_{-0.11}$ \\
150 & 37.661 & -4.992 & 0.292 & $2.0^{+0.3}_{-0.4}$ & $12.00\pm2.00$ & $1.09^{+0.47}_{-0.33}$ & $1.27^{+0.36}_{-0.28}$ & $0.94^{+0.37}_{-0.27}$ & $1.09^{+0.26}_{-0.21}$ \\
151 & 38.122 & -4.788 & 0.189 & $1.9^{+0.4}_{-0.3}$ & $7.00\pm1.00$ & $0.99^{+0.46}_{-0.31}$ & $1.01^{+0.31}_{-0.24}$ & $0.83^{+0.37}_{-0.25}$ & $0.85^{+0.20}_{-0.16}$ \\
154 & 38.502 & -4.826 & 0.179 & $1.2^{+0.1}_{-0.1}$ & $2.00\pm1.00$ & $0.58^{+0.26}_{-0.18}$ & $0.51^{+0.20}_{-0.15}$ & $0.48^{+0.17}_{-0.13}$ & $0.42^{+0.13}_{-0.10}$ \\
155 & 31.134 & -6.748 & 0.433 & $1.8^{+0.3}_{-0.3}$ & $16.00\pm3.00$ & $1.14^{+0.47}_{-0.33}$ & $1.32^{+0.39}_{-0.30}$ & $1.00^{+0.38}_{-0.27}$ & $1.17^{+0.27}_{-0.22}$ \\
157 & 30.865 & -6.929 & 0.585 & $3.2^{+0.8}_{-0.7}$ & $42.00\pm7.00$ & $2.00^{+0.93}_{-0.63}$ & $1.91^{+0.61}_{-0.46}$ & $1.85^{+0.73}_{-0.52}$ & $1.78^{+0.42}_{-0.34}$ \\
161 & 33.915 & -5.980 & 0.306 & $2.4^{+0.4}_{-0.5}$ & $12.00\pm2.00$ & $1.32^{+0.58}_{-0.40}$ & $1.27^{+0.36}_{-0.28}$ & $1.14^{+0.47}_{-0.33}$ & $1.09^{+0.26}_{-0.21}$ \\
163 & 32.463 & -6.117 & 0.283 & -- & -- & -- & -- & -- & -- \\
166 & 33.211 & -4.600 & 0.158 & $1.5^{+0.2}_{-0.2}$ & $3.00\pm1.00$ & $0.69^{+0.34}_{-0.23}$ & $0.64^{+0.23}_{-0.17}$ & $0.57^{+0.24}_{-0.17}$ & $0.53^{+0.14}_{-0.11}$ \\
167 & 32.479 & -4.630 & 0.298 & $1.8^{+0.3}_{-0.3}$ & $8.00\pm2.00$ & $1.01^{+0.43}_{-0.30}$ & $1.02^{+0.30}_{-0.23}$ & $0.86^{+0.33}_{-0.24}$ & $0.87^{+0.21}_{-0.17}$ \\
168 & 37.387 & -5.880 & 0.295 & $2.2^{+0.4}_{-0.4}$ & $19.00\pm2.00$ & $1.23^{+0.54}_{-0.37}$ & $1.61^{+0.45}_{-0.35}$ & $1.07^{+0.43}_{-0.31}$ & $1.40^{+0.33}_{-0.27}$ \\
169 & 37.538 & -5.679 & 0.498 & $4.7^{+1.1}_{-1.0}$ & $44.00\pm8.00$ & $2.86^{+1.40}_{-0.94}$ & $2.11^{+0.65}_{-0.50}$ & $2.66^{+1.18}_{-0.82}$ & $1.94^{+0.46}_{-0.37}$ \\
171 & 31.986 & -5.871 & 0.044 & -- & -- & -- & -- & -- & -- \\
173 & 31.251 & -5.931 & 0.413 & $4.3^{+0.3}_{-0.3}$ & $17.00\pm2.00$ & $2.80^{+1.32}_{-0.90}$ & $1.41^{+0.40}_{-0.31}$ & $2.55^{+1.11}_{-0.77}$ & $1.24^{+0.29}_{-0.23}$ \\
174 & 30.592 & -5.899 & 0.235 & $1.5^{+0.1}_{-0.1}$ & $8.00\pm1.00$ & $0.79^{+0.33}_{-0.23}$ & $1.05^{+0.31}_{-0.24}$ & $0.67^{+0.24}_{-0.18}$ & $0.90^{+0.21}_{-0.17}$ \\
176 & 32.490 & -4.980 & 0.141 & $1.4^{+0.2}_{-0.2}$ & $3.00\pm1.00$ & $0.64^{+0.31}_{-0.21}$ & $0.64^{+0.23}_{-0.17}$ & $0.52^{+0.22}_{-0.15}$ & $0.53^{+0.14}_{-0.11}$ \\
179 & 30.482 & -6.574 & 0.608 & -- & -- & -- & -- & -- & -- \\
180 & 33.863 & -5.556 & 0.289 & $2.7^{+0.2}_{-0.2}$ & $12.00\pm1.00$ & $1.59^{+0.67}_{-0.47}$ & $1.27^{+0.36}_{-0.28}$ & $1.38^{+0.56}_{-0.40}$ & $1.10^{+0.25}_{-0.20}$ \\
189 & 34.908 & -4.007 & 0.204 & $1.3^{+0.2}_{-0.2}$ & $3.00\pm1.00$ & $0.64^{+0.29}_{-0.20}$ & $0.63^{+0.22}_{-0.17}$ & $0.52^{+0.20}_{-0.15}$ & $0.52^{+0.14}_{-0.11}$ \\
190 & 36.748 & -4.589 & 0.07 & $1.1^{+0.1}_{-0.1}$ & $1.00\pm1.00$ & $0.44^{+0.23}_{-0.15}$ & $0.34^{+0.18}_{-0.12}$ & $0.35^{+0.14}_{-0.10}$ & $0.28^{+0.11}_{-0.08}$ \\
191 & 36.574 & -5.078 & 0.054 & $0.9^{+0.1}_{-0.1}$ & $1.00\pm1.00$ & $0.33^{+0.17}_{-0.11}$ & $0.31^{+0.18}_{-0.11}$ & $0.26^{+0.11}_{-0.08}$ & $0.26^{+0.12}_{-0.08}$ \\
196 & 30.728 & -7.652 & 0.136 & $1.3^{+0.1}_{-0.2}$ & $4.00\pm1.00$ & $0.56^{+0.27}_{-0.18}$ & $0.76^{+0.26}_{-0.20}$ & $0.46^{+0.18}_{-0.13}$ & $0.63^{+0.16}_{-0.13}$ \\
197 & 30.923 & -7.785 & 0.439 & $3.0^{+0.4}_{-0.5}$ & $76.00\pm9.00$ & $1.80^{+0.76}_{-0.53}$ & $2.95^{+0.95}_{-0.72}$ & $1.63^{+0.62}_{-0.45}$ & $2.68^{+0.64}_{-0.52}$ \\
198 & 33.496 & -5.186 & 0.356 & $1.3^{+0.1}_{-0.2}$ & $7.00\pm1.00$ & $0.77^{+0.32}_{-0.22}$ & $0.92^{+0.27}_{-0.21}$ & $0.65^{+0.25}_{-0.18}$ & $0.79^{+0.18}_{-0.15}$ \\
199 & 30.192 & -6.708 & 0.339 & $2.1^{+0.2}_{-0.3}$ & $32.00\pm3.00$ & $1.19^{+0.48}_{-0.34}$ & $2.01^{+0.55}_{-0.43}$ & $1.04^{+0.39}_{-0.28}$ & $1.81^{+0.42}_{-0.34}$ \\
200 & 30.331 & -6.830 & 0.333 & $2.1^{+0.3}_{-0.4}$ & $16.00\pm2.00$ & $1.19^{+0.49}_{-0.35}$ & $1.42^{+0.40}_{-0.31}$ & $1.04^{+0.40}_{-0.29}$ & $1.24^{+0.28}_{-0.23}$ \\
201 & 32.767 & -4.893 & 0.138 & $1.6^{+0.2}_{-0.3}$ & $10.00\pm2.00$ & $0.72^{+0.36}_{-0.24}$ & $1.26^{+0.39}_{-0.30}$ & $0.59^{+0.26}_{-0.18}$ & $1.05^{+0.26}_{-0.21}$ \\
501 & 348.873 & -53.063 & 0.333 & $3.22^{+0.27}_{-0.27}$ & $33.10\pm3.40$ & $1.97^{+0.86}_{-0.60}$ & $2.05^{+0.57}_{-0.45}$ & $1.75^{+0.73}_{-0.52}$ & $1.84^{+0.43}_{-0.35}$ \\
502 & 348.442 & -53.438 & 0.141 & $2.0^{+0.14}_{-0.14}$ & $8.15\pm0.53$ & $1.01^{+0.50}_{-0.34}$ & $1.15^{+0.34}_{-0.26}$ & $0.83^{+0.39}_{-0.26}$ & $0.95^{+0.23}_{-0.18}$ \\
503 & 350.646 & -52.747 & 0.3356 & $3.14^{+0.26}_{-0.26}$ & $31.10\pm2.61$ & $1.92^{+0.83}_{-0.58}$ & $1.98^{+0.54}_{-0.43}$ & $1.70^{+0.71}_{-0.50}$ & $1.78^{+0.41}_{-0.33}$ \\
504 & 351.930 & -52.425 & 0.243 & $2.33^{+0.24}_{-0.24}$ & $13.10\pm2.57$ & $1.29^{+0.56}_{-0.39}$ & $1.36^{+0.39}_{-0.30}$ & $1.11^{+0.46}_{-0.33}$ & $1.16^{+0.28}_{-0.23}$ \\
505 & 352.250 & -52.238 & 0.0552 & $1.71^{+0.12}_{-0.12}$ & $5.02\pm0.32$ & $0.79^{+0.39}_{-0.26}$ & $0.92^{+0.29}_{-0.22}$ & $0.62^{+0.29}_{-0.20}$ & $0.76^{+0.19}_{-0.15}$ \\
506 & 352.315 & -52.497 & 0.7167 & $4.05^{+0.61}_{-0.61}$ & $83.40\pm16.90$ & $2.49^{+1.19}_{-0.81}$ & $2.46^{+0.88}_{-0.65}$ & $2.39^{+0.90}_{-0.65}$ & $2.37^{+0.59}_{-0.47}$ \\
507 & 353.374 & -52.252 & 0.566 & $3.36^{+0.41}_{-0.41}$ & $45.00\pm7.31$ & $2.11^{+0.93}_{-0.65}$ & $2.02^{+0.65}_{-0.49}$ & $1.95^{+0.71}_{-0.52}$ & $1.88^{+0.44}_{-0.36}$ \\
509 & 356.461 & -54.044 & 0.6329 & $3.86^{+0.48}_{-0.48}$ & $68.50\pm7.94$ & $2.42^{+1.11}_{-0.76}$ & $2.36^{+0.81}_{-0.60}$ & $2.29^{+0.84}_{-0.62}$ & $2.26^{+0.52}_{-0.42}$ \\
510 & 357.539 & -55.334 & 0.3945 & $2.94^{+0.26}_{-0.26}$ & $27.20\pm2.18$ & $1.80^{+0.74}_{-0.53}$ & $1.80^{+0.51}_{-0.40}$ & $1.62^{+0.62}_{-0.45}$ & $1.61^{+0.36}_{-0.30}$ \\
511 & 357.753 & -55.371 & 0.13 & $1.62^{+0.12}_{-0.12}$ & $4.62\pm0.35$ & $0.77^{+0.37}_{-0.25}$ & $0.84^{+0.27}_{-0.20}$ & $0.64^{+0.27}_{-0.19}$ & $0.69^{+0.17}_{-0.13}$ \\
512 & 352.484 & -56.136 & 0.402 & $2.83^{+0.25}_{-0.25}$ & $24.90\pm1.70$ & $1.73^{+0.70}_{-0.50}$ & $1.73^{+0.49}_{-0.38}$ & $1.55^{+0.59}_{-0.42}$ & $1.55^{+0.35}_{-0.28}$ \\
513 & 349.221 & -54.902 & 0.3778 & $4.28^{+0.37}_{-0.37}$ & $72.70\pm4.49$ & $2.77^{+1.30}_{-0.88}$ & $2.97^{+0.90}_{-0.69}$ & $2.52^{+1.10}_{-0.77}$ & $2.71^{+0.64}_{-0.52}$ \\
514 & 351.396 & -54.722 & 0.1693 & $1.88^{+0.14}_{-0.14}$ & $7.00\pm0.54$ & $0.95^{+0.44}_{-0.30}$ & $1.03^{+0.31}_{-0.24}$ & $0.79^{+0.34}_{-0.24}$ & $0.86^{+0.20}_{-0.16}$ \\
515 & 351.416 & -54.741 & 0.1006 & $1.65^{+0.12}_{-0.12}$ & $4.73\pm0.30$ & $0.74^{+0.41}_{-0.26}$ & $0.89^{+0.28}_{-0.21}$ & $0.61^{+0.31}_{-0.20}$ & $0.72^{+0.17}_{-0.13}$ \\
517 & 350.449 & -55.971 & 0.6989 & $3.81^{+0.53}_{-0.53}$ & $70.20\pm10.40$ & $2.35^{+1.10}_{-0.75}$ & $2.29^{+0.80}_{-0.59}$ & $2.25^{+0.84}_{-0.61}$ & $2.18^{+0.52}_{-0.42}$ \\
518 & 349.822 & -55.325 & 0.1767 & $2.06^{+0.14}_{-0.14}$ & $9.01\pm0.54$ & $1.08^{+0.49}_{-0.34}$ & $1.16^{+0.34}_{-0.26}$ & $0.91^{+0.39}_{-0.27}$ & $0.99^{+0.23}_{-0.18}$ \\
519 & 353.019 & -55.212 & 0.27 & $2.23^{+0.2}_{-0.2}$ & $11.90\pm1.73$ & $1.25^{+0.53}_{-0.37}$ & $1.28^{+0.37}_{-0.29}$ & $1.08^{+0.43}_{-0.31}$ & $1.10^{+0.26}_{-0.21}$ \\
520 & 352.502 & -54.619 & 0.1752 & $2.94^{+0.19}_{-0.19}$ & $23.10\pm0.80$ & $1.73^{+0.85}_{-0.57}$ & $1.92^{+0.56}_{-0.43}$ & $1.47^{+0.71}_{-0.48}$ & $1.65^{+0.38}_{-0.31}$ \\
521 & 352.179 & -55.567 & 0.8072 & $5.16^{+0.78}_{-0.78}$ & $171.00\pm17.90$ & $3.00^{+1.65}_{-1.07}$ & $3.26^{+1.37}_{-0.97}$ & $3.03^{+1.17}_{-0.84}$ & $3.24^{+0.74}_{-0.60}$ \\
522 & 351.638 & -55.022 & 0.3948 & $2.89^{+0.25}_{-0.25}$ & $26.10\pm1.45$ & $1.77^{+0.72}_{-0.51}$ & $1.77^{+0.50}_{-0.39}$ & $1.59^{+0.60}_{-0.44}$ & $1.58^{+0.35}_{-0.29}$ \\
523 & 350.503 & -54.750 & 0.3429 & $3.0^{+0.25}_{-0.25}$ & $27.60\pm2.41$ & $1.82^{+0.77}_{-0.54}$ & $1.88^{+0.52}_{-0.41}$ & $1.60^{+0.66}_{-0.47}$ & $1.67^{+0.39}_{-0.31}$ \\
524 & 353.067 & -54.702 & 0.2701 & $2.21^{+0.18}_{-0.18}$ & $11.60\pm1.16$ & $1.24^{+0.52}_{-0.37}$ & $1.27^{+0.36}_{-0.28}$ & $1.07^{+0.42}_{-0.30}$ & $1.09^{+0.25}_{-0.21}$ \\
525 & 349.339 & -53.962 & 0.3793 & $4.51^{+0.39}_{-0.39}$ & $84.00\pm3.56$ & $2.94^{+1.40}_{-0.95}$ & $3.19^{+0.94}_{-0.72}$ & $2.67^{+1.19}_{-0.82}$ & $2.93^{+0.72}_{-0.58}$ \\
528 & 349.682 & -56.204 & 0.3017 & $2.54^{+0.2}_{-0.2}$ & $17.20\pm1.26$ & $1.48^{+0.62}_{-0.44}$ & $1.53^{+0.44}_{-0.34}$ & $1.28^{+0.50}_{-0.36}$ & $1.33^{+0.31}_{-0.25}$ \\
529 & 349.699 & -56.287 & 0.5474 & $3.84^{+0.42}_{-0.42}$ & $62.60\pm5.07$ & $2.44^{+1.10}_{-0.76}$ & $2.47^{+0.79}_{-0.60}$ & $2.24^{+0.85}_{-0.62}$ & $2.26^{+0.52}_{-0.42}$ \\
530 & 348.833 & -54.345 & 0.2029 & $2.28^{+0.16}_{-0.16}$ & $12.00\pm0.81$ & $1.22^{+0.56}_{-0.39}$ & $1.37^{+0.39}_{-0.30}$ & $1.04^{+0.46}_{-0.32}$ & $1.16^{+0.27}_{-0.22}$ \\
531 & 349.876 & -56.649 & 0.3906 & $2.95^{+0.27}_{-0.27}$ & $27.50\pm3.16$ & $1.81^{+0.75}_{-0.53}$ & $1.81^{+0.51}_{-0.40}$ & $1.63^{+0.62}_{-0.45}$ & $1.62^{+0.37}_{-0.30}$ \\
532 & 352.948 & -52.669 & 0.3917 & $3.32^{+0.29}_{-0.29}$ & $37.50\pm2.79$ & $2.06^{+0.89}_{-0.62}$ & $2.12^{+0.63}_{-0.49}$ & $1.87^{+0.73}_{-0.52}$ & $1.92^{+0.44}_{-0.35}$ \\
533 & 351.712 & -52.694 & 0.1073 & $3.01^{+0.18}_{-0.18}$ & $23.50\pm0.56$ & $1.75^{+0.90}_{-0.59}$ & $2.04^{+0.55}_{-0.43}$ & $1.51^{+0.77}_{-0.51}$ & $1.77^{+0.39}_{-0.32}$ \\
534 & 350.105 & -53.359 & 0.8523 & $4.86^{+0.77}_{-0.77}$ & $152.00\pm17.50$ & $2.86^{+1.53}_{-1.00}$ & $2.99^{+1.25}_{-0.88}$ & $2.82^{+1.07}_{-0.78}$ & $2.93^{+0.70}_{-0.56}$ \\
535 & 351.554 & -53.317 & 0.1723 & $3.06^{+0.2}_{-0.2}$ & $25.60\pm1.17$ & $1.83^{+0.93}_{-0.62}$ & $2.02^{+0.57}_{-0.44}$ & $1.54^{+0.78}_{-0.52}$ & $1.75^{+0.40}_{-0.33}$ \\
536 & 351.557 & -53.374 & 0.1699 & $2.02^{+0.15}_{-0.15}$ & $8.47\pm0.65$ & $1.04^{+0.49}_{-0.33}$ & $1.14^{+0.34}_{-0.26}$ & $0.87^{+0.38}_{-0.27}$ & $0.96^{+0.22}_{-0.18}$ \\
537 & 354.029 & -53.876 & 0.5148 & $4.0^{+0.42}_{-0.42}$ & $68.30\pm5.33$ & $2.53^{+1.13}_{-0.78}$ & $2.68^{+0.84}_{-0.64}$ & $2.34^{+0.92}_{-0.66}$ & $2.47^{+0.57}_{-0.47}$ \\
538 & 354.646 & -54.623 & 0.332 & $2.57^{+0.21}_{-0.21}$ & $18.10\pm1.57$ & $1.52^{+0.63}_{-0.45}$ & $1.51^{+0.43}_{-0.33}$ & $1.34^{+0.51}_{-0.37}$ & $1.33^{+0.30}_{-0.25}$ \\
539 & 355.797 & -55.881 & 0.1839 & $1.84^{+0.15}_{-0.15}$ & $6.66\pm0.85$ & $0.95^{+0.42}_{-0.29}$ & $0.99^{+0.30}_{-0.23}$ & $0.79^{+0.32}_{-0.23}$ & $0.83^{+0.20}_{-0.16}$ \\
540 & 355.632 & -56.353 & 0.4141 & $3.88^{+0.35}_{-0.35}$ & $57.90\pm3.71$ & $2.48^{+1.12}_{-0.77}$ & $2.65^{+0.79}_{-0.61}$ & $2.23^{+0.93}_{-0.66}$ & $2.42^{+0.56}_{-0.46}$ \\
541 & 355.431 & -55.965 & 0.1875 & $2.62^{+0.18}_{-0.18}$ & $17.10\pm0.98$ & $1.47^{+0.71}_{-0.48}$ & $1.64^{+0.46}_{-0.36}$ & $1.25^{+0.59}_{-0.40}$ & $1.40^{+0.32}_{-0.26}$ \\
542 & 353.113 & -53.976 & 0.4017 & $8.68^{+0.9}_{-0.9}$ & $484.00\pm12.50$ & $6.14^{+3.27}_{-2.13}$ & $7.92^{+2.29}_{-1.77}$ & $5.66^{+3.21}_{-2.05}$ & $8.14^{+2.48}_{-1.90}$ \\
543 & 354.863 & -55.843 & 0.3809 & $2.92^{+0.26}_{-0.26}$ & $26.60\pm2.24$ & $1.79^{+0.73}_{-0.52}$ & $1.78^{+0.50}_{-0.39}$ & $1.61^{+0.61}_{-0.44}$ & $1.59^{+0.36}_{-0.29}$ \\
544 & 349.816 & -53.534 & 0.0953 & $2.06^{+0.13}_{-0.13}$ & $8.44\pm0.32$ & $1.03^{+0.56}_{-0.36}$ & $1.22^{+0.38}_{-0.29}$ & $0.85^{+0.44}_{-0.29}$ & $1.03^{+0.25}_{-0.20}$ \\
546 & 352.416 & -53.249 & 0.792 & $4.53^{+0.67}_{-0.67}$ & $120.00\pm11.70$ & $2.69^{+1.39}_{-0.92}$ & $2.75^{+1.13}_{-0.80}$ & $2.66^{+1.01}_{-0.73}$ & $2.69^{+0.67}_{-0.54}$ \\
547 & 351.427 & -53.277 & 0.37138 & $3.31^{+0.29}_{-0.29}$ & $36.80\pm3.69$ & $2.04^{+0.89}_{-0.62}$ & $2.14^{+0.63}_{-0.49}$ & $1.84^{+0.74}_{-0.53}$ & $1.92^{+0.45}_{-0.36}$ \\
548 & 354.193 & -53.793 & 0.3212 & $2.01^{+0.2}_{-0.2}$ & $9.37\pm1.55$ & $1.15^{+0.47}_{-0.33}$ & $1.09^{+0.32}_{-0.25}$ & $1.00^{+0.37}_{-0.27}$ & $0.94^{+0.22}_{-0.18}$ \\
549 & 353.515 & -53.141 & 0.8085 & $4.13^{+0.63}_{-0.63}$ & $95.00\pm11.40$ & $2.48^{+1.25}_{-0.83}$ & $2.42^{+0.95}_{-0.68}$ & $2.42^{+0.91}_{-0.66}$ & $2.41^{+0.59}_{-0.48}$ \\
550 & 352.206 & -52.577 & 0.1087 & $1.45^{+0.18}_{-0.18}$ & $3.37\pm0.90$ & $0.62^{+0.33}_{-0.22}$ & $0.71^{+0.25}_{-0.19}$ & $0.50^{+0.24}_{-0.16}$ & $0.58^{+0.15}_{-0.12}$ \\
551 & 355.444 & -56.675 & 0.4749 & $3.34^{+0.38}_{-0.38}$ & $40.90\pm7.04$ & $2.10^{+0.89}_{-0.63}$ & $2.05^{+0.63}_{-0.48}$ & $1.91^{+0.71}_{-0.52}$ & $1.87^{+0.44}_{-0.35}$ \\
553 & 348.341 & -53.741 & 0.0464 & $0.62^{+0.11}_{-0.11}$ & $0.34\pm0.14$ & $0.23^{+0.09}_{-0.06}$ & $0.23^{+0.07}_{-0.05}$ & $0.19^{+0.05}_{-0.04}$ & $0.18^{+0.03}_{-0.03}$ \\
556 & 349.264 & -53.004 & 0.4887 & $2.78^{+0.3}_{-0.3}$ & $25.40\pm3.30$ & $1.75^{+0.73}_{-0.52}$ & $1.61^{+0.48}_{-0.37}$ & $1.57^{+0.57}_{-0.42}$ & $1.46^{+0.34}_{-0.27}$ \\
562 & 354.153 & -52.738 & 0.5356 & $2.78^{+0.38}_{-0.38}$ & $26.50\pm6.43$ & $1.75^{+0.74}_{-0.52}$ & $1.57^{+0.49}_{-0.38}$ & $1.59^{+0.59}_{-0.43}$ & $1.43^{+0.35}_{-0.28}$ \\
567 & 357.222 & -53.823 & 0.223 & $1.36^{+0.14}_{-0.14}$ & $3.08\pm0.57$ & $0.71^{+0.30}_{-0.21}$ & $0.63^{+0.20}_{-0.15}$ & $0.60^{+0.22}_{-0.16}$ & $0.53^{+0.13}_{-0.10}$ \\
573 & 353.623 & -54.608 & 0.3835 & $2.21^{+0.22}_{-0.22}$ & $12.60\pm1.76$ & $1.33^{+0.53}_{-0.38}$ & $1.22^{+0.34}_{-0.27}$ & $1.16^{+0.42}_{-0.31}$ & $1.07^{+0.25}_{-0.20}$ \\
577 & 351.386 & -55.740 & 0.2075 & $1.19^{+0.14}_{-0.14}$ & $2.16\pm0.49$ & $0.58^{+0.26}_{-0.18}$ & $0.54^{+0.19}_{-0.14}$ & $0.47^{+0.18}_{-0.13}$ & $0.44^{+0.11}_{-0.09}$ \\
581 & 352.416 & -54.789 & 0.1383 & $0.85^{+0.15}_{-0.15}$ & $0.84\pm0.35$ & $0.38^{+0.18}_{-0.12}$ & $0.34^{+0.13}_{-0.10}$ & $0.30^{+0.11}_{-0.08}$ & $0.27^{+0.07}_{-0.06}$ \\
583 & 352.040 & -55.838 & 0.8 & $3.86^{+0.57}_{-0.57}$ & $79.00\pm8.73$ & $2.32^{+1.15}_{-0.77}$ & $2.20^{+0.85}_{-0.61}$ & $2.27^{+0.84}_{-0.62}$ & $2.18^{+0.51}_{-0.41}$ \\
584 & 350.542 & -55.419 & 0.348 & $1.82^{+0.19}_{-0.19}$ & $7.42\pm1.30$ & $1.06^{+0.42}_{-0.30}$ & $0.94^{+0.28}_{-0.21}$ & $0.91^{+0.33}_{-0.24}$ & $0.82^{+0.19}_{-0.15}$ \\
585 & 353.696 & -55.274 & 0.3243 & $1.85^{+0.19}_{-0.19}$ & $7.59\pm1.32$ & $1.07^{+0.43}_{-0.31}$ & $0.95^{+0.28}_{-0.22}$ & $0.93^{+0.34}_{-0.25}$ & $0.82^{+0.19}_{-0.15}$ \\
586 & 351.848 & -55.066 & 0.2092 & $1.59^{+0.14}_{-0.14}$ & $4.67\pm0.54$ & $0.80^{+0.35}_{-0.24}$ & $0.82^{+0.25}_{-0.19}$ & $0.66^{+0.26}_{-0.19}$ & $0.68^{+0.16}_{-0.13}$ \\
587 & 349.935 & -54.640 & 0.5248 & $2.78^{+0.3}_{-0.3}$ & $26.10\pm2.56$ & $1.75^{+0.73}_{-0.51}$ & $1.57^{+0.48}_{-0.37}$ & $1.60^{+0.58}_{-0.43}$ & $1.44^{+0.32}_{-0.27}$ \\
588 & 352.659 & -55.727 & 0.2731 & $1.65^{+0.18}_{-0.18}$ & $5.39\pm1.06$ & $0.93^{+0.38}_{-0.27}$ & $0.83^{+0.25}_{-0.19}$ & $0.78^{+0.29}_{-0.21}$ & $0.70^{+0.17}_{-0.13}$ \\
592 & 349.594 & -55.833 & 0.237 & $1.92^{+0.15}_{-0.15}$ & $7.81\pm0.71$ & $1.04^{+0.44}_{-0.31}$ & $1.04^{+0.30}_{-0.24}$ & $0.88^{+0.34}_{-0.24}$ & $0.89^{+0.21}_{-0.17}$ \\
593 & 349.067 & -54.249 & 0.3316 & $1.87^{+0.18}_{-0.18}$ & $7.86\pm1.08$ & $1.09^{+0.43}_{-0.31}$ & $0.97^{+0.28}_{-0.22}$ & $0.94^{+0.34}_{-0.25}$ & $0.84^{+0.19}_{-0.16}$ \\
594 & 349.155 & -56.236 & 0.2319 & $1.31^{+0.15}_{-0.15}$ & $2.85\pm0.59$ & $0.69^{+0.29}_{-0.20}$ & $0.61^{+0.20}_{-0.15}$ & $0.57^{+0.21}_{-0.15}$ & $0.51^{+0.12}_{-0.10}$ \\
595 & 351.691 & -53.812 & 0.2132 & $1.7^{+0.14}_{-0.14}$ & $5.59\pm0.56$ & $0.88^{+0.38}_{-0.27}$ & $0.88^{+0.27}_{-0.21}$ & $0.74^{+0.29}_{-0.21}$ & $0.74^{+0.17}_{-0.14}$ \\
596 & 352.678 & -53.057 & 0.6775 & $3.3^{+0.45}_{-0.45}$ & $47.00\pm6.99$ & $2.06^{+0.95}_{-0.65}$ & $1.87^{+0.66}_{-0.49}$ & $1.96^{+0.72}_{-0.53}$ & $1.77^{+0.42}_{-0.34}$ \\
598 & 350.711 & -52.826 & 0.4436 & $2.36^{+0.27}_{-0.27}$ & $15.90\pm3.08$ & $1.45^{+0.59}_{-0.42}$ & $1.31^{+0.39}_{-0.30}$ & $1.29^{+0.47}_{-0.35}$ & $1.17^{+0.28}_{-0.22}$ \\
599 & 354.424 & -53.136 & 0.3146 & $2.0^{+0.18}_{-0.18}$ & $9.17\pm1.05$ & $1.15^{+0.46}_{-0.33}$ & $1.08^{+0.31}_{-0.24}$ & $0.99^{+0.37}_{-0.27}$ & $0.93^{+0.22}_{-0.18}$ \\
600 & 355.112 & -53.912 & 0.1676 & $1.26^{+0.15}_{-0.15}$ & $2.44\pm0.55$ & $0.59^{+0.28}_{-0.19}$ & $0.58^{+0.21}_{-0.15}$ & $0.48^{+0.19}_{-0.14}$ & $0.48^{+0.12}_{-0.09}$ \\
603 & 357.025 & -56.021 & 0.4264 & - & - & - & - & -& -\\
604 & 354.976 & -56.254 & 0.3806 & $1.95^{+0.21}_{-0.21}$ & $9.07\pm1.52$ & $1.18^{+0.47}_{-0.34}$ & $1.02^{+0.29}_{-0.23}$ & $1.03^{+0.37}_{-0.27}$ & $0.89^{+0.20}_{-0.17}$ \\
608 & 353.384 & -54.012 & 0.2118 & $1.68^{+0.19}_{-0.19}$ & $5.33\pm1.22$ & $0.85^{+0.38}_{-0.26}$ & $0.87^{+0.27}_{-0.21}$ & $0.70^{+0.28}_{-0.20}$ & $0.73^{+0.18}_{-0.14}$ \\
610 & 353.104 & -56.045 & 0.2752 & $1.75^{+0.16}_{-0.16}$ & $6.26\pm0.88$ & $0.98^{+0.40}_{-0.28}$ & $0.91^{+0.27}_{-0.21}$ & $0.84^{+0.31}_{-0.23}$ & $0.77^{+0.18}_{-0.14}$ \\
618 & 351.638 & -56.518 & 0.1699 & $1.48^{+0.13}_{-0.13}$ & $3.72\pm0.43$ & $0.71^{+0.33}_{-0.23}$ & $0.73^{+0.24}_{-0.18}$ & $0.59^{+0.24}_{-0.17}$ & $0.61^{+0.14}_{-0.12}$ \\
621 & 352.145 & -53.821 & 0.8701 & $3.7^{+0.6}_{-0.6}$ & $74.60\pm9.71$ & $2.21^{+1.17}_{-0.76}$ & $2.01^{+0.82}_{-0.58}$ & $2.18^{+0.83}_{-0.60}$ & $2.00^{+0.50}_{-0.40}$ \\
623 & 352.641 & -53.543 & 0.1713 & $1.16^{+0.14}_{-0.14}$ & $1.95\pm0.44$ & $0.56^{+0.25}_{-0.17}$ & $0.51^{+0.18}_{-0.13}$ & $0.46^{+0.17}_{-0.13}$ & $0.42^{+0.10}_{-0.08}$ \\
625 & 352.880 & -53.185 & 0.3336 & $1.69^{+0.22}_{-0.22}$ & $5.97\pm1.59$ & $0.98^{+0.40}_{-0.28}$ & $0.85^{+0.26}_{-0.20}$ & $0.85^{+0.31}_{-0.23}$ & $0.73^{+0.18}_{-0.14}$ \\
629 & 353.928 & -54.349 & 0.1727 & $0.88^{+0.19}_{-0.19}$ & $0.95\pm0.49$ & $0.45^{+0.21}_{-0.14}$ & $0.37^{+0.15}_{-0.11}$ & $0.37^{+0.15}_{-0.10}$ & $0.30^{+0.08}_{-0.06}$ \\
632 & 354.207 & -53.960 & 0.3222 & $1.47^{+0.22}_{-0.22}$ & $4.10\pm1.32$ & $0.85^{+0.35}_{-0.25}$ & $0.72^{+0.23}_{-0.17}$ & $0.72^{+0.27}_{-0.20}$ & $0.61^{+0.15}_{-0.12}$ \\
\end{longtable*}

\bibliography{sample631}{}
\bibliographystyle{aasjournal}



\end{document}